\documentclass[a4paper,onecolumn,11pt,accepted=2024-11-14]{quantumarticle}
\pdfoutput=1
\usepackage[utf8]{inputenc}
\usepackage[english]{babel}
\usepackage[T1]{fontenc}
\usepackage[numbers,sort&compress]{natbib}
\usepackage{url}

\usepackage{tikz}
\usepackage{graphicx}
\usepackage{amssymb,amsmath,amsthm}
\usepackage{hyperref}
\usepackage{accents,bm,bbm,braket,mleftright,mathtools,overpic,wasysym,xcolor,thmtools,thm-restate}
\usepackage{subcaption}
\usepackage[capitalize]{cleveref}

\graphicspath{{./figures/}}

\newtheorem{thm}{Theorem}
\newtheorem{lem}[thm]{Lemma}

\newtheorem{cor}[thm]{Corollary}

\newtheorem*{thm*}{Theorem}
\theoremstyle{definition}
\crefname{dfn}{Definition}{Definitions}
\crefname{exa}{Example}{Examples}

\DeclareMathOperator{\ghz}{GHZ}
\DeclareMathOperator{\tr}{tr}
\DeclareMathOperator{\epr}{EPR}
\DeclareMathOperator{\kag}{kagome}
\DeclareMathOperator{\fan}{fan}
\DeclareMathOperator{\trian}{tri}
\DeclareMathOperator{\GL}{GL}
\DeclareMathOperator{\stab}{Stab}
\DeclareMathOperator{\Span}{span}
\DeclareMathOperator{\poly}{poly}

\newcommand{\R}{\operatorname{R}}
\newcommand{\BR}{\underline{\operatorname{R}}}
\newcommand{\asR}{\undertilde{\operatorname{R}}}
\newcommand{\Q}{\operatorname{Q}}
\newcommand{\BQ}{\underline{\operatorname{Q}}}
\newcommand{\asQ}{\undertilde{\operatorname{Q}}}
\newcommand{\CR}{\operatorname{CR}}
\newcommand{\rk}{\operatorname{rk}}
\newcommand{\id}{\mathbbm{1}}
\newcommand{\CC}{\mathbb{C}}
\newcommand{\NN}{\mathbb{N}}
\newcommand{\ot}{\otimes}
\newcommand{\op}{\oplus}
\newcommand{\proj}[1]{\ket{#1}\!\bra{#1}}
\newcommand{\eps}{\varepsilon}
\newcommand{\geqdeg}{\trianglerighteq}
\newcommand{\geqas}{\gtrsim}

\newcommand{\bigO}{\mathcal O}
\newcommand{\tsp}{T}
\newcommand{\EPR}[1]{\epr_{#1}}

\renewcommand{\H}{\mathcal{H}}

\DeclarePairedDelimiter{\abs}{\lvert}{\rvert}
\DeclarePairedDelimiter{\ceil}{\lceil}{\rceil}
\DeclarePairedDelimiter{\floor}{\lfloor}{\rfloor}

\definecolor{blueish}{rgb}{0.0, 0.5, 0.69}
\definecolor{darkspringgreen}{rgb}{0.09, 0.45, 0.27}
\definecolor{mulberry}{rgb}{0.77, 0.29, 0.55}

\begin{document}

\title{The resource theory of tensor networks}

\author{Matthias Christandl}
\affiliation{Department of Mathematical Sciences, University of Copenhagen, Universitetsparken 5, 2100 Copenhagen, Denmark}
\author{Vladimir Lysikov}%
\affiliation{Faculty of Computer Science, Ruhr University Bochum, Universitätsstraße 150, 44801 Bochum, Germany}
\author{Vincent Steffan}%
\affiliation{Department of Mathematical Sciences, University of Copenhagen, Universitetsparken 5, 2100 Copenhagen, Denmark}
\author{Albert H. Werner}%
\affiliation{Department of Mathematical Sciences, University of Copenhagen, Universitetsparken 5, 2100 Copenhagen, Denmark}
\author{Freek Witteveen}%
\affiliation{Department of Mathematical Sciences, University of Copenhagen, Universitetsparken 5, 2100 Copenhagen, Denmark}
\affiliation{QuSoft, CWI, Science Park 123, 1098 XG Amsterdam, Netherlands}

\date{November 19 , 2024}

\maketitle

\begin{abstract}
    Tensor networks provide succinct representations of quantum many-body states and are an important computational tool for strongly correlated quantum systems. Their expressive and computational power is characterized by an underlying entanglement structure, on a lattice or more generally a (hyper)graph, with virtual entangled pairs or multipartite entangled states associated to (hyper)edges. Changing this underlying entanglement structure into another can lead to both theoretical and computational benefits.
    We study a natural resource theory which generalizes the notion of bond dimension to entanglement structures using multipartite entanglement.
    It is a direct extension of resource theories of tensors studied in the context of multipartite entanglement and algebraic complexity theory, allowing for the application of the sophisticated methods developed in these fields to tensor networks.
    The resource theory of tensor networks concerns both the local entanglement structure of a quantum many-body state and the (algebraic) complexity of tensor network contractions using this entanglement structure.
    We show that there are transformations between entanglement structures which go beyond edge-by-edge conversions, highlighting efficiency gains of our resource theory that mirror those obtained in the search for better matrix multiplication algorithms.
    We also provide obstructions to the existence of such transformations by extending a variety of methods originally developed in algebraic complexity theory for obtaining complexity lower bounds.
    The resource theory of tensor networks allows one to compare different entanglement structures and could lead to more efficient tensor network representations and contraction algorithms.
\end{abstract}

\section{Introduction}\label{sec:introduction}
What is the structure of quantum many-body states?
Physically relevant states, such as ground states of local Hamiltonians, typically have a very non-generic entanglement structure.
Indeed, such states often exhibit entanglement with a \emph{local} character, expressed by an area law for the entanglement entropy (as opposed to volume-law entanglement entropy for generic states) \cite{eisert2010colloquium}.
This observation has led to the Ansatz class of \emph{tensor network states} for representing quantum many-body states.
Such a tensor network state is created by first locally distributing states with bounded entanglement, and then applying local transformations.
Here, the amount of initial entanglement is captured by the \emph{bond dimension}.
Equivalently, the state is constructed by taking a collection of local tensors and contracting along a set of non-physical indices.
This encodes the global properties of the many-body state into local entangled states together with a local transformation.

Tensor network representations have become one of the main theoretical and numerical tools for understanding quantum many-body physics.
The first examples, now known as Matrix Product States (MPS), were discovered in the study of spin chains (in particular the AKLT model) as finitely correlated states \cite{fannes1992finitely}. Independently, around the same time White invented the Density Matrix Renormalization Group \cite{white1992density} as a numerical method, which in hindsight is a method to optimize over MPS.
Since its conception, tensor network research has developed these two complementary perspectives, one strand of research using tensor network states as a theoretical tool to construct interesting many-body states and to classify phases of matter, and another strand of research developing sophisticated numerical methods to simulate strongly interacting quantum many-body systems.

From a theoretical standpoint, tensor network states approximately parametrize ground states of local Hamiltonians.
Understanding phases of matter means that one would like to understand the set of ground states of local Hamiltonians under an appropriate equivalence relation.
Using sets of suitable tensor network states as a proxy for ground states allows one to reason more easily about phases of matter by reducing questions about the global quantum state to questions involving only the local tensors.
In one spatial dimension it is rigorously known that ground states of gapped Hamiltonians satisfy an area law \cite{hastings2007area,schuch2008entropy,landau2015polynomial} and can be approximated by MPS representations with bond dimensions growing polynomially with system size.
In two or more spatial dimensions it is widely believed that Projected Entangled Pair States (PEPS) are good ground state approximations \cite{verstraete2008matrix}, with area laws proven in special cases \cite{anshu2022area}.
This has amongst others been used to understand topological phases and symmetries in such ground states, since the global properties of the many-body state are encoded in the local tensors that make up the tensor network \cite{schuch2010peps,perez2010characterizing,pollmann2012symmetry,chen2013symmetry,chen2011two,chen2011classification,schuch2011classifying,ogata2021valued}, see also the reviews \cite{orus2019tensor,cirac2021matrix}.
Tensor networks are also a powerful numerical tool, since they hugely reduce the number of free parameters in the many-body state, allowing for variational methods for ground state approximation.
The ground state correlation functions, energies or other properties can then be extracted by tensor network contractions.
In one spatial dimension there exist rigorous polynomial time algorithms for finding ground state approximations using MPS with polynomial bond dimension for gapped Hamiltonians \cite{landau2015polynomial} and in practice DMRG provides an excellent simulation method for optimizing MPS representations \cite{white1992density,schollwock2005density}.
In two or more spatial dimensions, it is known that contraction of tensor networks is computationally hard \cite{schuch2007computational,haferkamp2020contracting}.
Nevertheless, there exist numerical methods for (approximately) contracting and optimizing tensor networks in two spatial dimensions \cite{jiang2008accurate,lubasch2014algorithms,banuls2023tensor} and these have been successfully applied to strongly interacting quantum systems \cite{zheng2017stripe,shi2022discovery}.

Tensor networks have been developed mostly in the context of condensed matter physics for the study of lattice systems.
However, they have found wide application in other many-body physics problems, for instance for simulating gauge theories and quantum field theories \cite{rico2014tensor,tilloy2019continuous}, as (toy) models for holographic quantum gravity \cite{pastawski2015holographic,hayden2016holographic,cheng2024random} and in quantum chemistry \cite{marti2010density,chan2011density,nakatani2013efficient,chen2022using}.
Besides this, tensor network methods can be used to simulate (small) quantum computers \cite{peng2020simulating,napp2022efficient,pan2022simulation,tindall2024efficient}.
The applications of tensor networks extend beyond quantum many-body physics: many mathematical and computational problems can be phrased in terms of tensors, and tensor networks provide general methods to decompose global tensors into a collection of local tensors.
A promising example is the use of tensor networks as a tool for machine learning \cite{stoudenmire2016supervised,cichocki2016tensor,cichocki2017tensor,stoudenmire2018learning,chen2018equivalence,huggins2019towards,liu2019machine,lu2021tensor,cheng2021supervised,pozas2022physics,vieijra2022generative} and graphical models \cite{robeva2019duality,glasser2019expressive,glasser2020probabilistic}.
Tensor networks also can encode counting problems (and therefore tensor network methods may be used for heuristic counting and optimization algorithms \cite{kourtis2019fast,liu2022computing}) and they are used in the design and decoding of quantum error correcting codes \cite{ferris2014tensor,farrelly2021tensor,chubb2021general}.

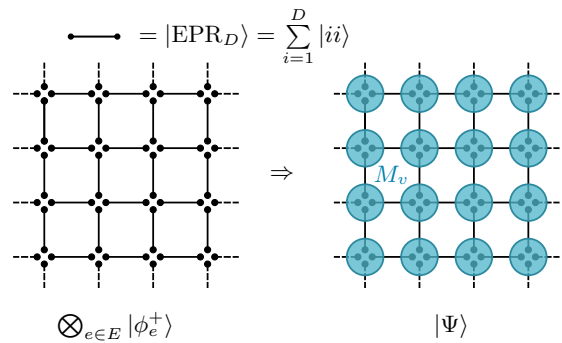
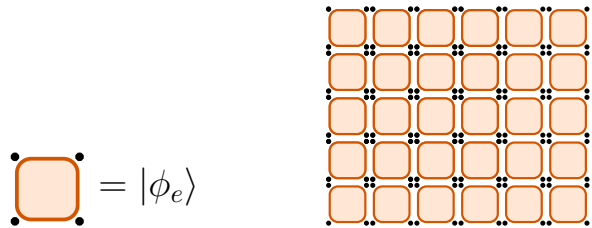
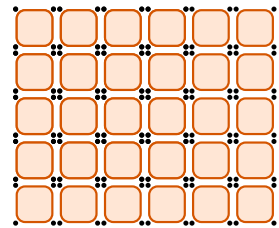
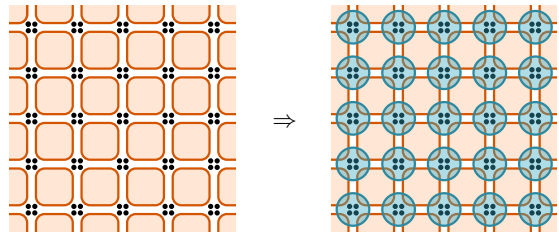
\begin{figure}
    \begin{subfigure}[t]{.55\linewidth}
        \begin{center}
            \begin{overpic}[width=.95\linewidth,grid=false]{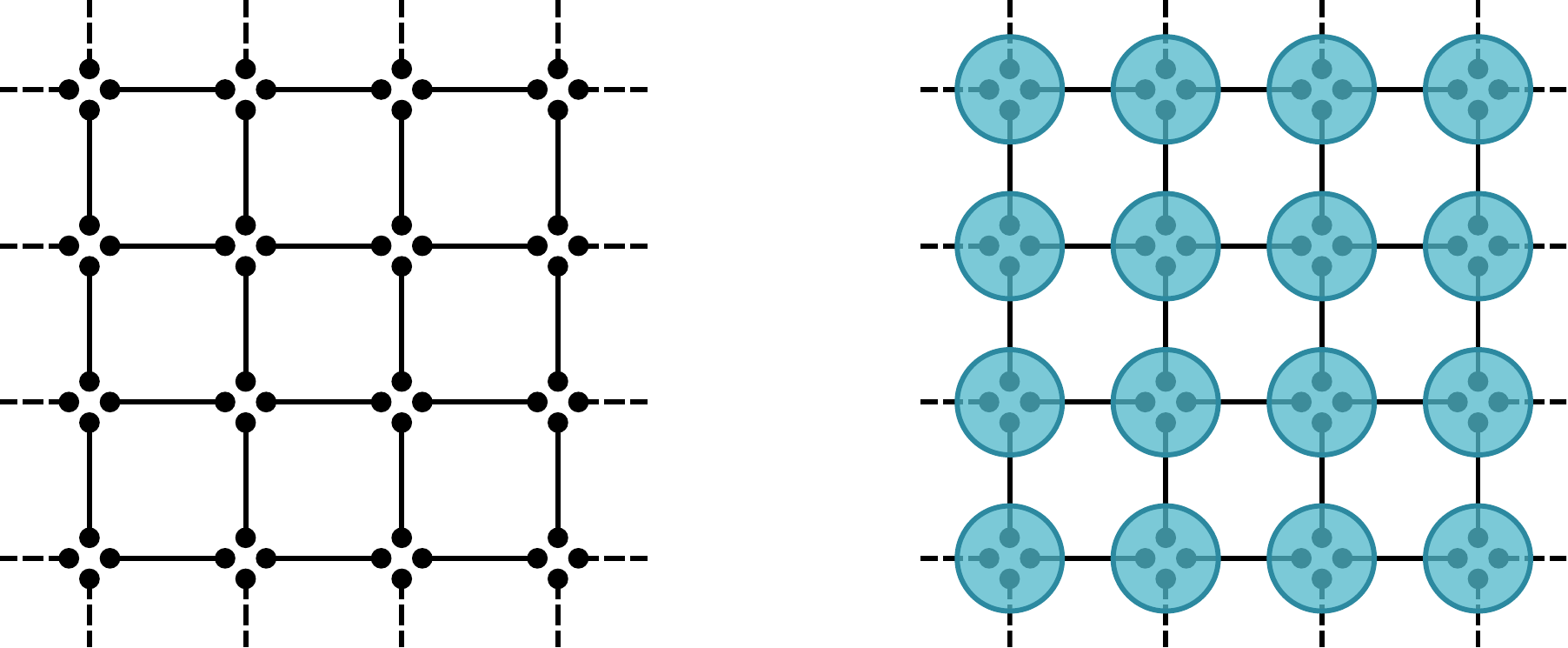}
                \put(47,20){$\Rightarrow$}
                \put(66,19.5){\small{$\color{blueish}{M_v}$}}
                \put(8,-8){$\bigotimes_{e \in E} \ket{\phi^+_{e}}$} \put(77,-8){$\ket{\Psi}$}
                \put(10,45.5){\includegraphics[width=.7cm]{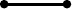}}
                \put(23,45){$ = \ket{\epr_D} = \sum\limits_{i=1}^D \ket{ii}$}
            \end{overpic}
            \vspace*{0.5cm}
        \end{center}
        \caption{A standard tensor network is built from $\epr$ pairs.}
        \label{fig:epr-plaquette}
    \end{subfigure}
    \begin{subfigure}[t]{.35\linewidth}
        \raggedright
        \begin{overpic}[width=0.25\linewidth]{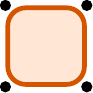}
            \put(120,40){\Large{$ = \ket{\phi_{e}}$}}
        \end{overpic}
        \vspace{0.5cm}
        \caption{Graphical notation for a plaquette state, here on four parties.}
        \label{fig:plaquette}
    \end{subfigure}%
    \vspace{0.5cm}
    \begin{subfigure}[t]{.4\linewidth}
        \begin{center}
            \includegraphics[width=0.6\linewidth]{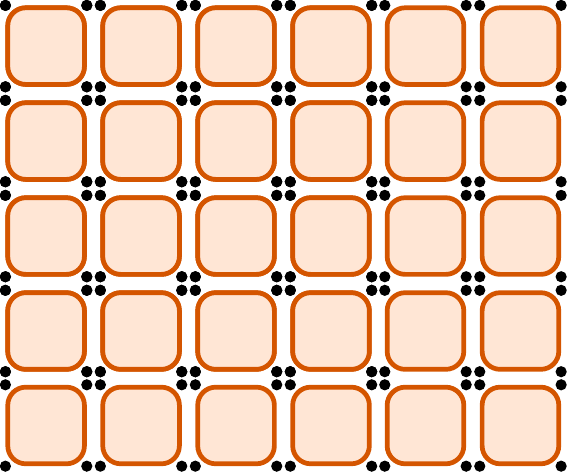}
        \end{center}
        \caption{A rectangular lattice with plaquette states on four parties.}
        \label{fig:rect-lattice}
    \end{subfigure}
    \hspace*{0.3cm}
    \begin{subfigure}[t]{.5\linewidth}
        \begin{center}
            \begin{overpic}[width=0.95\linewidth,grid=false]{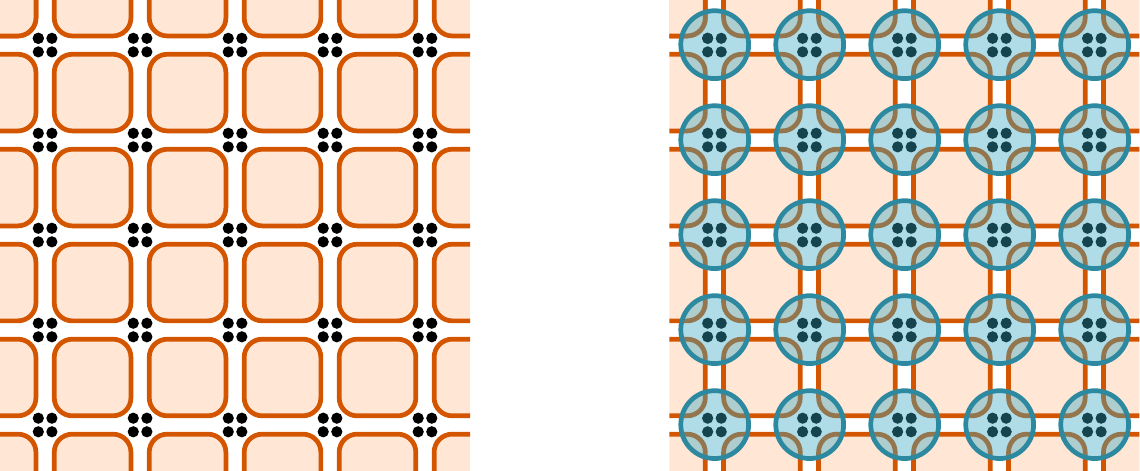}
                \put(48,19){$\Rightarrow$}
            \end{overpic}
        \end{center}
        \caption{A tensor network state constructed from an entanglement structure with four-party states.}
        \label{fig:ent structure tn}
    \end{subfigure}
    \caption{Graphical notation for entanglement structures. Standard tensor networks are constructed from an entanglement structure consisting of $\epr$ pairs of level $D$, as shown in \ref{fig:epr-plaquette}, corresponding to the special case of $2$-edges. In \ref{fig:plaquette} we illustrate a state on a single plaquette. We may then tile a lattice (or an arbitrary hypergraph) with such plaquette states to obtain an entanglement structure, as in \ref{fig:rect-lattice}. Parties which belong to the same node in the lattice are grouped together.
        From the entanglement structure one may construct a tensor network state by applying local linear maps, as shown in \ref{fig:ent structure tn}.}
    \label{fig:latticesvisualized}
\end{figure}

The essence of tensor network states is thus that they are quantum states exhibiting `local entanglement': they are obtained by applying local transformations to networks of bipartite entangled states. There are both theoretical and practical reasons, which we will review in \cref{sec:background}, for allowing a more general \emph{entanglement structure} based on local \emph{multipartite entanglement} \cite{christandl2020tensor}.
Indeed, the standard way to construct tensor network states is by placing maximally entangled pairs of dimension $D$ on the edges of a graph $G = (V,E)$ with vertex set $V$ and edge set $E$.
This is the origin of the nomenclature Projected Entangled Pair States (PEPS).
Typically, in the situation where we would like to simulate a condensed matter system, the graph will be a lattice.
The dimension $D$ of the maximally entangled states is the \emph{bond dimension} of the tensor network.
The actual tensor network state is now constructed by applying linear maps $M_v$ on each vertex of the graph.
The resulting state is given by
\begin{align}\label{eq:tensor network state}
    \ket{\Psi} = \left(\bigotimes_{v \in V} M_v\right) \bigotimes_{e \in E} \ket{\phi^+_{e}}
\end{align}
where $\ket{\phi^+_{e}} = \sum_{i=1}^D \ket{ii}$ is an (unnormalized) maximally entangled state, or \emph{EPR pair} at edge $e$.
This construction is illustrated in \cref{fig:epr-plaquette}.
One can think of the initial state $\bigotimes_{e \in E} \ket{\phi^+_{e}}$ as a resource for creating the state many-body state $\ket{\Psi}$: it is an \emph{entanglement structure} for $\ket{\Psi}$.

A natural generalization of \cref{eq:tensor network state} is to consider different local entanglement structures in this construction.
Here we are not restricted to only having states along edges, but we may also tile the lattice with states shared by more than two vertices.
Formally speaking, we may start with a \emph{hypergraph} $G = (V,E)$ where each (hyper)edge $e \in E$ is a subset of (possibly more than two) vertices in $V$.
Our main focus will be on two-dimensional lattices of `plaquettes' (we will use the terms plaquette and (hyper)edge interchangeably).

We then consider again states of the form
\begin{align*}
    \ket{\phi}_G = \bigotimes_{e \in E} \ket{\phi_{e}}
\end{align*}
as the entanglement structure, but now $\ket{\phi_e}$ is a $k$-party state if $e$ consists of $k$ vertices, visualized in \cref{fig:rect-lattice}.
For instance, we could take a rectangular lattice of plaquettes as depicted in \cref{fig:latticesvisualized} and tile the lattice with $\ghz$ states of level $r$
\begin{align*}
    \ket{\ghz_r} = \sum_{i=1}^r \ket{iiii}
\end{align*}
as a generalization of the usual maximally entangled states.
We then again obtain tensor network states by applying maps at each vertex as in \cref{eq:tensor network state}.
This is illustrated in \cref{fig:ent structure tn}.
We will provide a more precise definition of this construction in \cref{subsec:ent struc}.
A key feature is that for plaquettes containing more than two vertices there is more freedom of choice of the entanglement structure than in the usual tensor network approach.
In the usual tensor network approach the bond dimension is the only choice (as any other choice of state on the edge can be absorbed into the tensor).
This is not the case if the plaquette has more than two parties: in that case there are plenty of quantum states which are inequivalent under applying local linear maps. In particular, there might be various entanglement structures associated with the same lattice yielding a given target quantum state $\ket{\Psi}$.

In the usual PEPS picture of tensor network states, the parameters that determine how expressive the class of states is, is the set of bond dimensions on the edges.
Increasing the bond dimension allows one to represent a larger class of states.
In this work we study a \emph{resource theory} which allows one to compare different entanglement structures.

\begin{figure}[t]
    \centering
    \begin{overpic}[height=3.5cm,grid=false]{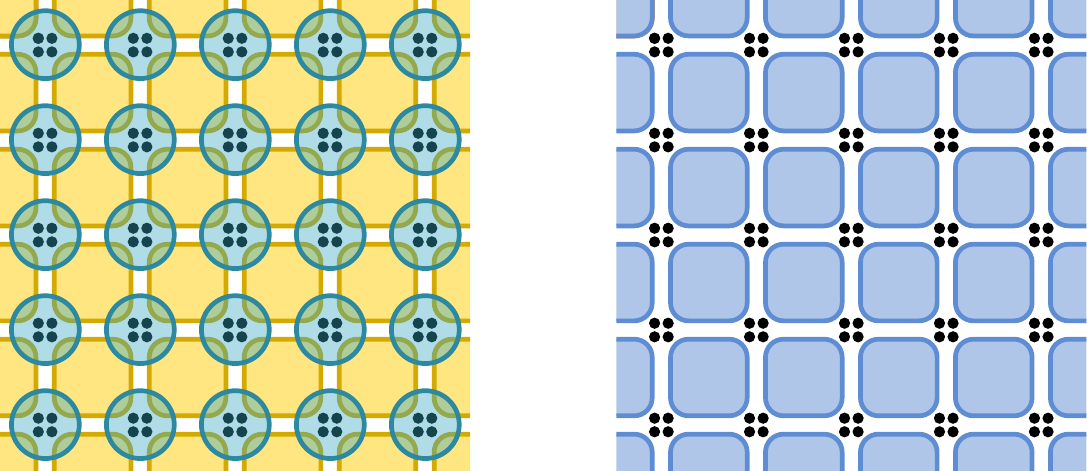}
        \put(4,-12){$\left( \bigotimes\limits_{v \in V} M_v \right) \ket{\phi}_G$}
        \put(47,21){$=$}
        \put(75,-12){$\ket{\psi}_G$}
    \end{overpic}
    \vspace{1cm}
    \caption{Relating different entanglement structures on a lattice through local transformations.}
    \label{fig:main question}
\end{figure}

Here, we will say that an entanglement structure $\ket{\phi}_G$ with states $\ket{\phi_e}$ on the edges is a stronger resource than an entanglement structure $\ket{\psi}_G$ which has states $\ket{\phi_e}$ on the edges if there exist local transformations at the vertices which map $\ket{\phi}_G$ to $\ket{\psi}_G$.
That is, there should exist linear maps $M_v$ at the vertices such that
\begin{align*}
    \left(\bigotimes_{v \in V} M_v\right) \bigotimes_{e \in E} \ket{\phi_{e}} = \bigotimes_{e \in E} \ket{\psi_e}
\end{align*}
as illustrated in \cref{fig:main question}.
In other words, the entanglement structure $\ket{\psi}_G$ can be written as a tensor network state using the resource $\ket{\phi}_G$, and it is clear that this implies that $\ket{\phi}_G$ is a more powerful resource than $\ket{\psi}_G$.
If there exists a local transformation on every single edge $e$ from $\ket{\phi_e}$ to $\ket{\psi_e}$ then it is clear that we can just apply these \emph{single-plaquette} transformations in parallel.
The main question we will address in this work is the following:
\begin{center}
    \emph{In the resource theory of tensor networks, what transformations between entanglement structures are possible that are not given by single-plaquette transformations?}
\end{center}

The question of how to transform tensors through local transformations has been extensively studied in the theory of multipartite entanglement as well as in algebraic complexity theory.
In complexity theory, finding certain transformations between tensors is closely related to finding faster algorithms for matrix multiplication.
In this context an extensive resource theory of tensors has been created.
One of the main goals of this work is to introduce certain powerful techniques, developed to better understand the complexity of matrix multiplication, to the theory of tensor networks.

As our first main result, we show that there indeed exist transformations of entanglement structures which go beyond single-plaquette restrictions on lattices.
That is, we construct examples where one can not transform $\ket{\phi_e}$ into $\ket{\psi_e}$, but once we place copies of these states on a lattice the transformation becomes possible.
There are examples of such transformations where EPR pairs are exchanged between different parties (and used as a resource).
A plausible intuition may be that since in a two-dimensional lattice the plaquettes are typically adjacent in at most two vertices, any lattice transformation can be performed by combining an exchange of EPR pairs with single-plaquette transformations.
We show that this is \emph{not} the case by giving an explicit example going beyond such transformations. This demonstrates the richness of the resource theory of entanglement structures.

On the other hand it is also important to be able to show that there do not exist transformations between different entanglement structures.
As our second main result we provide methods to find \emph{obstructions} for the existence of entanglement structure transformations.
The fact that there exist transformations between entanglement structures which go beyond single-plaquette transformations implies that it does not suffice to prove obstructions on the level of individual plaquettes.
We extend and apply various powerful methods from algebraic complexity theory to prove obstructions in prototypical examples.
As a first example we use \emph{flattening ranks} to prove nontrivial bounds for transforming entanglement structures with $\ghz$ states to entanglement structures using $\epr$ pairs.
Next, we use a version of the \emph{substitution method} to prove that for the $\lambda$-state on the kagome lattice (which is an entanglement structure for representing the Resonating Valence Bond state) there exists no representation of bond dimension two, even though there does exist an approximate representation of bond dimension two, answering one of the main open problems from \cite{christandl2020tensor}.
Finally, we study a class of asymptotic transformations where one has many copies of an entanglement structure.
Such transformations are characterized by the \emph{asymptotic spectrum} of tensors.
We show that certain points in this spectrum can be used as obstructions to entanglement structure transformations.

\subsection*{Organization of the paper}

We will provide background material on the relevance of entanglement structures beyond maximally entangled states in tensor network theory in \cref{sec:background}.
There we also provide an introduction to the resource theory of tensors.
We then introduce the resource theory of tensor networks, and in particular of entanglement structures, in \cref{sec:resource theory}.
We will discuss different types of local transformations between entanglement structure that can be considered.
We relate this resource theory to the (algebraic) complexity of tensor network contractions, observing along the way that this is a $VNP$-complete problem.
After having introduced the resource theory of tensor networks, we turn to the two main questions that can be answered in this resource theory.
Firstly, we provide a number of explicit transformation which reduce one entanglement structure to another in \cref{sec:constructions}.
Secondly, in \cref{sec:obstructions}, we study the converse question and give obstructions to the existence of transformations.
We present a number of general techniques for showing such obstructions and apply them to concrete examples.
In \cref{sec:sym and rank} we address structural questions relating to symmetries and ranks of entanglement structures.
We end with a summary and conclusion in \cref{sec:conclusion}.

\section{Background}\label{sec:background}

We will start by providing a detailed explanation of the notion of an entanglement structure and motivate its relevance for applications.
Then, since readers may not be familiar with the resource theory of tensors, we will give a brief introduction to this resource theory.
Finally, we give a concise overview of previous work which is relevant to the connection between the resource theory of tensors and entanglement structures for tensor networks.

\subsection{Entanglement structures}\label{subsec:ent struc}

We start by defining the notion of a tensor network and an entanglement structure on a hypergraph more carefully.
We start from some hypergraph $G$, which consists of a set of vertices, which we denote by $V$, and a set of hyperedges $E$.
Each hyperedge $e \in E$ consists of a subset of vertices, which are the vertices incident to $e$. In many examples we will assume that the cardinality $\abs{e}$ is a constant $k$, in which case we have a $k$-uniform hypergraph.
We will also assume that the vertices in any edge $e \in E$ are all different. The degree of a vertex is the number of edges $e$ such that $v \in e$.
We allow double edges (so strictly speaking $E$ is a multiset).
We consider Hilbert spaces of the form
\begin{align*}
    \H_G = \bigotimes_{e \in E} \bigotimes_{v \in e} \H_{e,v}
\end{align*}
where the $\H_{e,v}$ are Hilbert spaces. In other words, for each vertex $v \in V$ we have a Hilbert space for each edge incident to $v$.
We let
\begin{align*}
    \H_v = \bigotimes_{e : v \in e} \H_{e,v} \quad \text{ and } \quad \H_e = \bigotimes_{v \in e} \H_{e,v}
\end{align*}
be the Hilbert spaces at some fixed vertex or edge.

A tensor network state is now constructed from the following data: a collection of states $\ket{\phi_e} \in \H_e$ (not necessarily normalized) for $e \in E$ and a collection of linear maps
\begin{align*}
    M_v : \H_v \to \tilde \H_v \text{ for } v \in V.
\end{align*}
Here, $\tilde \H_v = \CC^{d_v}$ is the \emph{physical Hilbert space} and $d_v$ is the physical dimension at $v$.
Let
\begin{align*}
    \ket{\phi}_G = \bigotimes_{e \in E} \ket{\phi_e}.
\end{align*}
We call $\ket{\phi}_G$ an \emph{entanglement structure}.
In most examples the $\ket{\phi_e}$ are copies of the same $k$-party state.
We will sometimes assume that the tensors $\ket{\phi_e}$ are \emph{concise}, which means that the reduced density matrix on any single party has full rank.
While we have defined $\ket{\phi}_G$ as a tensor product over the edges, we can also regroup the Hilbert spaces along the vertices and think of $\ket{\phi}_G$ as a state in $\bigotimes_{v \in V} \H_v$.
We then get a tensor network state $\ket{\Psi}$ by applying the maps $M_v$ to each vertex
\begin{align}\label{eq:tensor network state 1}
    \ket{\Psi} = \left(\bigotimes_{v \in V} M_v\right)\ket{\phi}_G.
\end{align}
Tensor network states with entanglement structure $\ket{\phi}_G$ are precisely the states which are restrictions of the entanglement structure, where we have grouped according to the vertices $V$.
The usual version of tensor networks is the case where the hypergraph is a graph (so for each edge $e \in E$ we have $\abs{e} = 2$) and where we place (unnormalized) $\epr$ pairs $\ket{\phi_e} = \sum_{i = 1}^D \ket{ii} = \ket{\epr_D}$ on each edge.
Here $D$ is known as the \emph{bond dimension} of the tensor network.

\begin{figure}[t]
    \centering
    \begin{subfigure}[t]{.7\linewidth}
        \begin{overpic}[width=.6\linewidth,grid=false]{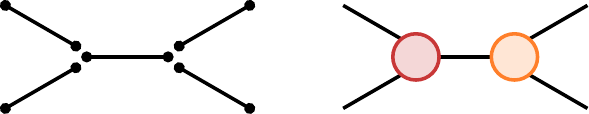}
            \put(110,8){$ \Rightarrow \sum\limits_{i = 1}^D T_{jki} S_{ilm}$}
            \put(60,21){\scalebox{0.8}{$j$}} \put(60,5){\scalebox{0.8}{$k$}} \put(78,11){\scalebox{0.8}{$i$}} \put(97,21){\scalebox{0.8}{$l$}} \put(97,5){\scalebox{0.8}{$m$}}
            \put(69,8){\small{$T$}} \put(85.5,8){\small{$S$}}
        \end{overpic}
        \vspace{0.3cm}
        \caption{Standard tensor contraction.}
        \label{fig:epr-contraction}
    \end{subfigure}
    \vspace{0.3cm}
    \begin{subfigure}[t]{.7\linewidth}
        \begin{overpic}[width=.6\linewidth,grid=false]{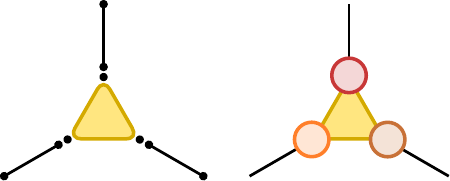}
            \put(110,20){$ \Rightarrow \sum\limits_{i = 1}^r S_{ij} T_{ik} U_{il}$}
            \put(57,5){\scalebox{0.8}{$j$}} \put(80,35){\scalebox{0.8}{$k$}} \put(77,12){\scalebox{0.8}{$i$}} \put(97,5){\scalebox{0.8}{$l$}}
            \put(76,22){\scalebox{0.8}{$T$}} \put(67.5,7.5){\scalebox{0.8}{$S$}} \put(84.5,7.5){\scalebox{0.8}{$U$}}
            \put(30,16){\scalebox{0.8}{$\ket{\ghz_r}$}}
        \end{overpic}
        \vspace{0.3cm}
        \caption{Contraction with a $\ghz$ state.}
        \label{fig:ghz-contraction}
    \end{subfigure}
    \begin{subfigure}[t]{.7\linewidth}
        \begin{overpic}[width=.6\linewidth,grid=false]{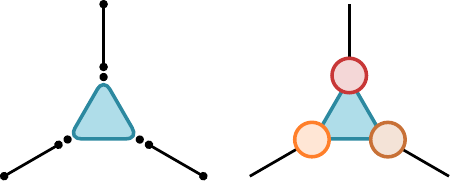}
            \put(100,20){$\Rightarrow  S_{1j} T_{0k} U_{0l} + S_{0j} T_{1k} U_{0l}$}
            \put(125,10){$+ S_{0j} T_{0k} U_{1l}$}
            \put(57,5){\scalebox{0.8}{$j$}} \put(80,35){\scalebox{0.8}{$k$}} \put(97,5){\scalebox{0.8}{$l$}}
            \put(76,22){\scalebox{0.8}{$T$}} \put(67.5,7.5){\scalebox{0.8}{$S$}} \put(84.5,7.5){\scalebox{0.8}{$U$}}
            \put(30,16){\scalebox{0.8}{$\ket{W}$}}
        \end{overpic}
        \caption{Contraction with a $W$ state.}
        \label{fig:w-contraction}
    \end{subfigure}
    \caption{Illustration of how to interpret different entanglement structures as tensor network contractions.}
    \label{fig:contractions}
\end{figure}
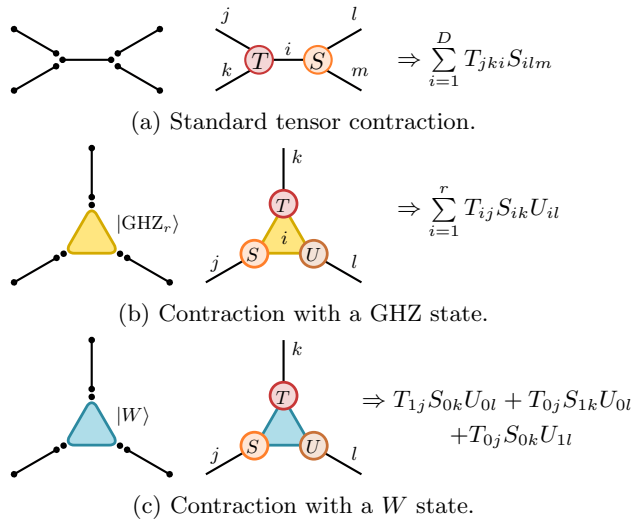

There are different perspectives on tensor network states.
Another perspective is that one takes a collection of tensors, and contracts along edges in a graph.
From the perspective of tensor network states as contractions of local tensors, we can also interpret different entanglement structures as different contraction rules, see \cref{fig:contractions}.

The idea to use entanglement structures different from maximally entangled states has first come up in the construction of concrete states \cite{rico20082d,chen2011two} and its theory has been developed more systematically in \cite{christandl2020tensor} (which introduced the terminology of entanglement structures) and \cite{molnar2018generalization}.
A basic observation is that given some entanglement structure on the lattice one can always represent the resulting tensor network states as tensor network states using only maximally entangled states, provided one allows sufficiently large bond dimension. For instance, suppose we have a tensor network state $\ket{\Psi}$ as given by \cref{eq:tensor network state 1}.
One can then take the graph $\tilde G$ with vertex set $\tilde V$ consisting of the union of $V$ and $E$ (so we introduce an additional vertex for each hyperedge), and with edge set $\tilde E$ consisting of all pairs $(e,v)$ for which $v \in e$.
If we take bond dimension $\dim(\H_{e,v})$ on the edge $(e,v) \in \tilde E$, and apply the transpose of $\ket{\phi_e}$ at each $e$  (with trivial physical Hilbert space at $e$) and $M_v$ at $v \in V$ we recover $\ket{\Psi}$.

There are various important reasons to study the generalization of the usual notion of tensor network states to more general entanglement structures.
The main reason to allow different entanglement structures is that this can lead to more efficient representations of tensor network states, with smaller bond dimensions (i.e. an entanglement structure with lower Hilbert space dimensions).
More efficient representations with minimal bond dimension are crucial for numerics in two or more spatial dimensions and developing the theory of entanglement structures beyond maximally entangled states could lead to improved numerical methods \cite{christandl2021optimization}.
The special case with 3-tensors on a kagome lattice has been proposed under the name Projected Entangled Simplex States (PESS) and one can extend PEPS optimization algorithms to this class of states, achieving superior approximation in frustrated lattice models with the appropriate entanglement structure, especially for spin liquids \cite{xie2014tensor,liao2017gapless,kshetrimayum2020tensor}.
In such methods, one fixes an entanglement structure and then optimizes variationally over the choice of local tensors.

A prominent theoretical model for spin liquid behavior is the Resonating Valence Bond (RVB) state and the closely related orthogonal dimer state, which is a superposition over dimers of the lattice, where a dimer is a subset of edges such that each edge is adjacent to exactly one edge in the dimer.
The state is then a uniform superposition over dimers, where singlet states are distributed on the edges in the dimer
\begin{center}
    \begin{overpic}[height=3.5cm,grid=false]{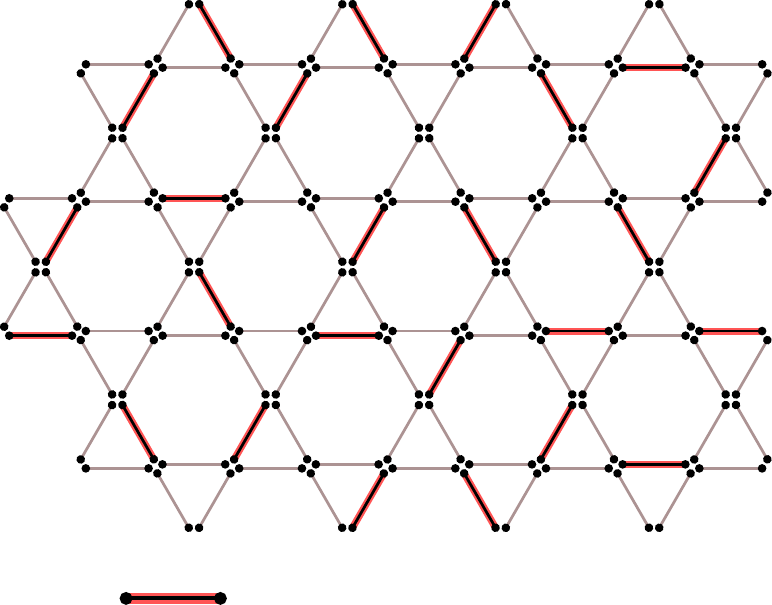}
        \put(-55,35){\large{$\ket{\Psi}$} \Large{$= \sum$}} \put(-28,24){\footnotesize{dimers}}
        \put(33,-1){\scalebox{0.8}{$= \ket{01} - \ket{10}$}}
    \end{overpic}
\end{center}
In this case it is most interesting to study this on a \emph{frustrated} lattice, such as the \emph{kagome lattice} above.
This state can be obtained from an entanglement structure placing a tensor $\ket{\lambda}$ at each plaquette
\begin{center}
    \hspace{-3cm}
    \begin{overpic}[height=3.5cm,grid=false]{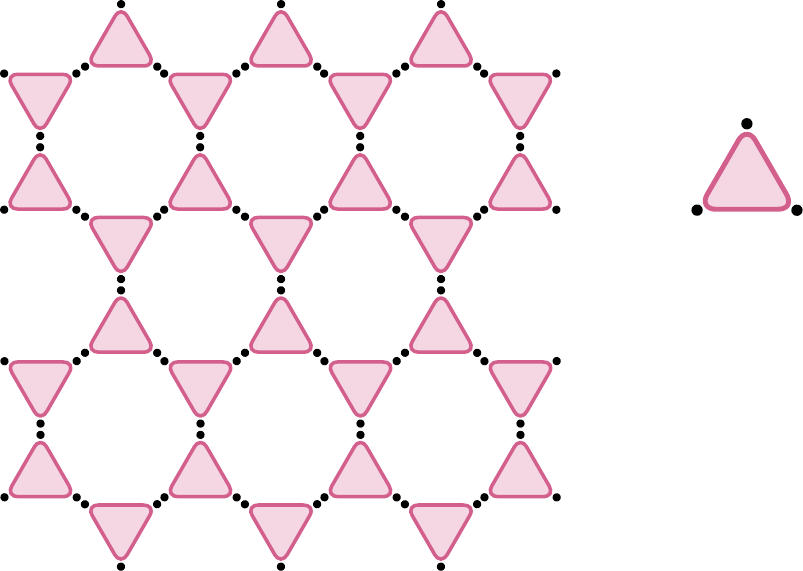}
        \put(91,46){\scalebox{0.9}{$\lambda$}}
        \put(80,28){$\ket{\lambda} = \sum\limits_{i,j,k = 0}^2 \epsilon_{ijk} \ket{ijk} + \ket{222}$}
    \end{overpic}
\end{center}
where $\epsilon_{ijk}$ is the antisymmetric tensor, as shown in \cite{schuch2012resonating}.
This perspective can be used to derive a PEPS representation of the orthogonal dimer state, and therefore of the RVB state, with reduced bond dimension \cite{christandl2020tensor}.
One should think of these examples as representing situations where the local entanglement structure of the many-body state is not accurately represented by pairwise entanglement between neighboring sites, but rather by some locally shared multipartite entanglement.

\begin{figure}[t]
    \centering
    \begin{subfigure}[t]{.45\linewidth}
        \begin{overpic}[width=.55\linewidth,grid=false]{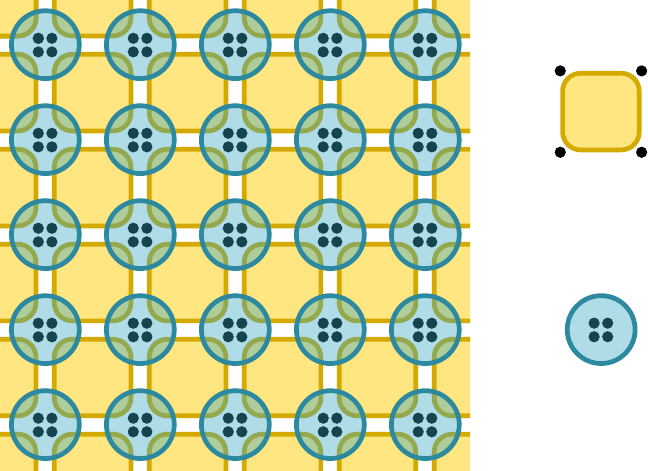}
            \put(102,53){$= \ket{\ghz_2(4)}$}
            \put(102,41){$= \ket{0000} + \ket{1111}$}
            \put(102,20){$= X^{\ot 4} \prod_{i,j} CZ_{ij}$}
        \end{overpic}
        \caption{An injective representation for the CZX model. Here $X$ is the Pauli $X$ matrix, and we act with controlled Pauli $Z$ operators on the pairs of qubits.}
        \label{fig:czx}
    \end{subfigure}
    \hspace*{0.5cm}
    \begin{subfigure}[t]{.45\linewidth}
        \begin{overpic}[width=.8\linewidth,grid=false]{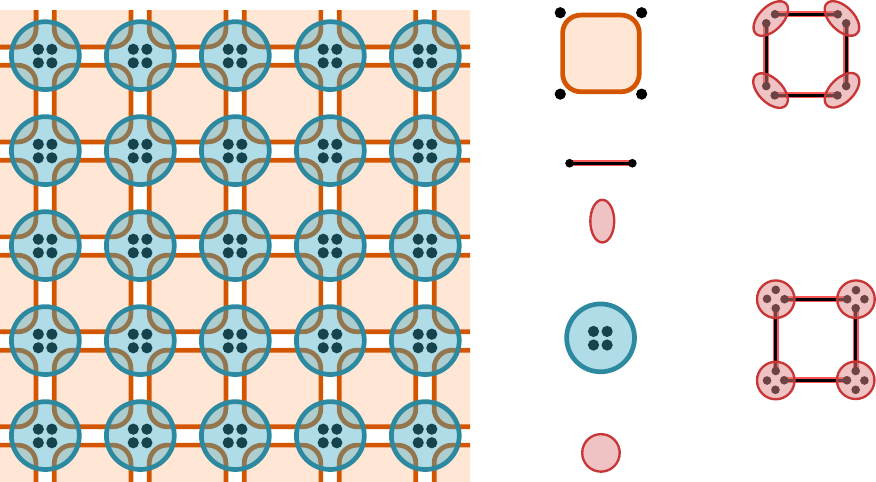}
            \put(78,47.5){$=$}
            \put(75,35){\scalebox{0.8}{$= \ket{01} - \ket{10}$}}
            \put(75,28){\scalebox{0.8}{$= \Pi_2$}}
            \put(78,15){$=$}
            \put(75,2){\scalebox{0.8}{$= \Pi_4$}}
        \end{overpic}
        \caption{An injective representation of the AKLT model. The operator $\Pi_n$ is the projector onto the symmetric subspace on $n$ qubits. Each party of an edge state consists of two qubits; the parties share singlet states to which one applies the projector $\Pi_2$.}
        \label{fig:aklt}
    \end{subfigure}
    \caption{Examples of states which are injective when using the right entanglement structure, see \cite{molnar2018generalization} for a derivation.}
    \label{fig:injective}
\end{figure}

Beyond this motivation there are also important theoretical reasons to study different entanglement structures.
A central notion in the study of tensor network states is that of \emph{injectivity}, which means that the state as constructed in \cref{eq:tensor network state} is with injective maps $M_v$.
In this case one can essentially invert the map $M_v$ and the properties of $\ket{\Psi}$ are very closely related to that of the entanglement structure.
Many theoretical results for tensor networks are only valid for injective tensor networks (or \emph{normal} tensor networks, which become injective upon grouping together multiple copies of the tensor).
For example, for an injective tensor network state one can always find a local Hamiltonian (the so-called \emph{parent Hamiltonian}) which has $\ket{\Psi}$ as its unique ground state.
There is a clear generalization of injectivity to states where we allow different entanglement structures.
This concept has been introduced in \cite{molnar2018generalization} as the class of states where one considers tensor network states with an arbitrary entanglement structure, and the maps $M_v$ are invertible (or more generally injective).
A number of physically important models have injective PEPS representations upon choosing the right entanglement structure, while they are not injective with respect to the standard construction using maximally entangled states.
This is important in the classification of two-dimensional symmetry protected topological phases \cite{molnar2018generalization,chen2011two}.
A first example is the CZX model \cite{chen2011two}, using GHZ states on a square lattice an applying controlled Pauli $Z$ operations as well as Pauli $X$.
Another prominent example is the AKLT model on the square lattice \cite{affleck1988valence}. Its ground state has an injective representation, illustrated in \cref{fig:injective}, by using an entanglement structure with a 4-tensor given by four singlet states where at each vertex one projects onto the symmetric subspace (in other words, it is the one-dimensional AKLT state on a periodic chain of length 4) \cite{molnar2018generalization}.

Finally, an especially natural example is using $\ghz$ states as entanglement structure: this simply corresponds to contracting multiple indices at the same time as illustrated in \cref{fig:contractions}. This is relevant for tensor network contractions in quantum circuits with controlled gates \cite{gray2021hyper}.
Entanglement structures using $\ghz$ states for tensor networks have also been used in relating tensor networks to graphical models \cite{robeva2019duality} and in studying random tensor networks \cite{walter2021hypergraph}.

\subsection{Resource theory of tensors}\label{subsec:tensors}
Tensors are widely studied in mathematics, physics and computer science.
Two domains where tensors are a central object of interest is in \emph{algebraic complexity theory} and \emph{quantum information theory}.
In both these fields one of the main questions is when and how two tensors can be transformed into one another using local operations.
Throughout this work we will identify tensors with quantum states shared between different parties where we do not necessarily normalize the quantum states, so a $k$-party quantum state is just a $k$-party tensor.
As an example of a transformation between quantum states, suppose we have three parties Alice, Bob and Charlie, and we have quantum states $\ket{\phi_{ABC}}$ and $\ket{\psi_{ABC}}$.
Then we can ask whether there exist linear maps $M_A$, $M_B$ and $M_C$ which Alice, Bob and Charlie can locally apply to convert the states, i.e. such that
\begin{align}\label{eq:slocc}
    \bigl(M_A \ot M_B \ot M_C\bigr) \ket{\phi_{ABC}} = \ket{\psi_{ABC}}.
\end{align}
In quantum information theory such transformations can be implemented using local operations and classical communication, if we additionally allow postselection on measurement outcomes and are therefore known as Stochastic Local Operations and Classical Communication (SLOCC).
This leads to a \emph{resource theory} of entanglement in multiparty quantum states, under the class of SLOCC operations.
For instance, entanglement with respect to SLOCC for three or four qubit systems has been completely classified \cite{dur2000three,verstraete2002four}.

The same resource theory has been studied in the context of theoretical computer science with the goal of understanding the complexity of matrix multiplication.
Here the resource theory, as introduced by Strassen \cite{strassen1987relative}, is formulated in terms of tensors rather than quantum states.
In this context, one says that $\ket{\phi}$ \emph{restricts} to $\ket{\psi}$ if there exists a transformation as in \cref{eq:slocc} (then called a \emph{restriction}).
More generally, if we have a collection of $k$ parties and
\begin{align*}
    \ket{\phi} \in \bigotimes_{i=1}^k \H_i, \qquad \ket{\psi} \in \bigotimes_{i=1}^k \H'_i
\end{align*}
then $\ket{\phi}$ restricts to $\ket{\psi}$ if there exist linear maps
\begin{align*}
    M_i : \H_i \to \H'_i
\end{align*}
such that
\begin{align}\label{eq:restriction}
    \left(\bigotimes_{i=1}^k M_i \right) \ket{\phi} = \ket{\psi}.
\end{align}
We write $\ket{\phi} \geq \ket{\psi}$ and this actually defines a partial order on the set of $k$-party states.
The interpretation justifying the $\geq$ sign is that $\ket{\phi}$ as a resource is at least as powerful as $\ket{\psi}$.

The resource theory of tensors turns out to have intimate connections to \emph{algebraic complexity theory}.
One of the most important outstanding open problems in computer science is to understand the computational complexity of matrix multiplication.
One would like to know how many \emph{multiplication operations} are required in order to multiply $n \times n$ matrices.
The naive algorithm uses $n^3$ multiplication operations.
However, a surprising realization by Strassen was that one can multiply two $2 \times 2$ matrices with only 7 (rather than 8) multiplications \cite{strassen1969gaussian}.
By recursively applying this construction one sees that asymptotically one can perform matrix multiplication with only $\bigO(n^\alpha)$ multiplications where $\alpha = \log_2(7) \approx 2.81$.
The study of the complexity of $n \times n$ matrix multiplication can be recast as a problem about tensors.

When computing $C = AB$ for $n \times n$ matrices $A$ and $B$, from
\begin{align*}
    C_{ij} = \sum_{k=1}^n A_{ik}B_{kj}
\end{align*}
we see that matrix multiplication is closely related to the tensor
\begin{align*}
    \ket{\EPR{n}}_{\triangle} = \sum_{i,j,k = 1}^n \ket{ik}_A \ket{jk}_B \ket{ij}_C
\end{align*}
which is a 3-party tensor where each pair shares an EPR pair (i.e. maximally entangled state) of dimension $n$ (in the algebraic complexity literature this is known as the \emph{matrix multiplication tensor} $\langle n, n, n\rangle$).
The precise relation is that one can show that every restriction of a 3-party GHZ state of $r$ levels
\begin{align*}
    \ket{\ghz_{r}(3)} = \sum_{i=1}^r \ket{i}_A\ket{i}_B\ket{i}_C
\end{align*}
to $\ket{\EPR{n}}_{\triangle}$ corresponds to an algorithm to perform $n \times n$ matrix multiplication with $r$ multiplication operations.

This motivates one to compute the \emph{rank} of a $k$-tensor $\ket{\phi}$ as
\begin{align*}
    \R(\ket{\phi}) = \min \left(r : \ket{\ghz_{r}(k)} \geq \ket{\phi} \right)
\end{align*}
where the GHZ state of level $r$ on $k$ parties is defined as
\begin{align*}
    \ket{\ghz_{r}(k)} = \sum_{i=1}^r \underbrace{\ket{i} \ot \dots \ot \ket{i}}_{k \text{ times }}.
\end{align*}
Equivalently, the rank $\R(\ket{\phi})$ is the minimal number of terms $r$ needed to write $\ket{\phi}$ as a sum of product states:
\begin{align}\label{eq:rank as product sum}
    \ket{\phi} = \sum_{i=1}^r \ket{e_{i,1}} \ot \dots \ot \ket{e_{i,k}}.
\end{align}
Indeed, if we have a restriction $\ket{\ghz_{r}(k)} \geq \ket{\phi}$ with restriction maps $M_j$, for $j = 1,\dots, k$, then we may take $\ket{e_{i,j}} = M_j \ket{i}$ (and vice versa one can define the $M_j$ from the decomposition \eqref{eq:rank as product sum}).
This definition is such that we can do $n \times n$ matrix multiplication using $\R(\ket{\EPR{n}}_{\triangle})$ multiplication operations.
For example, the insight of \cite{strassen1969gaussian} is that $\R(\ket{\EPR{2}}_{\triangle}) = 7$.

There is also an approximate version of restriction known as \emph{degeneration}.
If $\ket{\phi}$ and $\ket{\psi}$ are $k$-tensors, then $\ket{\phi} \geqdeg \ket{\psi}$ if there exist maps $M_i(\eps)$ for $i = 1,\dots, k$ continuously depending on a parameter $\eps$ such that
\begin{align}\label{eq:degeneration1}
    \lim_{\eps \to 0} \left(\bigotimes_{i=1}^k M_i(\eps) \right) \ket{\phi} = \ket{\psi}
\end{align}
so for each $\eps > 0$ we have a restriction, and its limit as $\eps$ goes to zero is the target tensor $\ket{\psi}$.
Accordingly, one may define the \emph{border rank}
\begin{align*}
    \BR(\ket{\phi}) = \min \left(r : \ket{\ghz_{r}(k)} \geqdeg \ket{\phi} \right).
\end{align*}
It turns out that if $\ket{\phi} \geqdeg \ket{\psi}$, there exist $T_i(\eps)$ which are polynomial in $\eps$ and a positive integer $d$ such that
\begin{align}\label{eq:degeneration}
    \left(\bigotimes_{i=1}^k T_i(\eps) \right) \ket{\phi} = \eps^d \ket{\psi} + \sum_{l=1}^e \eps^{d+l} \ket{\psi_l}
\end{align}
for some degree $e$ and tensors $\ket{\psi_l}$.
In this case, we will write $\ket{\phi} \geqdeg^e \ket{\psi}$ to indicate the degree.

Complexity theory motivates another type of transformations.
In complexity theory one is typically interested in the asymptotic behavior as the instance size grows.
For this reason one may investigate the \emph{asymptotic rank}
\begin{align*}
    \asR(\ket{\phi}) = \lim_{n \to \infty}  \R(\ket{\phi}^{\ot n})^{\frac{1}{n}}.
\end{align*}
If we let $\omega = \log_2(\asR(\ket{\EPR{2}}_{\triangle}))$ then the complexity of matrix multiplication is given by $\bigO(n^{\omega + o(1)})$.
The current best upper bound on the \emph{matrix multiplication exponent} $\omega$ is approximately $2.37$ \cite{alman2021refined,duan2023faster} while it is possible that $\omega = 2$ (which coincides with the best known, and trivial, lower bound).
The asymptotic rank is closely related to \emph{asymptotic restriction}; for tensors $\ket{\phi}$ and $\ket{\psi}$ we say that $\ket{\phi} \geqas \ket{\psi}$ if $\ket{\phi}^{\ot(n + o(n))} \geq \ket{\psi}^{\ot n}$ for all $n \in \NN$.
In other words, for all $n$ there exist maps $M_i^{(n)}$ acting on $n + f(n)$ copies of the Hilbert space of $\ket{\phi}$ for some $f(n) = o(n)$ such that
\begin{align}\label{eq:as restriction}
    \left(\bigotimes_{i=1}^k M_i^{(n)}\right) \ket{\phi}^{\ot(n + f(n))} = \ket{\psi}^{\ot n}.
\end{align}
Asymptotic restriction is also natural from the perspective of entanglement theory: it simply corresponds to \emph{asymptotic SLOCC} conversions!
In this perspective, $\log_2(\asR(\ket{\phi}))$ is the optimal rate at which one can convert level-2 $\ghz$ states to the state $\ket{\phi}$ using SLOCC.
Note that when we write $\ket{\phi}^{\ot n}$ in this context for a $k$-tensor $\ket{\phi}$, we really mean that we group together the $n$ copies of the $k$ systems into a single party so we consider $\ket{\phi}^{\ot n}$ as a $k$-tensor again \footnote{In the resource theory of tensors one typically gives this product its own name (the \emph{Kronecker product}) and its own symbol $\boxtimes$. We will be a bit loose in notation and not explicitly distinguish between tensor product and Kronecker product.}.
This is of course the usual way of thinking about asymptotics in quantum information theory.

How are the three different transformations (restriction, degeneration and asymptotic restriction, in equations \eqref{eq:restriction}, \eqref{eq:degeneration1} and \eqref{eq:as restriction} respectively) and the corresponding ranks related?
In \cref{eq:degeneration}, we see that the degeneration can be written as a polynomial in $\eps$, where the target tensor is the coefficient of the lowest degree term of order $\eps^d$.
The highest degree in the polynomial is $d + e$, so by evaluating the polynomial at $e + 1$ values of $\eps$ we can reconstruct the target tensor.
This polynomial interpolation idea can be used to show that from $\ket{\phi} \geqdeg^e \ket{\psi}$, we can get $\ket{\psi}$ from restriction of a \emph{direct sum} of $e + 1$ copies of $\ket{\phi}$ \cite{bini1980relations,christandl2018tensor}
\begin{center}
    \includegraphics[height=0.5cm]{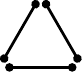}  \, \raisebox{0.2cm}{$\geqdeg$} \, \includegraphics[height=0.5cm]{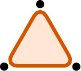} \quad \raisebox{0.2cm}{$\Rightarrow$} \quad \raisebox{0.15cm}{$\bigoplus\limits_{i=0}^e$} \, \includegraphics[height=0.5cm]{epr-triangle} \, \raisebox{0.2cm}{$\geq$} \, \includegraphics[height=0.5cm]{plaquette-triangle}
\end{center}
or equivalently
\begin{align*}
    \ket{\phi} \ot \ket{\ghz_{e+1}(k)} \geq \ket{\psi}.
\end{align*}
This can be used to show that $\ket{\phi} \geqdeg \ket{\psi}$ implies $\ket{\phi} \geqas \ket{\psi}$, and hence $\asR(\ket{\phi}) \leq \BR(\ket{\phi}) \leq \R(\ket{\phi})$.

The rank, asymptotic and border rank measure optimal conversions of GHZ states to a tensor of interest.
One can also study the converse direction, and define \emph{subrank} as
\begin{align*}
    \Q(\ket{\phi}) = \max \left(r : \ket{\phi} \geq \ket{\ghz_{r}(k)} \right),
\end{align*}
the largest GHZ state which can be extracted by SLOCC from $\ket{\phi}$.
Similarly, we may define asymptotic subrank as
\begin{align*}
    \asQ(\ket{\phi}) = \lim_{n \to \infty}  \Q(\ket{\phi}^{\ot n})^{\frac{1}{n}}
\end{align*}
and border subrank as
\begin{align*}
    \BQ(\ket{\phi}) = \left(r : \ket{\phi} \geqdeg \ket{\ghz_{r}(k)} \right).
\end{align*}
All these different notions of rank are related as
\begin{align*}
    \Q(\ket{\phi}) \leq \BQ(\ket{\phi}) \leq \asQ(\ket{\phi}) \leq \asR(\ket{\phi}) \leq \BR(\ket{\phi}) \leq \R(\ket{\phi}).
\end{align*}
While for $2$-tensors all these notions collapse to the same (standard) notion of rank, for $k \geq 3$ all inequalities can be strict.
In summary, understanding the computational complexity of matrix multiplication naturally leads to a \emph{resource} theory of tensors, involving the notions of \emph{restriction}, \emph{degeneration} and \emph{asymptotic restriction}. In algebraic complexity theory one typically either shows by some \emph{construction} that a restriction (or degeneration, or asymptotic restriction) exists, possibly leading to faster algorithms; or one shows that there are \emph{obstructions} to the existence of a restriction, which corresponds to lower bounds for certain algorithms.

See \cite{blaser2013fast} for an accessible introduction to algebraic complexity theory with a focus on the complexity of matrix multiplication, as well as the reference work \cite{burgisser2013algebraic} for rigorous definitions and proofs of the statements in this review section.

\subsection{Prior work}\label{subsec:prior work}
As alluded to in \cref{subsec:ent struc}, the idea of using entanglement structures beyond maximally entangled states has been explored in various works, both for exactly constructing interesting many-body ground states and for numerical purposes. While the general theory of entanglement structures has remained relatively underexplored, we will here highlight some relevant previous work.

The connection between tensor network states and the resource theory of tensors was first studied in \cite{christandl2020tensor}.
A first observation is that a tensor network state is precisely a restriction of an initial entanglement structure, by applying linear maps at all the vertices of the network.
As a demonstration of the applicability of the resource theory of tensors to tensor networks, the authors show that degenerations between plaquettes are a useful tool to get lower bond dimension representations of tensor network states.
By an interpolation argument, one can prove that if one has a degeneration $\ket{\phi} \geqdeg \ket{\psi}$ and one considers the entanglement structures $\ket{\phi}_G$ and $\ket{\psi}_G$ which have these states on each edge of some hypergraph $G$, then one can compute observables for any tensor network state $\ket{\Psi}$ constructed from the entanglement structure $\ket{\psi}_G$ from observables in $\bigO(\abs{E})$ tensor network states using $\ket{\phi}$ as entanglement structure.
For instance, given a degeneration on a single plaquette from an entanglement structure of level-$D$ $\epr$ pairs
\begin{center}
    \includegraphics[height=0.5cm]{epr-triangle}  \, \raisebox{0.2cm}{$\geqdeg$} \, \includegraphics[height=0.5cm]{plaquette-triangle}
\end{center}
this gives a representation using $\bigO(\abs{E})$ bond dimension $D$ tensor network states
\begin{center}
    \begin{overpic}[height=2.7cm,grid=trufalsee]{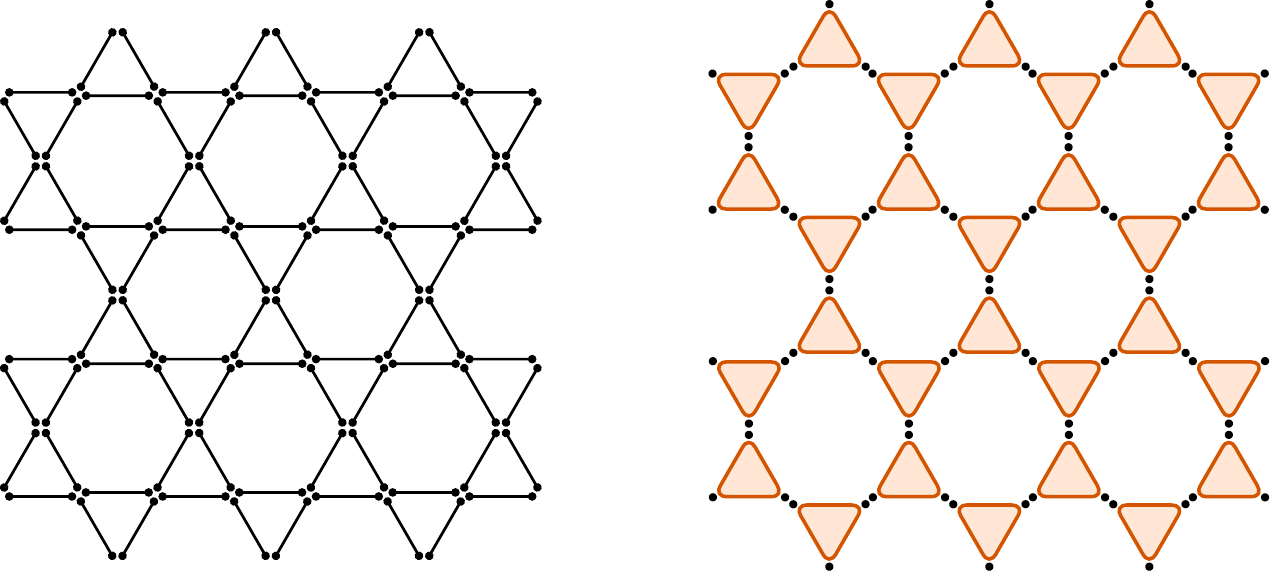}
        \put(48,20){$\geq$}
        \put(-20,21){$\bigoplus\limits_{\bigO(\abs{E})}$}
    \end{overpic}
\end{center}
Importantly, the overhead from the interpolation which turns the degeneration into a restriction (i.e. a proper tensor network state) is only \emph{linear} in the system size, while contraction algorithms for tensor networks typically scale in the bond dimension as $\bigO(D^m)$ where $m$ grows with the system size.
In other words, when computing tensor network observables, the potential savings in bond dimension from a plaquette degeneration are much more significant than the overhead from the interpolation.
Since the notion of degenerations is central in this work as well, we briefly review this result in \cref{sec:degenerations appendix}.
As an example of the techniques in \cite{christandl2020tensor}, the border subrank of $\ket{\EPR{n}}_{\triangle}$ gives rise to low bond dimension representations of entanglement structures of $\ket{\ghz_{r}(3)}$, by computing observables in a linear number (in the system size) of tensor network states.
Another application is that there is a degeneration $\ket{\epr_2}_{\triangle} \geqdeg \ket{\lambda}$, where $\ket{\lambda}$ is the tensor which can be used to construct the RVB state.
Therefore, one can compute expectation values for the RVB state using a linear number (in the system size) of tensor network states with bond dimension 2.

One can also investigate states which are the limit of tensor network states. In the language of tensors, these are states which are degenerations of an entanglement structure, an idea which has been explored in \cite{christandl2021optimization}.

Another systematic study of entanglement structures investigated entanglement structures on two-dimensional rectangular lattices \cite{molnar2018generalization}.
The results of this work apply to the translation invariant case with periodic boundary conditions.
The main results concern the question when two entanglement structures are \emph{equivalent}, i.e. they are related by an \emph{invertible} transformation $M_v$ on each of the vertices, moreover assuming that each $M_v$ is equal.
Under these assumptions, an invertible transformation between two entanglement structures exists for some lattice of size $n \times m$ for $n,m \geq 3$ if and only if it exists for all sizes.
Moreover, in that case the map $M_v$ can be taken to be a composition of maps only acting pairwise:

\begin{center}
    \begin{overpic}[height=3.5cm,grid=false]{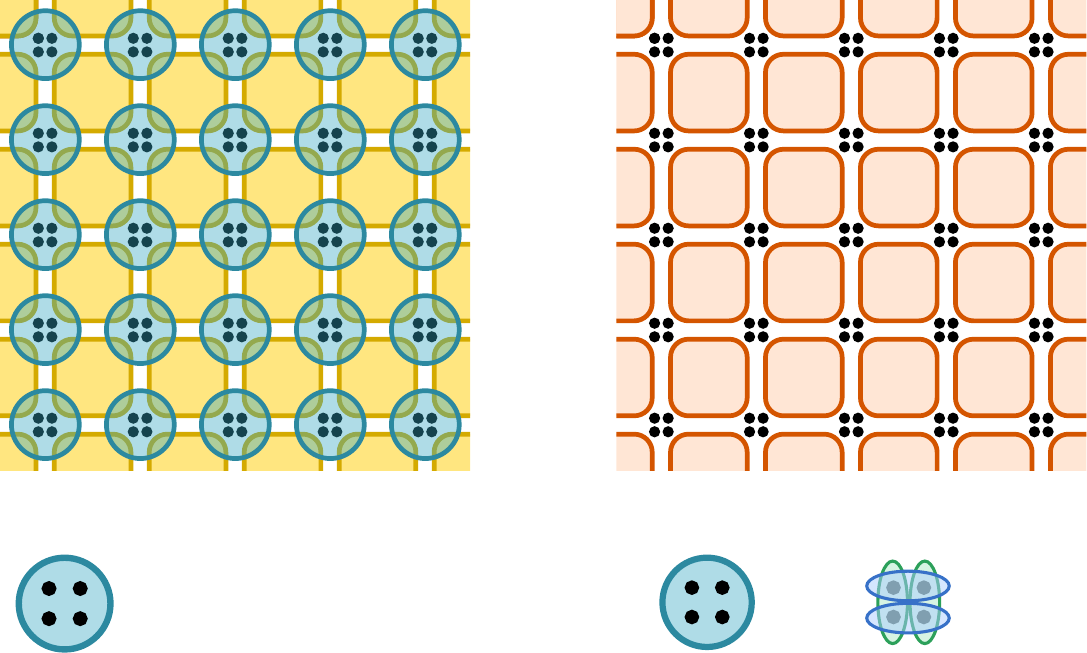}
        \put(47,37){$=$}
        \put(14,3){\small{invertible}} \put(46,3){$\Rightarrow$} \put(72,3){$=$}
    \end{overpic}
\end{center}

These results are then used to show that the classification of symmetry protected topological phases using group cohomology is also valid for injective tensor networks using translation invariant entanglement structures.

While the resource theory of entanglement structures on lattices and other hypergraphs have not been studied in generality, there are various known results in the study of tensors which are closely related.
There are a number of important results in the resource theory of tensors which can be formulated as computing tensor ranks and subranks of entanglement structures of hypergraphs.
These results provided evidence that the resource theory of tensor networks is highly nontrivial.
To make this concrete, consider the $W$ state
\begin{align*}
    \ket{W} = \ket{100} + \ket{010} + \ket{001}.
\end{align*}
It is known that it has tensor rank $\R(\ket{W}) = 3$, so there exists a restriction from $\ket{\ghz_3(3)}$
\begin{center}
    \begin{overpic}[height=1cm,grid=false]{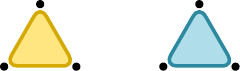}
        \put(45,11){\large{$\geq$}}
        \put(14,7.5){$3$} \put(77.5,6.5){$W$}
    \end{overpic}
\end{center}
and there does not exist such a restriction for $\ket{\ghz_2(3)}$ (see also \cref{sec:substitution}).
However, if we take two copies and take this again as a 3-tensor (i.e. we take the Kronecker product of two copies of $\ket{W}$, which is again a 3-party tensor), the resulting object only has rank $7$ rather than the naive $9 = 3^2$ \cite{yu2010tensor}.
In other words, the tensor rank is not multiplicative under the Kronecker product, a fact already observed in the context of matrix multiplication \cite{blaser2013fast}.
One can think of this situation as placing two copies of the $W$ state on top of each other
\begin{center}
    \begin{overpic}[height=1cm,grid=false]{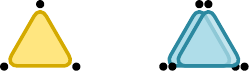}
        \put(45,11){\large{$\geq$}}
        \put(14,7){$7$} \put(79,6){$W$}
    \end{overpic}
\end{center}
We can also think of the following scenario: we place the two $W$ states on the completely disconnected hypergraph consisting of two 3-edges (i.e. this is the usual tensor product and we now have six parties).
It is known that this state has tensor rank 8 \cite{christandl2018tensor,chen2018tensor}, so there exists a restriction from a six-party $\ket{\ghz_8(6)}$ state
\begin{center}
    \begin{overpic}[height=1.2cm,grid=false]{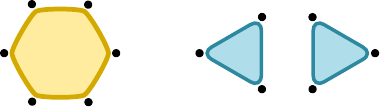}
        \put(40,11){\large{$\geq$}}
        \put(13,11){$8$} \put(59.5,11.5){\scalebox{0.9}{$W$}} \put(84,11.5){\scalebox{0.9}{$W$}}
    \end{overpic}
\end{center}
and the tensor rank is also not multiplicative for the tensor product.
Similar strict submultiplicativity under the tensor product is known for the border rank \cite{christandl2019border}.
The (asymptotic) tensor rank of other entanglement structures, especially for states which are EPR pairs distributed over a graph \cite{christandl2019asymptotic,christandl2019tensor}, have also been studied.
One can pose the converse question as well where one starts with some arbitrary multiparty tensor and tries to transform to a global $\ghz$ state. That is, one tries to compute the subrank.
This question first came up in \cite{strassen1988asymptotic}, lower bounding the border subrank $\BQ(\ket{\epr_n}_{\triangle})$.
For general hypergraphs with $\ghz$ states, the asymptotic subrank was determined in \cite{vrana2017entanglement}.

The picture that emerges is that understanding the tensor (sub)rank of entanglement structures is a rich subject which does \emph{not} reduce to the ranks of the individual edge states, suggesting that the resource theory of entanglement structures could be similarly nontrivial.
However, the above works do not study the resource theory of tensor networks, as the resource state (or target state) is a global $\ghz$ state.
We will rather consider scenarios where both the resource and the target state are a tensor product of $\ghz$ or other states over the edges of the hypergraph.
This resource theory has so far not been studied.

There is a number of other works studying tensor networks using tools and ideas from algebraic complexity theory.
For example, \cite{ye2018tensor} defines the $G$-rank of a tensor with respect to a graph $G$ as the minimal bond dimensions required to write the tensor as a tensor network state on the graph $G$.
This is in similar spirit as our work, but it does not take into account entanglement structures beyond maximally entangled states and does not formulate a resource theory of tensor networks.
We also mention that there is a line of works, inspired by algebraic complexity theory, studying sets of tensor network states as algebraic varieties \cite{landsberg2012geometry,
    bernardi2023dimension,delazzari2022linear,czaplinski2023uniform}.

\section{The resource theory of tensor networks}\label{sec:resource theory}
We proceed to define our object of study: the resource theory of tensor networks and entanglement structures.
This resource theory is the natural extension of the resource theory of tensors in algebraic complexity and the theory of entanglement under SLOCC transformations.
We will introduce the resource theory from the entanglement perspective, where we define tensor network states as quantum states arising from SLOCC transformations of strictly local networks of quantum states.
However, tensor networks are also a computational tool, and we show that the resource theory of tensor networks relates directly to the algebraic complexity of tensor network contraction.
Additionally, we investigate the computational complexity of tensor network contractions from the algebraic perspective and observe that tensor network contraction is a $VNP$-complete problem.

\subsection{The resource theory of tensor networks}\label{subsec:resource-tn}
It is clear that the formalism of tensor network states closely aligns with the notions of tensor restriction in \cref{subsec:tensors}.
Indeed, a tensor network state is nothing else than a restriction of a distribution of EPR pairs, where the parties are the vertices of the graph.
When we consider the order induced by restrictions (i.e. SLOCC transformations on the vertices), a state $\ket{\Psi}$ has a representation as a tensor network state using an entanglement structure $\ket{\phi}_G$ as in \cref{eq:tensor network state 1} if and only if $\ket{\phi}_G \geq \ket{\Psi}$.

We will use this perspective to compare different entanglement structures.
If we are given a lattice (or more generally some hypergraph) $G = (V,E)$, then $\ket{\phi}_G \geq \ket{\psi}_G$ if $\ket{\psi}_G$ has a tensor network representation using $\ket{\phi}_G$ as entanglement structure.
Concretely, $\ket{\phi}_G \geq \ket{\psi}_G$ if there exist maps $M_v$ on each of the vertices of $G$ such that
\begin{align*}
    \left(\bigotimes_{v \in V} M_v \right) \bigotimes_{e \in E} \ket{\phi_e} = \bigotimes_{e \in E} \ket{\psi_e}.
\end{align*}
If we have some many-body state $\ket{\Psi}$ with a tensor network representation by some entanglement structure $\ket{\psi}_G$, then the existence of a restriction $\ket{\phi}_G \geq \ket{\psi}_G$ directly implies that we also obtain a tensor network representation for $\ket{\Psi}$ using $\ket{\phi}_G$ as our initial entanglement structure:

\begin{equation*}
    \begin{tikzpicture}
        \draw[rounded corners, thick, color =black!10!white, fill = black!10!white] (0,0) rectangle (1.7,1);
        \node at (.9,.7) {physical};
        \node at (.9,.3) {state};
        \node at (2, .5){$\leq$};
        \begin{scope}[shift={(2.7,0)}]
            \draw[rounded corners, thick, color =black!10!white, fill = black!10!white] (-.35,-0.1) rectangle (2.2,1.1);
            \node at (.9,.9) {current};
            \node at (.9,.5) {entanglement};
            \node at (.9,.1) {structure};
            \node at (2.5, .5){$\leq$};
        \end{scope}
        \begin{scope}[shift={(5.9,0)}]
            \draw[rounded corners, thick, color =black!10!white, fill = black!10!white] (-.35,-0.1) rectangle (2.2,1.1);
            \node at (.9,.9) {desired};
            \node at (.9,.5) {entanglement};
            \node at (.9,.1) {structure};
        \end{scope}
    \end{tikzpicture}
\end{equation*}
For instance, if we have an \emph{injective} representation of a many-body state using GHZ states on each plaquette, the question of finding a minimal bond dimension representation of the state is equivalent to finding an optimal restriction from the entanglement structure which has EPR pairs along all the edges.

It is now clear that this defines a \emph{resource theory for entanglement structures}.
This is the resource theory of entanglement structures induced by SLOCC (in information theory terminology), or restrictions (in algebraic complexity theory terminology).
Note that while in quantum information theory the notion of SLOCC as the class of allowed operations for a theory of multipartite entanglement is not completely natural (operationally one would prefer LOCC, but this is too complicated to classify in practice), for tensor networks it is the natural class of allowed operations.

Based on the three different notions of tensor transformations (restrictions, degenerations, asymptotic degenerations) we can study different transformations of entanglement structures.
\begin{enumerate}
    \item Given two entanglement structures we can ask whether we can transform one into another on each individual \emph{plaquette}. Whether this is possible then can be formulated as asking whether there is a restriction $\ket{\phi_e} \geq \ket{\psi_e}$ for each edge \cite{christandl2020tensor}.
          We will call such transformations \emph{single-plaquette restrictions}.
    \item We can also ask whether on the plaquette level there exists a degeneration, so $\ket{\phi_e} \geqdeg \ket{\psi_e}$ for all $e \in E$.
          The consequences of the existence of both single-plaquette restrictions and degenerations have been investigated in \cite{christandl2020tensor}.
          Given single-plaquette degenerations, one can represent the tensor network state with only \emph{linear} overhead in the system size, see \cref{sec:degenerations appendix}.
    \item We know that there exist transformations of tensors which are only possible asymptotically. Motivated by this fact we can ask whether there exists a restriction on the global entanglement structure, i.e. $\ket{\phi}_G \geq \ket{\psi}_G$, or a global degeneration $\ket{\phi}_G \geqdeg \ket{\psi}_G$.
    \item Also motivated by asymptotic restrictions we can ask, given some entanglement structure $\ket{\phi}_G$ and a tensor network target state $\ket{\Psi}$, at what \emph{rate} we can produce copies of $\ket{\Psi}$ from copies of $\ket{\phi}_G$? More generally, given $\ket{\phi}_G^{\ot n}$, what is the optimal number $m$ such that $\ket{\phi}_G^{\ot n} \geq \ket{\Psi}^{\ot m}$?
\end{enumerate}
It is the third and fourth questions on which we focus on in this work, in order to develop a comprehensive resource theory of tensor networks, in particular of the underlying entanglement structures.
While in principle one can study restrictions, degenerations and asymptotic restrictions between arbitrary many-body states, from the perspective of tensor networks the main interest is the situation where we look for restrictions
\begin{align*}
    \ket{\phi}_G \geq \ket{\Psi},
\end{align*}
so we have a restriction from an entanglement structure (which means that $\ket{\Psi}$ is by definition a tensor network state).
The focus of our work is the situation where \emph{both} many-body states are entanglement structures, that is, we investigate restrictions
\begin{align*}
    \ket{\phi}_G \geq \ket{\psi}_G.
\end{align*}
The existence of such a restriction means that the class of tensor network states using the entanglement structure $\ket{\phi}_G$ encompasses the class of tensor network states using the entanglement structure $\ket{\psi}_G$, justifying the terminology of a resource theory.
Moreover, this encompasses the important class of injective tensor network states (which are by definition the states which are equivalent to an entanglement structure).
Similarly, we study this resource theory with respect to degenerations, so
\begin{align*}
    \ket{\phi}_G \geqdeg \ket{\psi}_G.
\end{align*}
This relation is less standard in the tensor network literature, but we emphasize that (as argued in \cite{christandl2020tensor}) if the degeneration has low degree, tensor network computations for $\ket{\psi}_G$ reduce to tensor network computations with $\ket{\phi}_G$ with only small overhead.
An important reason to consider degenerations is that they can allow for significant savings in bond dimensions and degenerations are in practice often easier to find (and again, the small overhead is not relevant for the asymptotic scaling of the complexity of the algorithm); for these reasons degenerations play a crucial role in the search for faster matrix multiplication algorithms.

We may for instance take a rectangular lattice and consider 4-party states $\ket{\phi}$ and $\ket{\psi}$ and investigate whether there exist maps at the vertices such that
\begin{align*}
    \left( \bigotimes\limits_{v \in V} M_v \right) \ket{\phi}_G = \ket{\psi}_G
\end{align*}
as illustrated in \cref{fig:main question}.
On first sight, one might think that since the initial and final states are tensor product states over the edges, the best one can do is perform transformations on each plaquette.
However, it is well known that there exist examples for which $\ket{\phi} \ngeq \ket{\psi}$ but $\ket{\phi}^{\ot n} \geq \ket{\psi}^{\ot n}$ for some $n$.
One can think of this as the case where the hypergraph is such that the edges are all stacked on top of each other, so we may have
\begin{center}
    \begin{overpic}[height=0.7cm,grid=false]{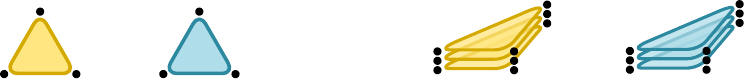}
        \put(40,3){while}
        \put(14,3){$\ngeq$} \put(76,3){$\geq$}
    \end{overpic}
\end{center}
The question is whether such phenomena still occur when we use a lattice hypergraph (which is a much sparser structure), which is the main question we address in this work.

\begin{figure}[t]
    \centering
    \vspace{0.1cm}
    \begin{overpic}[height=3.5cm,grid=false]{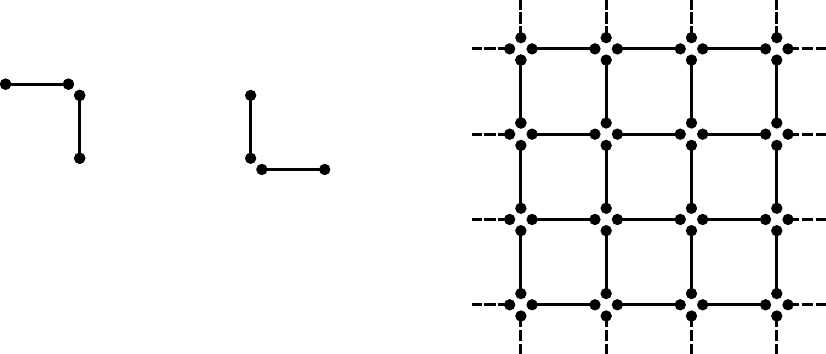}
        \put(2,14){$\ket{\phi}$} \put(31,14){$\ket{\psi}$}
        \put(47,20){$\Rightarrow$}
    \end{overpic}
    \caption{An example of two states which can not be transformed into one another on a single plaquette, but which give the same entanglement structure.}
    \label{fig:epr lattice}
\end{figure}

We will indeed construct examples where $\ket{\phi}_G \geq \ket{\psi}_{G}$ while $\ket{\phi} \ngeq \ket{\psi}$ in \cref{sec:constructions}.
There are obvious examples of such transformations where EPR pairs are exchanged between different parties (and used as a resource).
A basic example is shown in \cref{fig:epr lattice} where we have tensors $\ket{\phi}$ and $\ket{\psi}$ consisting of two EPR pairs which are clearly not equivalent (or related by a restriction) on a single plaquette, but when placed on a periodic lattice they give rise to the same state, as already observed in \cite{molnar2018generalization}.
Given this example, a plausible intuition is that since in a two-dimensional lattice the plaquettes are typically adjacent in at most two vertices, any lattice transformation can be performed by combining an exchange of EPR pairs with single plaquette transformations.
This is however not the case, as we will show in \cref{thm:ghz5-to-epr}.
This demonstrates that the resource theory of tensors is highly nontrivial.
On the other hand, we will also provide \emph{obstructions} for the existence of entanglement structure transformations in \cref{sec:obstructions}, applying various methods from algebraic complexity for proving complexity lower bounds.

The physically most interesting case for tensor networks are two- and higher-dimensional lattices.
In one spatial dimension, using multipartite entanglement is less natural, although one can in principle consider entanglement structures on a `strip' of plaquettes.
However, our concepts and methods apply to arbitrary hypergraphs.
From a mathematical perspective this poses new interesting questions in the theory of tensors.
We know that in general $\ket{\phi} \geqas \ket{\psi}$ does not imply $\ket{\phi} \geq \ket{\psi}$.
We can think of the asymptotic restriction as closely related to restrictions on the hypergraph where we stack all edges on top of each other.
This leads to the general question how connected the hypergraph $G$ has to be in order for a restriction $\ket{\phi}_G \geq \ket{\psi}_G$ to be possible.
In the special case where the hypergraph is acyclic, we show in \cref{cor:restriction tree} that any global transformation reduces to a local one, acting on each plaquette separately, so $\ket{\phi}_G \geq \ket{\psi}_G$ implies $\ket{\phi} \geq \ket{\psi}$.

\subsection{Algebraic complexity theory of tensor networks}\label{subsec:complexity}
One of the main motivations for studying the resource theory of tensors under restrictions, degenerations and asymptotic restrictions was to understand the complexity of matrix multiplication.
A similar natural question is to study the algebraic complexity of tensor network contractions, and we will now explain how the resource theory of tensor networks relates to the (algebraic) complexity of tensor network contractions.
This discussion is not needed to understand the remainder of this paper.
The question we would like to understand is the following: given an entanglement structure $\ket{\phi}_G$, what is the number of operations needed to compute a coefficient of a tensor network state using the entanglement structure $\ket{\phi}_G$?
Here the model of complexity is given by \emph{arithmetic circuits}, which perform addition, scalar multiplication, multiplication and division.
To measure the complexity of computing a certain function, one assigns weights to these different operations.
A common choice (when studying for instance the complexity of matrix multiplication) is to make addition and scalar multiplication `free' and only count the number multiplications and divisions.
This is a different model than the usual Turing machine model for computation.
In particular, we work directly over some field of numbers (such as the complex numbers $\CC$) and we do not use a binary representation of elements in the field.
In the case of matrix multiplication, this is the complexity which is closely related to the tensor rank.

Given an entanglement structure $\ket{\phi}_G$, we may obtain a tensor network state by applying local linear maps $M_v$ for each vertex $v \in V$.
Choose a basis $\ket{i_v}$, for the physical Hilbert space at vertex $v$, and let $T^{(i_v)}_v = \bra{i_v}M_v$.
Then we may expand the tensor network state as
\begin{align*}
    \sum_{\{i_v\}_{v \in V}} \left( \left(\bigotimes_{v \in V} T_v^{(i_v)}\right)\ket{\phi}_G \right) \bigotimes_{v \in V} \ket{i_v}
\end{align*}
The terms
\begin{align*}
    \left(\bigotimes_{v \in V} T_v^{(i_v)}\right)\ket{\phi}_G
\end{align*}
are therefore the coefficients of the state when expanded in the chosen product basis.
In general, we define the following map, which assigns a value to any collection of linear maps $\{T_v\}_{v \in V}$, $T_v : \H_v \to \CC$
\begin{align}\label{eq:tensor network coeff}
    \begin{split}
        f_{\ket{\phi}_G} : \bigoplus_{v \in V} \H_v^* &\to \CC \\
        (T_v)_{v \in V} &\mapsto \left(\bigotimes_{v \in V} T_v \right)\ket{\phi}_G.
    \end{split}
\end{align}
We call this map a \emph{tensor network coefficient}.
We will think of $f_{\ket{\phi}_G}$ as a polynomial in the entries of the $T_v$ (so it has $\prod_v \dim(\H_v)$ variables).
We will also refer to the task of computing the value of this polynomial on some input (i.e. computing a coefficient of some choice of tensors) as \emph{tensor network contraction}. It is the basic computational primitive in any tensor network based algorithm.
For instance, when performing variational optimization over the class of tensor network states, the energy expectation value reduces to computing a tensor network coefficient.
We will now investigate the hardness of computing $f_{\ket{\phi}_G}$ as a polynomial in the arithmetic circuit model.
Given an entanglement structure $\ket{\phi}_G$, we define $C(\ket{\phi}_G, G)$ to be the minimal size of an arithmetic circuit computing $f_{\ket{\phi}_G}$.
The resource theory for tensor networks we have defined now relates back to algebraic complexity theory by the following simple observation.

\begin{thm}\label{thm:restriction vs complexity}
    \begin{enumerate}
        \item\label{it:complexity restriction} If $\ket{\phi}_G \geq \ket{\psi}_G$, it holds that
        \begin{align*}
            C(\ket{\psi}_G, G) \leq C(\ket{\phi}_G, G).
        \end{align*}
        \item\label{it:complexity degeneration} Similary, if $\ket{\phi}_G \geqdeg^e \ket{\psi}_G$, it holds that
        \begin{align*}
            C(\ket{\psi}_G, G) \leq (e+1) C(\ket{\phi}_G, G)
        \end{align*}
    \end{enumerate}
\end{thm}

\begin{proof}
    Suppose that we have an arithmetic circuit that computes $f_{\ket{\phi}_G}$, and we have maps $M_v$ such that
    \begin{align*}
        \left(\bigotimes_{v \in V} M_v \right) \ket{\phi}_G = \ket{\psi}_G.
    \end{align*}
    Then
    \begin{align*}
        f_{\ket{\psi}_G}((T_v)_{v \in V}) = f_{\ket{\phi}_G}((T_v M_v)_{v \in V})
    \end{align*}
    so we can compute $f_{\ket{\psi}_G}((T_v)_{v \in V})$ by first computing $(T_v)_{v \in V} \mapsto (T_v M_v)_{v \in V}$ and then use the circuit for $f_{\ket{\phi}_G}$.
    The cost of the first step is zero in the model we consider (as it only requires multiplications with fixed numbers and additions, which are free), proving assertion \ref{it:complexity restriction}.
    We conclude that the cost of evaluating $f_{\ket{\psi}_G}((T_v)_{v \in V})$ is at most $C(\ket{\phi}_G, G)$.
    For \ref{it:complexity degeneration} we note that by a standard polynomial interpolation argument \cite{bini1980approximate,christandl2018tensor,christandl2020tensor} we can compute $f_{\ket{\psi}_G}((T_v)_{v \in V})$ from $e+1$ different \emph{restrictions} of $\ket{\phi}_G$, giving the desired result.
\end{proof}

How hard are tensor network computations?
The standard algorithms contract along edges one by one. Restricting to such computations on a graph (and an entanglement structure with EPR pairs), one finds that the complexity is exponential in the \emph{treewidth} of the graph \cite{markov2008simulating}. Finding optimal contraction orders is NP-hard in general, so one typically resorts to heuristics to finding good contraction orders \cite{pan2020contracting,gray2021hyper,gray2024hyperoptimized}.
It is well-known that in general tensor network contraction is indeed a hard problem.
For instance, one can show that tensor network contraction is a $\# P$-hard problem \cite{schuch2007computational,haferkamp2020contracting}, see \cite{gharibian2015tensor,scarpa2020projected} for further complexity-theoretic investigations of tensor networks.
One can also study the hardness of tensor network contraction in the arithmetic circuit model.
Recall that in this model of computation, the goal is to compute some polynomial through an arithmetic circuit.
The analog of the class $P$ is the class $VP$, which consists of families of polynomials $f_n$ for which there is a family of polynomial-sized circuit computing the desired polynomials.
The arithmetic analog of $NP$ is the class $VNP$, which is, informally speaking, the class of families of polynomials $f_n$ of polynomial degree which is such that for each monomial one can determine its coefficient in $f_n$ by a polynomial-sized circuit.
In \cref{sec:vnp} we prove that on arbitrary hypergraphs the computation of tensor network coefficients is in $VNP$ and tensor network contraction on a  two-dimensional square lattice with constant bond dimension $D = 2$ is $VNP$-hard. This is not very surprising (given the corresponding $\# P$-hardness) but we are not aware of previous work explicitly making this observation.

\begin{thm*}
    The problem of computing tensor network contraction coefficients, given by the polynomial $f_{\ket{\phi}_G}$ as in \cref{eq:tensor network coeff} on hypergraphs with $n$ edges and constant degree is in $VNP$. The problem of computing tensor network contraction coefficients with bond dimension $D = 2$ on an $n \times n$ square lattice is $VNP$-hard.
\end{thm*}

In \cref{sec:vnp} we state this result more formally as \cref{thm:hardness} and provide the proof.
While for two-dimensional lattices tensor network contraction is $VNP$-hard, it is easy to see that for acyclic hypergraphs the problem of computing $f_{\ket{\phi}_G}$ is in $VP$.
Finally, we mention that from a complexity theory perspective, tensor network contractions are closely related to an approach to counting problems known as \emph{Holant problems}, see \cite{backens2017new, backens2021full} for a discussion relating to quantum states and multiparty entanglement. The resource theory of tensor networks should therefore also be relevant to such Holant problems.

\section{Constructions}\label{sec:constructions}
Recall that the central question of our work is to study what transformations of entanglement structures are possible \emph{beyond transformations that act on the individual plaquettes}.
In this section we will give explicit constructions to demonstrate that there indeed exist transformations of entanglement structures which go beyond plaquette by plaquette restrictions.
This can be studied on arbitrary hypergraphs, but we will focus on examples which are relevant to lattices, as these are of primary interest for many-body physics applications.
We will consider periodic lattices, to avoid boundary terms (but this is not crucial).
We will focus on two-dimensional structures, since these are important in tensor network applications, and in one spatial dimension entanglement structures beyond EPR pairs are somewhat artificial (although one can study `strips' of plaquettes to obtain nontrivial examples).
We also provide classes of examples in higher spatial dimensions, and we end by discussing the asymptotic resource theory of tensor networks.

\subsection{Two-dimensional entanglement structures}\label{subsec:construction-2d}

The first and most basic class of examples are examples where we just distribute EPR pairs of some dimension over the plaquettes.
When arranged on the lattice, we may move EPR pairs to neighboring plaquettes
\begin{center}
    \begin{overpic}[height=1cm,grid=false]{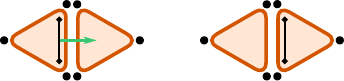}
        \put(47,10){$\Rightarrow$}
    \end{overpic}
\end{center}
which is an invertible transformation.
For example, as we already saw
\begin{center}
    \begin{overpic}[height=2.5cm,grid=false]{epr-lattice}
        \put(2,14){$\ket{\phi}$} \put(31,14){$\ket{\psi}$}
        \put(47,20){$\Rightarrow$}
    \end{overpic}
\end{center}
which is an example where on the lattice the entanglement structures are \emph{equivalent} while there are no restrictions (or degenerations) between $\ket{\phi_e}$ and $\ket{\psi_e}$ for single plaquettes $e$.

This principle can be used to construct many more examples!
One way is to use the $\epr$ pairs for teleportation.
Here is an example on a cyclic graph of four vertices
\begin{center}
    \begin{overpic}[height=1cm,grid=false]{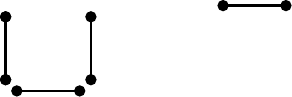}
        \put(50,15){$\geq$}
    \end{overpic}
\end{center}

In general, in a lattice one can move an $\epr$ pair to a neighboring plaquette, and there use it as a resource.
For example, consider plaquettes
\begin{center}
    \begin{overpic}[height=1cm,grid=false]{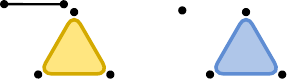}
        \put(45,10){$\ngeq$}
        \put(18,-12){$\ket{\phi_e}$}
        \put(80,-12){$\ket{\psi_e}$}
    \end{overpic}
    \vspace{0.4cm}
\end{center}
where we place a $\ghz_r(3)$ state and a level-$p$ $\epr$ pair as plaquette state $\ket{\phi_e}$ and $\ket{\psi_e}$ is given by an arbitrary 3-party state $\ket{\psi}$ distributed over three of the four parties.
When we place these on a lattice

\begin{center}
    \begin{overpic}[height=2.3cm,grid=false]{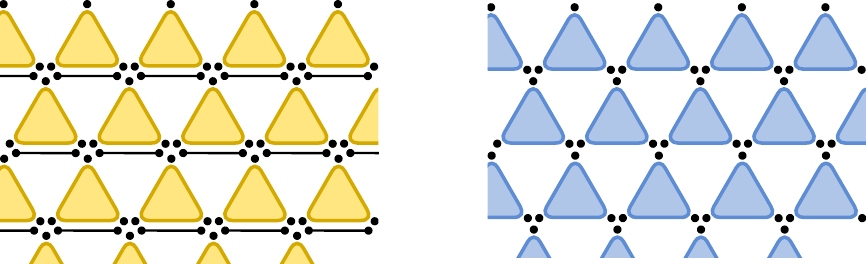}
        \put(48,13){$\geq$}
    \end{overpic}
\end{center}
we see there exists a restriction if the restriction
\begin{center}
    \begin{overpic}[height=1cm,grid=false]{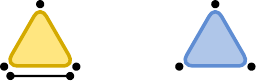}
        \put(48,12){$\geq$}
        \put(13,12){\scalebox{0.8}{$r$}}
        \put(12,-5){\scalebox{0.7}{$p$}}
    \end{overpic}
    \vspace{0.1cm}
\end{center}
exists.
In fact, the existence of such a restriction is (by definition) the statement that the $p$-aided rank of $\ket{\psi}$ is at most $r$.
See \cite{christandl2024partial} for a study of the aided rank, and example computations of the aided rank of tensors.

Next, for a slightly more involved example we consider a triangular lattice.
Let us suppose that we start with $\ket{\ghz_r(3)}$ on a sublattice as entanglement structure $\ket{\phi}_G$ and take $\ket{\psi}_G$ to be the entanglement structure where we place $\ket{\ghz_n(3)}$ at \emph{each} edge, and we look for a degeneration $\ket{\phi}_G \geqdeg \ket{\psi}_G$:
\begin{center}
    \begin{overpic}[height=2cm,grid=false]{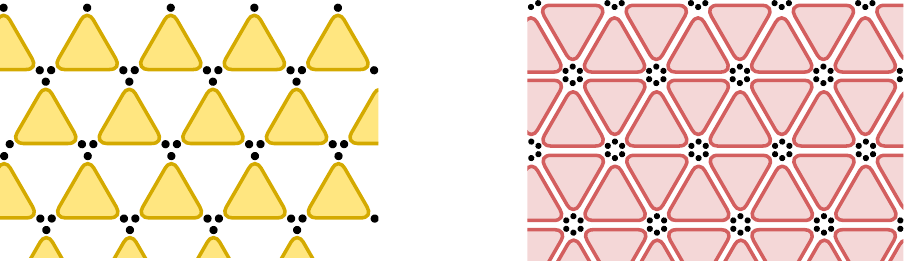}
        \put(48,12){$\geqdeg$}
        \put(13.5,14.5){\scalebox{0.9}{$r$}} \put(71.5,14){\scalebox{0.9}{$n$}}
    \end{overpic}
\end{center}
For which values of $r$ can we perform such a transformation?
It is clear that this is not an edge-wise transformation as in $\ket{\phi}_G$ we only have entangled states on a sublattice and the other sublattice is `empty' (i.e. has only product states).
A good strategy to perform this conversion is as follows.
First, convert each
\begin{align*}
    \ket{\ghz_r(3)} \geqdeg \ket{\ghz_n(3)} \ot \ket{\epr_D}_{\triangle}
\end{align*}
in $\ket{\phi}_G$.
This is certainly possible if we take $r$ to be the border rank times the remaining level $n$, so
\begin{align*}
    r = n\BR(\ket{\epr_D}_{\triangle}).
\end{align*}
Next, for each of the empty edges, we may now take EPR pairs from the surrounding edges, so we get an $\ket{\epr_D}_{\triangle}$ state at this edge.
From this, we can now extract a $\ghz_m(3)$ state, where the optimal $m$ (by definition) is given by the border subrank $m = \BQ(\ket{\epr_D}_{\triangle})$.
\begin{center}
    \begin{overpic}[height=1.5cm,grid=false]{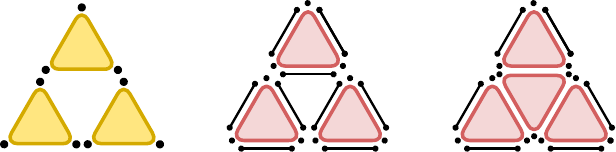}
        \put(30,10){$\geqdeg$} \put(66,10){$\geqdeg$}
        \put(12.5,16){\scalebox{0.9}{$r$}} \put(85,16){\scalebox{0.9}{$n$}} \put(48,9){\scalebox{0.7}{$D$}}
    \end{overpic}
\end{center}
It is known that
\begin{align}\label{eq:border subrank epr}
    \BQ(\ket{\epr_D}_{\triangle}) = \ceil[\bigg]{\frac{3D^2}{4}}
\end{align}
by a construction in \cite{strassen1987relative} and a lower bound in \cite{kopparty2020geometric}.
So, we can achieve the transformation with
\begin{align}\label{eq:ghz sublattice conversion}
    r = n \BR(\ket{\epr_{D}}_{\triangle}) \text{ such that } n \geq \ceil[\bigg]{\frac{3D^2}{4}}
\end{align}
An interesting question is whether this transformation is actually optimal, or whether a smaller $r$ would also suffice.

While nontrivial, the above example still had as a crucial ingredient that we somehow `moved' EPR pairs from one edge to another.
In the lattice hypergraphs we consider, two edges have at most two overlapping vertices.
This suggests the intuition that all `interaction' between different edges may be mediated by two-party entanglement, i.e. by moving an EPR pair from one edge to another.
We will now show that this intuition is \emph{not} correct and that the resource theory of tensor networks has a richer variety of possible transformations.

We will show that there exists a degeneration on a hypergraph with three 3-edges, where we place $\ghz_5(3)$ states on the three edges, and end up with an entanglement structure with EPR pairs.

\begin{thm}\label{thm:ghz5-to-epr}
    There exists a degeneration
    \begin{center}
        \begin{overpic}[height=1.5cm,grid=false]{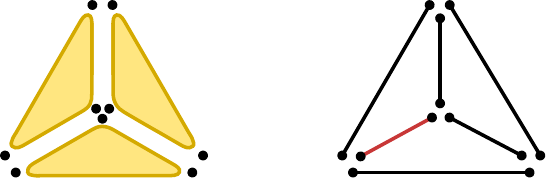}
            \put(11,15){\scalebox{0.7}{$5$}} \put(24,15){\scalebox{0.7}{$5$}} \put(17.5,2){\scalebox{0.7}{$5$}}
            \put(65,16){\scalebox{0.7}{$2$}} \put(76,20){\scalebox{0.7}{$2$}} \put(94,16){\scalebox{0.7}{$2$}}
            \put(79,-5){\scalebox{0.7}{$2$}} \put(87,10){\scalebox{0.7}{$2$}}
            \put(73,3){\scalebox{0.7}{$3$}}
            \put(48,13){$\geqdeg$}
        \end{overpic}
    \end{center}
\end{thm}
Note that there is one level-3 EPR pair.
The key point is that the final structure consists just of EPR pairs, but there is no way to distribute them over the three plaquettes such that we have an edge-to-edge degeneration.
Indeed, if we distribute the EPR pairs over the edges, at least one of them will have both a level-3 and a level-2 EPR pair, and there is no possible degeneration
\begin{align}\label{no-degeneration-ghz5}
    \begin{overpic}[height=1cm,grid=false]{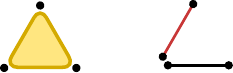}
        \put(48,13){$\geqdeg$} \put(49,13){$/$}
        \put(69,15.5){\scalebox{0.7}{$3$}} \put(83,-5){\scalebox{0.7}{$2$}}
    \end{overpic}
\end{align}
The impossibility of this degeneration follows from an easy flattening rank bound (see \cref{sec:obstructions}).

Nevertheless, the degeneration in \cref{thm:ghz5-to-epr} does exist.
The main ingredient is the three-party \emph{Bini tensor} $\ket{\beta} \in \H_A \ot \H_B \ot H_C$ with each Hilbert space equal to $(\CC^2)^{\ot 2}$, which is given by taking three level-2 EPR pairs, that is the tripartite state $\ket{\epr_2}_{\triangle}$, projecting out $\ket{11}$ on $C$, so
\begin{center}
    \vspace*{0.3cm}
    \begin{overpic}[height=2.5cm,grid=false]{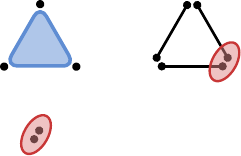}
        \put(13,44){$\beta$}
        \put(46,48){$=$}
        \put(56,33){\color{blue}{\scalebox{0.8}{$A$}}} \put(77,66){\color{orange}{\scalebox{0.8}{$B$}}} \put(100,33){\color{red}{\scalebox{0.8}{$C$}}}
        \put(28,8){$= \id - \proj{11}$}
    \end{overpic}
\end{center}

We may write this out in the standard basis
\begin{align*}
    \ket{\beta_{ABC}} & = \ket{00}_A\ket{00}_B\ket{00}_C + \ket{00}_A \ket{01}_B\ket{10}_C         \\
                      & \qquad + \ket{01}_A \ket{10}_B\ket{00}_C + \ket{10}_A \ket{00}_B\ket{01}_C \\
                      & \qquad + \ket{01}_A \ket{11}_B\ket{10}_C + \ket{11}_A \ket{10}_B\ket{01}_C
\end{align*}
so it is clear that $\R(\beta) \leq 6$.
In \cite{bini1979complexity} it was shown that there exists a degeneration $\ket{\ghz_5(3)} \geqdeg \ket{\beta}$ (so we have border rank $\BR(\ket{\beta}) \leq 5$), given by
\begin{align*}
     & \bigl(\ket{01} + \eps \ket{11}\bigr)\ket{10}\bigl(\ket{01} + \eps\ket{00}\bigr)                               \\
     & \quad + \ket{00}\bigl(\ket{00} + \eps\ket{01}\bigr)\bigl(\ket{10} + \eps\ket{00}\bigr)                        \\
     & \quad - \ket{01}(\ket{00} + \ket{10} + \eps\ket{11})\ket{01}                                                  \\
     & \quad - \bigl(\ket{00} + \ket{01} + \eps\ket{10}\bigr)\ket{00}\ket{10}                                        \\
     & \quad + \bigl(\ket{01} + \eps\ket{10}\bigr)\bigl(\ket{00} + \eps\ket{11}\bigr)\bigl(\ket{01} + \ket{10}\bigr) \\
     & = \eps \ket{\beta} + \bigO(\eps^2)
\end{align*}
The authors of \cite{bini1979complexity} were motivated by the complexity of matrix multiplication, as $\ket{\epr_2}_{\triangle}$ is the $2 \times 2$ matrix multiplication tensor, and they used a direct sum of two copies of $\ket{\beta}$ to perform $2 \times 3$ matrix multiplication.

\begin{proof}[Proof of \cref{thm:ghz5-to-epr}]
    We start by using the degeneration $\ket{\ghz_5(3)} \geqdeg \ket{\beta}$ on all three $\ket{\ghz_5(3)}$ states
    \begin{center}
        \begin{overpic}[height=1.5cm,grid=false]{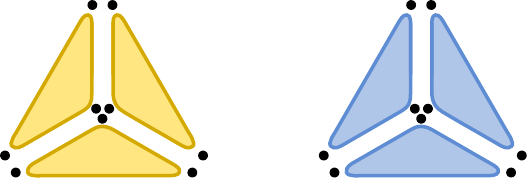}
            \put(47,16){$\geqdeg$}
            \put(11.5,16){\scalebox{0.7}{$5$}} \put(24,16){\scalebox{0.7}{$5$}} \put(18,2){\scalebox{0.7}{$5$}}
            \put(71,16){\scalebox{0.7}{$\beta$}} \put(84,16){\scalebox{0.7}{$\beta$}} \put(77,2){\scalebox{0.7}{$\beta$}}
        \end{overpic}
    \end{center}
    Let us call the three outer parties $A$, $B$ and $C$, and the interior party $D$.
    With the Bini tensors, on the outer edges we now have level-2 EPR pairs (between $AB$, $BC$ and $AC$), and it remains to extract EPR pairs between $D$ and $A$, $B$ and $C$ respectively:
    \begin{center}
        \vspace*{0.4cm}
        \begin{overpic}[height=1.7cm,grid=false]{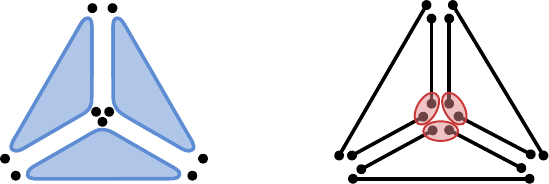}
            \put(50,15){$=$}
            \put(10,15){\scalebox{0.9}{$\beta$}} \put(23,15){\scalebox{0.9}{$\beta$}} \put(16,2.5){\scalebox{0.9}{$\beta$}}
            \put(56,-2){\color{blue}{\scalebox{0.8}{$A$}}} \put(78,35){\color{orange}{\scalebox{0.8}{$B$}}} \put(100,-2){\color{red}{\scalebox{0.8}{$C$}}}
            \put(86,13){\color{darkspringgreen}{\scalebox{0.7}{$D$}}}
        \end{overpic}
    \end{center}
    We divide $D$ into parties $A'$, $B'$ and $C'$, and apply a projection operator on $D$
    \begin{center}
        \vspace*{0.4cm}
        \begin{overpic}[height=2.2cm,grid=false]{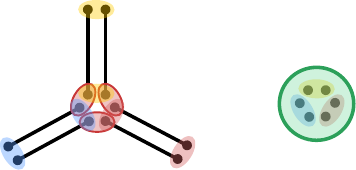}
            \put(-3,-1.5){\color{blue}{\scalebox{0.8}{$A$}}} \put(25,49){\color{orange}{\scalebox{0.8}{$B$}}} \put(53,-1.5){\color{red}{\scalebox{0.8}{$C$}}}
            \put(19.5,5){\color{blue}{\scalebox{0.8}{$A'$}}} \put(17,24){\color{orange}{\scalebox{0.8}{$B'$}}} \put(36,18){\color{red}{\scalebox{0.8}{$C'$}}}
            \put(103,18){$ = P$}
        \end{overpic}
    \end{center}
    where $P$ projects onto the span of the states
    \begin{align*}
        \Bigl\{\ket{0a}_{A'}\ket{0b}_{B'}\ket{0c}_{C'}\Bigr\}_{a,b,c \in \{0,1\}}
    \end{align*}
    together with
    \begin{align*}
        \Bigl\{\ket{10}_{A'}\ket{b0}_{B'}\ket{c0}_{C'}\Bigr\}_{b,c \in \{0,1\}}.
    \end{align*}
    Note that all of these states are in the support of the tensor on $D$, indeed, the only elements which are not in the support are
    \begin{align*}
        \ket{a_1 a_2}_{A'}\ket{b_1 b_2}_{B'}\ket{c_1 c_2}_{C'}
    \end{align*}
    with either $a_2 = b_1 = 1$, or $b_2 = c_1 = 1$ or $c_2 = a_1 = 1$.
    Finally, we apply the linear map
    \begin{align*}
        \ket{0}\bra{00} + \ket{1}\bra{01} + \ket{2}\bra{10}
    \end{align*}
    on $A$ and $A'$, and on $B$, $B'$, $C$ and $C'$ we apply the map
    \begin{align*}
        \ket{0}\bra{00} + \ket{1}\bigl(\bra{01} + \bra{10}\bigr)
    \end{align*}
    which results in the transformation
    \begin{center}
        \vspace{0.2cm}
        \begin{overpic}[height=2.3cm,grid=false]{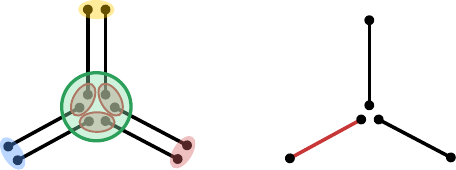}
            \put(-3,-1.5){\color{blue}{\scalebox{0.8}{$A$}}} \put(20,39){\color{orange}{\scalebox{0.8}{$B$}}} \put(42,-1.5){\color{red}{\scalebox{0.8}{$C$}}}
            \put(50,20){$\Rightarrow$}
            \put(70,9){\scalebox{0.8}{$3$}} \put(83,23){\scalebox{0.8}{$2$}} \put(90,9){\scalebox{0.8}{$2$}}
        \end{overpic}
    \end{center}
    which gives the target state of \cref{thm:ghz5-to-epr}.
\end{proof}

\begin{figure}[t]
    \centering
    \begin{subfigure}[t]{.45\linewidth}
        \begin{center}
            \begin{overpic}[width=.95\linewidth,grid=false]{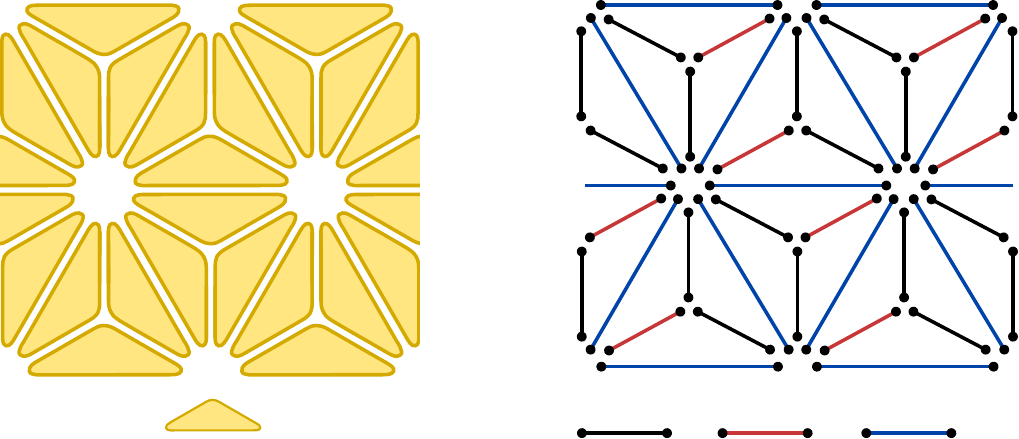}
                \put(48,23){$\geqdeg$}
                \put(61,2){\scalebox{0.6}{$2$}} \put(75,2){\scalebox{0.6}{$3$}} \put(89,2){\scalebox{0.6}{$4$}}
                \put(20,1){\scalebox{0.6}{$5$}}
            \end{overpic}
        \end{center}
        \caption{One can place the degeneration of \cref{thm:ghz5-to-epr} on a lattice.}
        \label{fig:bini lattice}
    \end{subfigure}
    \hspace*{0.2cm}
    \begin{subfigure}[t]{.45\linewidth}
        \begin{center}
            \begin{overpic}[width=.95\linewidth,grid=false]{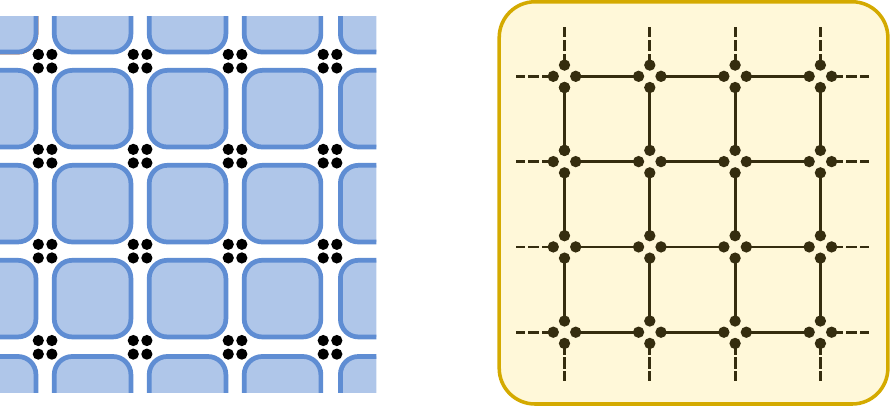}
                \put(47,20){$\geq$}
                \put(72,-4){\scalebox{0.8}{$\ket{\ghz_4(\abs{V})}$}}
            \end{overpic}
        \end{center}
        \caption{There is a restriction from the entanglement structure using the state in \eqref{eq:global-ghz-plaquette} to an entanglement structure with level-$n$ $\epr$ pairs and a global level-4 $\ghz$ state.}
        \label{fig:global ghz}
    \end{subfigure}
    \caption{Transformations of lattice entanglement structures beyond exchanging $\epr$ pairs.}
    \label{fig:lattice constructions}
\end{figure}
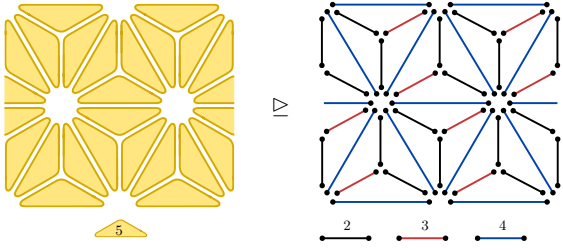
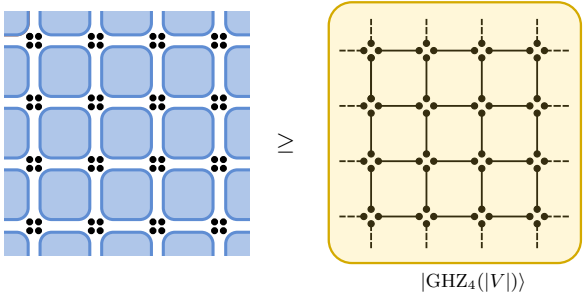

We can also put this degeneration on a lattice, see \cref{fig:bini lattice}. An interesting open question is whether this is optimal for this lattice, or whether one can do better.

We give one more example. This time we will see a restriction which is only possible on the lattice, which is such that it extracts a global $\ghz$ state.
We consider the square lattice, and consider the entanglement structure where we place the state
\begin{align}\label{eq:global-ghz-plaquette}
    \begin{overpic}[height=2cm,grid=false]{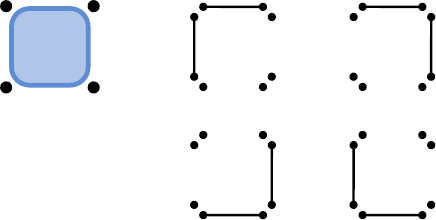}
        \put(-30,38){$\ket{\phi_e} = $} \put(28,38){$=$}
        \put(69,38){$+$} \put(32,8){$+$} \put(69,8){$+$}
        \put(-7,25){\scalebox{0.8}{$A$}} \put(-7,50){\scalebox{0.8}{$B$}}
        \put(24,25){\scalebox{0.8}{$D$}} \put(24,50){\scalebox{0.8}{$C$}}
    \end{overpic}
\end{align}
at each plaquette.
Here, the four parties of the plaquette each have Hilbert spaces $(\CC^{n+1})^{\ot 2}$, with basis $\ket{0},\dots,\ket{n}$.
Unconnected dots mean that the two parties have a product state $\ket{0}\ket{0}$, connected dots mean that they share a level-$n$ EPR pair $\sum_{i=1}^n \ket{i}\ket{i}$.
For example, the first term on the right-hand side in \cref{eq:global-ghz-plaquette} is given by
\begin{align*}
    \sum_{i,j = 1}^n \ket{0i}_A \ket{ij}_B \ket{j0}_C \ket{00}_D.
\end{align*}
We will show that there exists a restriction to the entanglement structure where we have level-$n$ $\epr$ pairs on the square lattice (i.e. the entanglement structure for a standard bond dimension $n$ tensor network state) together with a global level-$4$ $\ghz$ state shared by all parties $v \in V$, as illustrated in \cref{fig:global ghz}.

\begin{thm}
    Let $G$ be the square lattice, and let $\ket{\phi}_G$ be the entanglement structure with edge states given by \cref{eq:global-ghz-plaquette}.
    Let $\ket{\psi}_G$ be the entanglement structure with level-$n$ $\epr$ pairs on the square lattice as in \cref{fig:epr lattice}.
    Then
    \begin{align*}
        \ket{\phi}_G \geq \ket{\psi}_G \ot \ghz_4(\abs{V}).
    \end{align*}
\end{thm}

\begin{proof}
    We will now argue that we can apply a restriction to get the entanglement structure with level-$n$ EPR pairs on the edges with additionally a level-$4$ $\ghz$ state shared by all vertices.
    We apply the projection given by $P = P_1 + P_2 + P_3 + P_4$
    \begin{center}
        \begin{overpic}[height=3cm,grid=false]{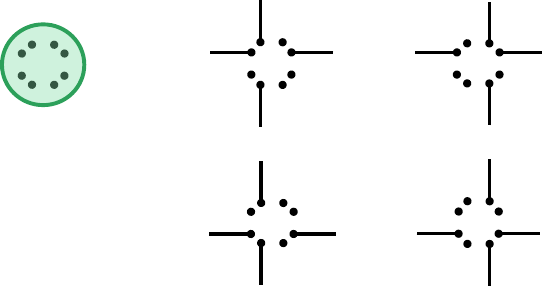}
            \put(-18,39){$P = $} \put(24,39){$=$}
            \put(67,39){$+$} \put(30,10){$+$} \put(67,10){$+$}
            \put(-2,32){\scalebox{0.6}{$A$}} \put(-2,47){\scalebox{0.6}{$B$}}
            \put(14,32){\scalebox{0.6}{$D$}} \put(14,47){\scalebox{0.6}{$C$}}
        \end{overpic}
    \end{center}
    where the figure should be interpreted similar to \cref{eq:global-ghz-plaquette}; for instance the first term is the projection onto the subspace spanned by
    \begin{align*}
        \{\ket{i0}_A \ket{jk}_B\ket{0l}_C\ket{00}_D; i,j,k,l = 1, \dots, n\}
    \end{align*}
    It is clear that each of the four terms in this projection gives a lattice of EPR pairs, and the images of the $P_i$ are orthogonal so we extract a global $\ghz_4(\abs{V})$ state.
\end{proof}

Note that such a global $\ghz$ state may in turn be used for the purpose of transforming degenerations of entanglement structures into restrictions \cite{christandl2020tensor} as explained in \cref{subsec:prior work}.
Finally, an interesting open question is the following: do there exist entanglement structures (with the same state at each plaquette) where the existence of a transformation depends on the size of the lattice? So far, our lattice examples exist for any lattice size, while there may exist transformations that only become possible when the size of the lattice is sufficiently large.

\subsection{Higher dimensional entanglement structures}\label{sec:construction-higher-d}

The construction principles which are demonstrated in \cref{subsec:construction-2d} may be applied in similar fashion for higher dimensional lattices.
The most basic constructions in two spatial dimensions used that two adjacent plaquettes can exchange $\epr$ pairs (even though we show that, perhaps surprisingly, there exist transformations that go beyond this mechanism).
In higher spatial dimensions it is significantly easier to construct nontrivial examples of lattice transformations since adjacent plaquettes now may be neighboring in more than two parties. This means that plaquettes can exchange multipartite entangled states.

It allows for a new construction principle, which does not yet show up in two spatial dimensions.
This is based on the fact that there exist states $\ket{\phi}$, $\ket{\psi}$ on more than two parties which are such that when we take two copies of the state on the hypergraph which consists of two edges on top each other there exists a transformation while this is not the case for a single copy.
That is,
\begin{align*}
    \ket{\phi} \ngeq \ket{\psi} \text{ while } \ket{\phi}^{\ot 2} \geq \ket{\psi}^{\ot 2}
\end{align*}
(where the tensor product is the Kronecker product).
The same phenomenon occurs for degenerations.
Let us see in a concrete example how this can be used.
We will again use that we know the precise value of the border subrank $\BQ(\ket{\epr_D}_{\triangle})$ as in \cref{eq:border subrank epr}.
Now, as plaquettes we take tetrahedra (so with four parties), where for $\ket{\phi_e}$ we place level-$D^2$ $\epr$-pairs on the 2-edges of the tetrahedron, which corresponds to placing $\ket{\epr_{D}}_{\triangle}$ states on the faces of the tetrahedron.
For $\ket{\psi_e}$ we place level-$r$ $\ghz$ states on the faces of the tetrahedron:
\begin{center}
    \begin{overpic}[height=2.1cm,grid=false]{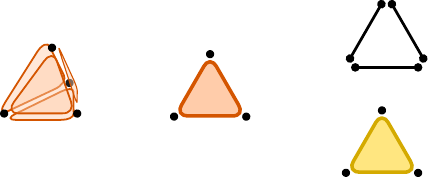}
        \put(65,17){$\Rightarrow$}
        \put(88.5,5){\scalebox{0.8}{$r$}} \put(97,35){\scalebox{0.8}{$D$}}
        \put(110,30){$\ket{\phi_e}$}
        \put(110,4){$\ket{\psi_e}$}
    \end{overpic}
\end{center}

Then, on an individual plaquette, we can use a $\ket{\epr_D}_{\Delta}$ state on each of the faces to perform a degeneration to a $\ket{\ghz_{r}(3)}$ state for $r = \ceil{\frac{3D^2}{4}}$, so for this value of $r$ we have a degeneration $\ket{\phi_e} \geqdeg \ket{\psi_e}$.
Now, if we consider a three-dimensional lattice where the tetrahedra have adjacent faces, we can group together $\epr$ pairs from adjacent plaquettes and apply a degeneration (and at the end redistribute the resulting $\ghz$ states to the two plaquettes).
This allows a transformation for
\begin{align*}
    r = \floor[\Big]{\sqrt{\ceil[\Big]{\frac{3D^4}{4}}}} \geq \BQ\mleft(\ket{\epr_D}_{\triangle}\mright).
\end{align*}
To give a concrete example, for $D = 4$ the single plaquette transformation is possible for $r = 12$, while on the lattice we can achieve $r  = 13$.

\subsection{Asymptotic conversions of tensor network states}\label{sec:asymptotic tensor network conversions}
We now address a question of a different nature, where we consider many copies of an entanglement structure to see how it can be used \emph{asymptotically} as a resource.
Let $\ket{\phi}_G = \bigotimes_{e \in E} \ket{\phi_e}$ be an entanglement structure on a hypergraph $G = (V,E)$.
Then we let $\ket{\phi}_G^{\ot n}$ be the entanglement structure on the same hypergraph $G$ where we place $\ket{\phi_e}^{\ot n}$ at edge $e$
\begin{align*}
    \ket{\phi}_G^{\ot n} = \bigotimes_{e \in E} \ket{\phi_e}^{\ot n}.
\end{align*}
So, at each plaquette we place $n$ copies of the state $\ket{\phi_e}$:
\begin{center}
    \includegraphics[height=2.3cm]{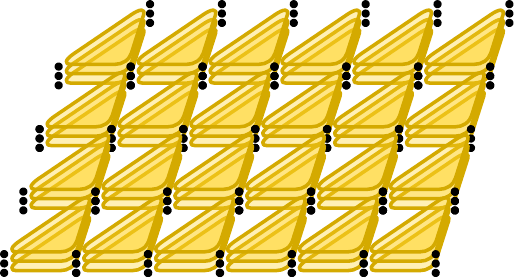}
\end{center}

Given a target state $\ket{\Psi}$ (which could be another entanglement structure, but in principle any state) we would like to know how useful $\ket{\phi}_G$ is as a resource for $\ket{\Psi}$.
More precisely, we would like to know for which $m$ and $n$ it is true that
\begin{align*}
    \ket{\phi}_G^{\ot n} \geq \ket{\Psi}^{\ot m}.
\end{align*}
To study the asymptotics, we say that $\ket{\phi}_G \geqas \ket{\Psi}$ if
\begin{align*}
    \ket{\phi}_G^{\ot (n + o(n))} \geq \ket{\Psi}^{\ot n}.
\end{align*}
In the particular case where $\ket{\Psi} = \ket{\psi}_G$ is some entanglement structure
\begin{center}
    \begin{overpic}[height=1.8cm,grid=false]{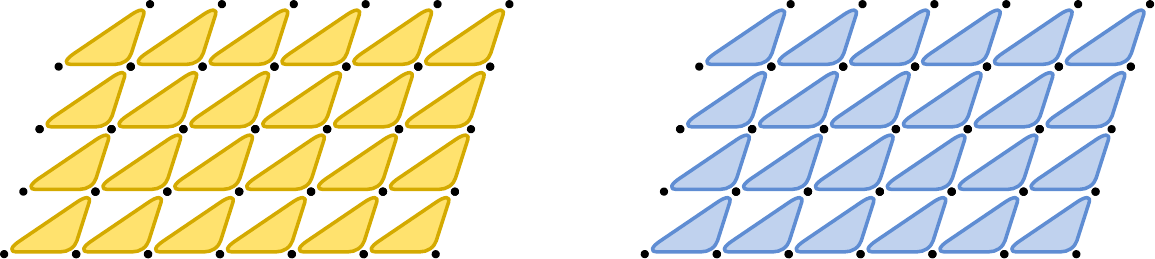}
        \put(49,9){$\geqas$}
    \end{overpic}
\end{center}
means that we have restrictions
\begin{center}
    \begin{overpic}[height=2.1cm,grid=false]{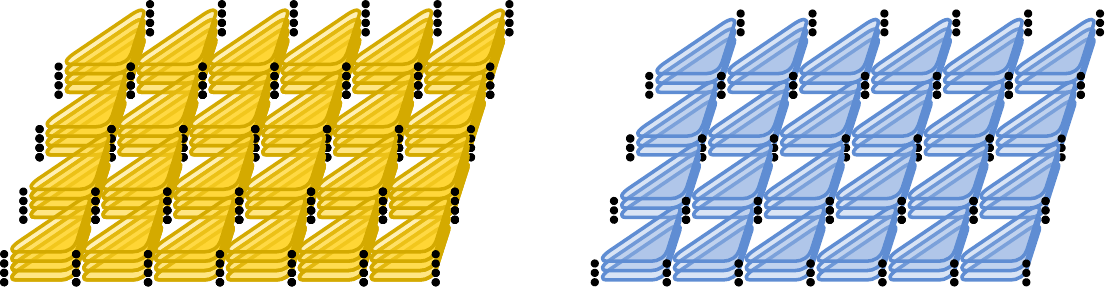}
        \put(50,11){$\geq$}
        \put(47,23.5){\footnotesize{$\} n + o(n)$}}
        \put(95,0){\footnotesize{$\} n$}}
    \end{overpic}
\end{center}
for $n \to \infty$.

To find an example, it suffices to have $k$-party states $\ket{\phi}$ and $\ket{\psi}$ such that on the one hand if we consider the entanglement structures where we place these states on each hyperedge, there is no restriction, so $\ket{\phi}_G \ngeq \ket{\psi}_G$, but we do have $\ket{\phi} \geqas \ket{\psi}$.
It is clear that by applying the asymptotic restrictions plaquette-wise in that case we obtain $\ket{\phi}_{G} \geqas \ket{\psi}_G$.
There are many examples of tensors for which $\ket{\phi} \geqas \ket{\psi}$ but $\ket{\phi} \ngeq \ket{\psi}$.
However, it is nontrivial to show that $\ket{\phi}_G \ngeq \ket{\psi}_G$.
In \cref{sec:substitution} we investigate the case where on the kagome lattice we have $\ket{\phi} = \ket{\epr_2}_{\triangle}$ and $\ket{\psi} = \ket{\lambda}$, the tensor from which we may obtain the RVB state.
It is known that $\ket{\epr_2}_{\triangle} \geqdeg \ket{\lambda}$ and hence $\ket{\epr_2}_{\triangle} \geqas \ket{\lambda}$. We prove that on the kagome lattice $\ket{\phi}_G \ngeq \ket{\psi}_G$ \cite{christandl2020tensor}.
Note that this example also directly implies that for the RVB state $\ket{\Psi}$ on the kagome lattice we have $\ket{\epr_2}_{\triangle} \geqas \ket{\Psi}$ (although we do not prove that $\ket{\epr_2}_{\triangle} \ngeq \ket{\Psi}$).
By a similar method we prove that when we take $\ket{\phi} = \ket{\ghz_2(3)}$ and $\ket{\psi} = \ket{W(3)}$ on the kagome lattice there is no restriction $\ket{\phi}_G \geq \ket{\psi}_G$, while already on the level of a single plaquette $\ket{\ghz_2(3)} \geqas \ket{W(3)}$.

For a final example, the value of the asymptotic subrank of $\ket{\epr_D}$ is given by \cite{strassen1987relative}
\begin{align*}
    \asQ(\ket{\epr_D}_{\triangle}) = D^2
\end{align*}
\begin{center}
    \begin{overpic}[height=1cm,grid=false]{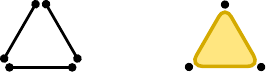}
        \put(48,10){$\geqas$}
        \put(-1,14){\scalebox{0.6}{$D$}} \put(13,-6){\scalebox{0.6}{$D$}} \put(27,14){\scalebox{0.6}{$D$}}
        \put(82,7.5){\scalebox{0.6}{$D^2$}}
    \end{overpic}
    \vspace{0.1cm}
\end{center}
This means that we have an asymptotic restriction
\begin{center}
    \begin{overpic}[height=1.5cm,grid=false]{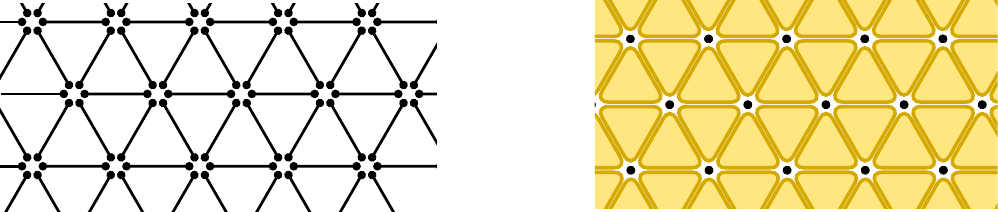}
        \put(50,10){$\geqas$}
        \put(10,13){\scalebox{0.6}{$D$}} \put(78,12){\scalebox{0.5}{$D$}}
    \end{overpic}
\end{center}
On the other hand, a simple flattening rank bound (see \cref{sec:obstructions}) shows that this is optimal.
Determining the optimal bond dimension for finite $n$ is an interesting open problem.

\section{Obstructions}\label{sec:obstructions}
In \cref{sec:constructions} we saw explicit examples where the conversion between entanglement structures was possible only by putting the tensors on the plaquettes of a lattice. This clearly demonstrates that in order to show that no conversion can exist we must take into account the entire lattice.
That is, if one wants to show $\ket{\phi}_G \ngeq \ket{\psi}_G$ it does \emph{not} suffice to show that for the edges $\ket{\phi_e} \ngeq \ket{\psi_e}$.
Whereas previous work proved the optimality of certain single-plaquette transformations \cite{christandl2020tensor}, our work provides for the first time general tools to show the impossibility of the conversion of entanglement structures.

The fundamental insight is that we may coarse-grain (or \emph{`fold'}) the network, and that any restriction on the initial graph should also be a restriction on the folded graph.
This allows one to reduce the problem to showing that there does not exist a restriction on the folded graph.
We will explain two important methods from algebraic complexity theory for obtaining obstructions for the existence of restrictions, namely (generalized) \emph{flattenings} and the \emph{substitution method} and we adapt them to prove the nonexistence of a restriction in interesting and nontrivial examples.

Next, we will discuss the question of asymptotic conversion of entanglement structures.
Asymptotic restriction is completely characterized by a compact topological space of functionals (the so-called \emph{asymptotic spectrum}).
We will explain how to use this in the context of entanglement structures.

\subsection{Folding the tensor network}
If $G$ is a hypergraph, we can obtain a new hypergraph by grouping some vertices together to a single vertex.
We call this procedure \emph{folding}, and if $H$ is obtained in such a way from $G$ then $H$ is a folding of $G$.
To be precise, $H$ has vertex set $V'$ and we have a surjective map $f : V \to V'$.
The edge set $E'$ of $H$ is given by edges
\begin{align*}
    f(e) = \{v' \, \text{ such that } v' = f(v) \text{ for } v \in e\}
\end{align*}
for each edge $e = \{v_1,\dots,v_k\} \in E$.
Note that $\abs{f(e)}$ may be smaller than $\abs{e}$ since it is possible that $f$ sends vertices in the same edge to the same image.
If we have an entanglement structure $\ket{\phi}_G = \bigotimes_{e \in E} \ket{\phi_e}$ on $G$, then this can naturally also be interpreted as an entanglement structure on $H$, by grouping the Hilbert spaces as
\begin{align*}
    \H_{v'} = \bigotimes_{v : f(v) = v'} \H_v \qquad \text{ and } \qquad \H_{f(e)} = \H_e
\end{align*}
and simply reinterpreting the state as
\begin{align*}
    \ket{\phi}_H = \bigotimes_{e' \in E'} \ket{\phi_{e'}}.
\end{align*}
The only thing we have done here is that we have grouped parties (vertices) together into a single parties.
It is clear that folding preserves restrictions.

\begin{lem}\label{lem:folding}
    Let $G = (V,E)$ be a hypergraph and $H = (W, F)$ a folding of $G$.
    Let $\ket{\phi}_G$ and $\ket{\psi}_G$ be entanglement structures for $G$.
    Then $\ket{\phi}_G \geq \ket{\psi}_G$ implies $\ket{\phi}_H \geq \ket{\psi}_H$.
    Similarly, $\ket{\phi}_G \geqdeg \ket{\psi}_G$ implies $\ket{\phi}_H \geqdeg \ket{\psi}_H$.
\end{lem}
\begin{proof}
    If the restriction is given by maps $M_v$ for $v \in V$, then it is clear that $\bigotimes_{v : f(v) = v'} M_v$ for $v' \in V'$ defines a restriction on $H$.
\end{proof}

While \cref{lem:folding} is obvious, it is the key to proving obstructions!
An important special case is where we fold to a graph with only two vertices.
For 2-tensors $\ket{\phi}$ and $\ket{\psi}$ we know that $\ket{\phi} \geq \ket{\psi}$ (and in fact $\ket{\phi} \geqdeg \ket{\psi}$) if and only if $\ket{\phi}$ has Schmidt rank at least as large as $\ket{\psi}$.
For a 2-tensor $\ket{\phi_{AB}} \in \H_{AB}$ we denote by $\rk(\ket{\phi_{AB}})$ its Schmidt rank (or entanglement rank, we will just call this the \emph{rank}), which is equal to the rank of its reduced density matrix on either $A$ or $B$, or the rank of $\ket{\phi_{AB}}$ as a linear map $\H_A \to \H_B^*$.
The theory of restrictions (or SLOCC conversion) and degenerations for 2-tensors is completely determined by the rank and the Schmidt rank.

With \cref{lem:folding} this directly yields the following:

\begin{cor}\label{cor:bipartition}
    Let $G = (V,E)$ be a hypergraph and let $\ket{\phi}_G$ and $\ket{\psi}_G$ be entanglement structures for $G$. If $\ket{\phi}_G \geqdeg \ket{\psi}_G$, then we must have that for any bipartitioning of the vertices $V = A \sqcup B$ the rank along $A$ and $B$ of $\ket{\phi}_G$ is at least that of $\ket{\psi}_G$.
\end{cor}

As an example, if we have the matrix multiplication tensor for $n \times n$ matrix multiplication (i.e. level-$n$ EPR pairs shared between three parties) and make the bipartitioning

\begin{center}
    \begin{overpic}[height=1.4cm,grid=false]{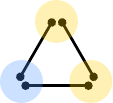}
        \put(-25,10){\color{blue}{$A$}} \put(70,80){\color{orange}{$B$}} \put(105,10){\color{orange}{$C$}}
        \put(15,45){\scalebox{0.8}{$n$}} \put(45,0){\scalebox{0.8}{$n$}} \put(75,45){\scalebox{0.8}{$n$}}
    \end{overpic}
\end{center}

then this has Schmidt rank $n^2$ between $A$ and $BC$. If we compare with the 3-party GHZ state of level $r$ we get Schmidt rank $r$, and therefore $\ket{\ghz_r(3)} \geq \ket{\EPR{n}}_{\triangle}$ implies $r \geq n^2$, so
\begin{align}\label{eq:rank mamu lower bound}
    \R(\ket{\EPR{n}}_{\triangle}) \geq n^2.
\end{align}
This gives a lower bound on the matrix multiplication exponent of $\omega \geq 2$.
While this lower bound is obvious, it is highly nontrivial to find lower bounds improving \cref{eq:rank mamu lower bound}!

Another example is the impossibility of degeneration in \cref{no-degeneration-ghz5}, where we see that if we flatten, the $\ghz_5(3)$ tensor has rank 5, while the two EPR pairs together have rank 6.

In the algebraic complexity literature, grouping parties such that we get a 2-tensor is often called a \emph{flattening}, since the resulting 2-tensor can be thought of as a matrix (which is a two-dimensional array, as opposed to a $k$-tensor, which is a $k$-dimensional array).

Let us now look at an example of an entanglement structure on a lattice. We take a rectangular lattice of size $n_1 \times n_2$ with $n_1, n_2$ even, with periodic boundary conditions and rectangular plaquettes as in \cref{fig:rect-lattice}.
For $\ket{\phi}_G$ we tile plaquettes with GHZ states on four parties, so $\ket{\phi_e} = \ket{\ghz_{r}(4)}$ for each edge.
For $\ket{\psi}_G$ we place maximally entangled pairs at the boundary of the plaquette, so $\ket{\psi_e} = \ket{\EPR{D}}_{\square}$ for each edge.
We now take the following bipartitioning:

\begin{center}
    \begin{overpic}[height=3cm,grid=false]{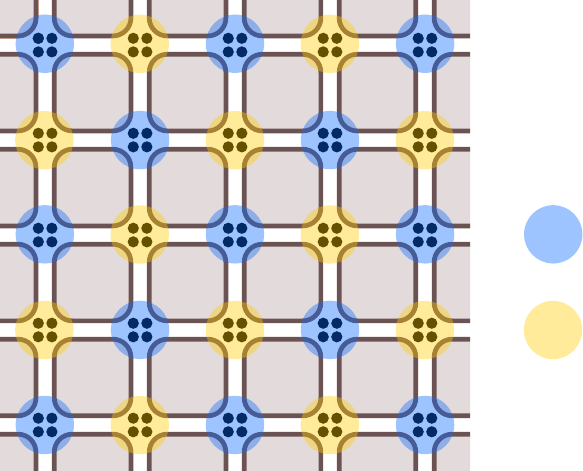}
        \put(105,38){\color{blue}{$A$}} \put(105,21){\color{orange}{$B$}}
    \end{overpic}
\end{center}

It is easy to see that this gives Schmidt rank of $\abs{E}r$ for $\ket{\phi}_G$ and $\abs{E}D^4$ for $\ket{\psi}_G$.
Therefore, by \cref{cor:bipartition}, $\ket{\phi}_G \geq \ket{\psi}_G$ implies that $r \geq D^4$.
On the other hand, it is easy to see that on the level of a single plaquette, for $r \geq D^4$ we have $\ket{\ghz_{r}(4)} \geq \ket{\EPR{D}}_{\square}$.
The lower bound is sharp, and there is no gain from placing the states on the lattice (i.e. in this case $\ket{\phi}_G \geq \ket{\psi}_G$ is only possible when $\ket{\phi_e} \geq \ket{\psi_e}$).

A next observation is that if the folding $f$ is such that for some edge $e = {v_1,\dots,v_k}$ all vertices are mapped to the same vertex $v'$, then we may remove this edge for the purpose of finding obstructions.
To see this, let $\tilde H$ be the hypergraph with the same vertex set as $H$, and with the edge $f(e)$ removed from the edge set.
An entanglement structure $\ket{\phi}_H$ then defines an entanglement structure $\ket{\phi}_{\tilde H}$ on $\tilde H$ by leaving out the state $\ket{\phi_e}$ on the edge $e$ that gets removed.
It holds that
\begin{align*}
    \ket{\phi}_H \geq \ket{\psi}_H \text{ if and only if } \ket{\phi}_{\tilde H} \geq \ket{\psi}_{\tilde H}
\end{align*}
and the same statement holds with degenerations rather than restrictions.
This is the case since $\ket{\phi_e}$ is at a single party $v'$, and we have the equivalences $\ket{\phi}_H \cong \ket{\phi}_{\tilde{H}}$ and $\ket{\psi}_H \cong \ket{\psi}_{\tilde{H}}$ (since we can locally prepare the states $\ket{\phi_e}$ and $\ket{\psi_e}$ at $v'$).
This is often helpful to reduce a problem on a large hypergraph such as a lattice, to a problem involving only a small number of edges.
As a special case one can show that if the hypergraph has no cycles, $\ket{\phi}_G \geq \ket{\psi}_H$ implies $\ket{\phi_e} \geq \ket{\psi_e}$ for each edge (so the graph structure does not allow additional transformations).
Here a cycle is any path of vertices $v_1, v_2, \dots, v_k$  for $k \geq 2$ such that $v_k = v_1$ and for each $i = 1,\dots,k=1$, $v_i$ and $v_{i+1}$ are not equal and are connected by an edge, so there is an edge $e$ with $v_i \in e$ and $v_{i+1} \in e$.
A hypergraph is \emph{acyclic} if it has no cycles (this notion is also known as Berge-acyclic \cite{berge1973graphs}). Note that in this definition if a graph is acyclic, any two different edges share at most one vertex.
If $G$ is an acyclic hypergraph, then it is easy to see that for any edge $e$ there is a folding from $G$ to the hypergraph which consists just of the single edge $e$.
Thus, \cref{lem:folding} directly implies the following result.

\begin{cor}\label{cor:restriction tree}
    Let $G$ be an acyclic hypergraph and let $\ket{\phi}_G$ and $\ket{\psi}_G$ be entanglement structures for $G$. Then $\ket{\phi}_G \geq \ket{\psi}_G$ if and only if $\ket{\phi_e} \geq \ket{\psi_e}$ for each edge $e \in E$. Similarly, $\ket{\phi}_G \geqdeg \ket{\psi}_G$ if and only if $\ket{\phi_e} \geqdeg \ket{\psi_e}$ for each edge $e \in E$.
\end{cor}

Finally, we consider the situation where we have a $k$-uniform hypergraph $G$ which is such that there exists a folding onto $k$ vertices $\{v_1',\dots,v_k'\}$, such that each edge $e$ gets mapped to $\{v_1',\dots,v_k'\}$. This means that edges are all folded on top of each other (and this is often possible for a lattice).
We will say that the hypergraph is \emph{foldable} in this case, such as in the following example:

\begin{center}
    \begin{overpic}[height=3cm,grid=false]{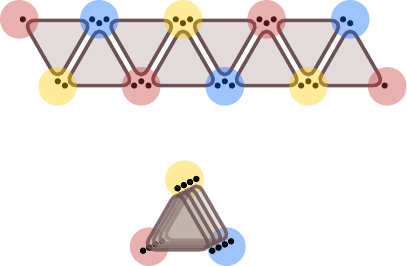}
        \put(43,32){$\Downarrow$}
    \end{overpic}
\end{center}

Consider entanglement structures $\ket{\phi}_G$ and $\ket{\psi}_G$ where we place the same $k$-party states $\ket{\phi}$ and $\ket{\psi}$ at each edge, and where we moreover assume that $\ket{\phi}$ and $\ket{\psi}$ are invariant under permutation of the $k$ parties.
Then we directly see by folding that if $\ket{\phi}_G \geq \ket{\psi}_G$ then we also find a restriction $\ket{\phi}^{\ot \abs{E}} \geq \ket{\psi}^{\ot \abs{E}}$.
In particular, this implies the following result.

\begin{thm}\label{thm:folding asymptotic}
    Suppose $G$ is a foldable $k$-uniform hypergraph, and $\ket{\phi}_G$ and $\ket{\psi}_G$ are entanglement structures using permutation invariant $k$-party states $\ket{\phi}$ and $\ket{\psi}$.
    Then if $\ket{\phi}_G \geqdeg \ket{\psi}_G$ it holds that
    \begin{align*}
        \ket{\phi} \geqas \ket{\psi}.
    \end{align*}
\end{thm}

\subsection{Folding and generalized flattenings}

In this section we describe another useful way of grouping vertices in a hypergraph by folding some vertices onto each other, leaving other vertices separate, leading to a fan-like structure.
Using this grouping, we can get obstructions for the conversion of entanglement structures using the ideas of~\cite{christandl2018tensor} about multiplicativity of rank lower bounds obtained via generalized flattenings of tensors.

Let us first explain the idea of using generalized flattenings of tensors.
For convenience, we will restrict to the case of 3-tensors (although this is not crucial).
Let $\ket{\phi} \in \H_A \ot \H_B \ot \H_C$.
A `usual' flattening bound would be to group two out of $A$, $B$ and $C$ together and compute the rank of the resulting 2-tensor (i.e. matrix) and use this as a lower bound for the tensor rank. However, we may also try to `split' by some arbitrary linear map.
Let $\ket{\phi} \in \H_A \ot \H_B \ot \H_C$ and let
\begin{align*}
    P : \H_A \ot \H_B \ot \H_C \to \H_X \ot \H_Y
\end{align*}
be a linear map.
Then we say that the 2-tensor
\begin{align*}
    \ket{\phi^{(P)}} = P \ket{\phi} \in \H_X \ot \H_Y
\end{align*}
is a \emph{generalized flattening} of $\ket{\phi}$.

We need to take into account the map $P$ when we are using this relation to bound the rank.
To this end we define the \emph{commutative rank} to be the maximum rank in the image of $P$ of any rank-1 tensor
\begin{align*}
    \CR(P) = \max_{\R(\ket{\psi_{ABC}}) = 1} \; \rk(P(\ket{\psi_{ABC}})),
\end{align*}
so we maximize over
\begin{align*}
    \ket{\psi_{ABC}} = \ket{\psi_A}\ket{\psi_B}\ket{\psi_C} \in  \H_A \ot \H_B \ot \H_C.
\end{align*}
It is then easy to see that one can bound the (border) rank as follows (see e.g. \cite{christandl2018tensor}):
\begin{align*}
    \BR(\ket{\phi}) \geq \dfrac{\rk(\ket{\phi^{(P)}})}{\CR(P)}.
\end{align*}

Lower bound methods of this type have been studied in \cite{derksen2018non,efremenko2017barriers} where it has been shown that the power of such bounds when applied to the border rank of the matrix multiplication tensor is limited.

An important special case is where one applies a linear map $P_C : \H_C \to \H_{X'} \ot \H_{Y'}$ only to the $C$-system, and let $P = \id_{AB} \ot P_C$ and group together $A X'$ as $X$ and $BY'$ as $Y$, so
\begin{align*}
    \ket{\phi^{(P)}} = (\id_{AB} \ot P_C)\ket{\phi}.
\end{align*}
\begin{center}
    \vspace*{0.2cm}
    \begin{overpic}[height=1.5cm,grid=false]{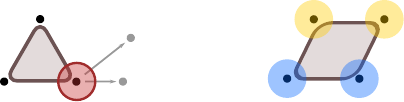}
        \put(50,10){$\Rightarrow$}
        \put(-7,3){\color{blue}{$A$}} \put(2,19){\color{orange}{$B$}} \put(18,12){\color{red}{$P_C$}}
        \put(32,3){\color{gray}{$X'$}} \put(35,14){\color{gray}{$Y'$}}
        \put(68,-6){\color{blue}{$A$}} \put(75,26){\color{orange}{$B$}}
        \put(86,-6){\color{blue}{$X'$}} \put(93,26){\color{orange}{$Y'$}}
        \put(77,11){\scalebox{0.8}{$\ket{\phi^{(P)}}$}}
    \end{overpic}
    \vspace*{0.2cm}
\end{center}

The most useful generalized flattenings for (border) rank bounds are the \emph{Koszul flattenings} arising from the antisymmetrization map viewed as a linear map $P_C \colon \H_C \to \Lambda^p \H_C^* \otimes \Lambda^{p + 1} \H_C$.
In this case $\CR(P) = \binom{c - 1}{p}$ where $c = \dim \H_C$.
There are also the more general \emph{Young flattenings}, based on other representations of the symmetric group.
See \cite{landsberg2017geometry} for extensive discussion of Young flattenings and associated lower bounds.

Here, we generalize this to the case where the system $C$ is replaced by multiple systems, each of which we split up.
Let
\begin{align*}
    \ket{\phi} \in \H_A \ot \H_B \ot \bigotimes_{i=1}^m \H_{C_i}
\end{align*}
and let $P_i : \H_{C_i} \to \H_{X_i} \ot \H_{Y_i}$ be linear maps.
Then the 2-tensor obtained by grouping together $AX_1\dots X_m$ and $BY_1\dots Y_m$ from
\begin{align*}
    \ket{\phi^{(P)}} = \underbrace{(\id_{AB} \ot P_1 \ot \dots \ot P_m) \ket{\phi}}_{\in \H_A \ot \H_B \ot \bigotimes_{i=1}^m \H_{X_i} \ot \H_{Y_i}}
\end{align*}
is a \emph{generalized multiflattening} of $\ket{\phi}$.

\begin{lem}\label{lem:multiflattening}
    \begin{align*}
        \BR(\ket{\phi}) \geq \dfrac{\rk(\ket{\phi^{(P)}})}{\prod_{i=1}^m \CR(P_i)}.
    \end{align*}
\end{lem}

\begin{proof}
    Suppose that $\R(\ket{\phi}) = r$, so there exists a decomposition
    \begin{align*}
        \ket{\phi} = \sum_{i = 1}^r \underbrace{\ket{a_i}\ket{b_i}\ket{c_{1,i}} \dots \ket{c_{m,i}}}_{:= \ket{\psi_i}}.
    \end{align*}
    Then $\ket{\phi^{(P)}}$ can be written as a sum of terms $\ket{\psi_i^{(P)}}$.
    Now, it is clear that since $\ket{\psi_i}$ is a product tensor
    \begin{align*}
        \rk(\ket{\psi_i^{(P)}}) \leq \prod_{i=1}^m \CR(P_i)
    \end{align*}
    and therefore
    \begin{align*}
        \rk(\ket{\phi^{(P)}}) & \leq \sum_{i=1}^r \rk(\ket{\psi_i^{(P)}}) \leq r \prod_{i=1}^m \CR(P_i)
    \end{align*}
    so we get
    \begin{align*}
        \R(\ket{\phi}) \geq \dfrac{\rk(\ket{\phi^{(P)}})}{\prod_{i=1}^m \CR(P_i)}.
    \end{align*}
    By semicontinuity of the rank the same inequality holds for tensors $\ket{\phi}$ with $\BR(\ket{\phi}) = r$.
\end{proof}

Next, we will combine this idea with a folding.
Consider some hypergraph $G$, which we assume to be 3-uniform.
We assume that we can divide the vertices together into subsets $A$, $B$ and a collection of subsets $C_i$ in such a way that the resulting folded hypergraph has a \emph{fan} structure, meaning that there are edges $\{A,B,C_i\}$, but no edges involving multiple $C_i$.
We will denote by $\fan(n)$ the fan hypergraph, with vertices $V = \{A, B, C_1,\dots, C_m\}$ and edges $e_i = \{A,B,C_i\}$.

\begin{center}
    \begin{overpic}[height=1.7cm,grid=false]{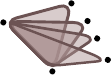}
        \put(-15,42){\color{blue}{$A$}} \put(30,-10){\color{orange}{$B$}}
        \put(70,65){\color{red}{$C_1$}} \put(84,48){\color{red}{$C_2$}} \put(92,33){\color{red}{$C_3$}} \put(102,13){\color{red}{$C_4$}}
    \end{overpic}
\end{center}

The fan hypergraph is useful for two reasons.
First of all, many interesting lattice graphs can be folded to a fan.
Secondly, the generalized multiflattenings behave well with respect to the fan structure!

\begin{lem}\label{lem:fan flattening rank}
    Let $\ket{\phi}_{\fan(m)}$ be an entanglement structure with state $\ket{\phi_{i}}$ on edge $\{A,B,C_i\}$.
    Then for any generalized flattening maps $P_i : \H_C \to \H_X \ot \H_Y$ we have
    \begin{align*}
        \BR(\ket{\phi}_G) \geq \prod_{i=1}^m \dfrac{\rk(\ket{\phi_i^{(P)}})}{\CR(P_i)}
    \end{align*}
\end{lem}

\begin{proof}
    If $\ket{\psi} = \ket{\phi}_{\fan(m)}$ we see that $\ket{\psi^{(P)}}$ is a tensor product of the states $\ket{\phi^{P_i}_i}$, so
    \begin{align*}
        \rk(\ket{\phi}_{\fan(m)}^{(P)}) = \prod_{i=1}^m \rk(\ket{\phi_i^{(P)}}).
    \end{align*}
    We conclude by \cref{lem:multiflattening}.
\end{proof}

\begin{figure}
    \centering
    \begin{subfigure}{.45\linewidth}
        \begin{center}
            \begin{overpic}[width=.9\linewidth,grid=false]{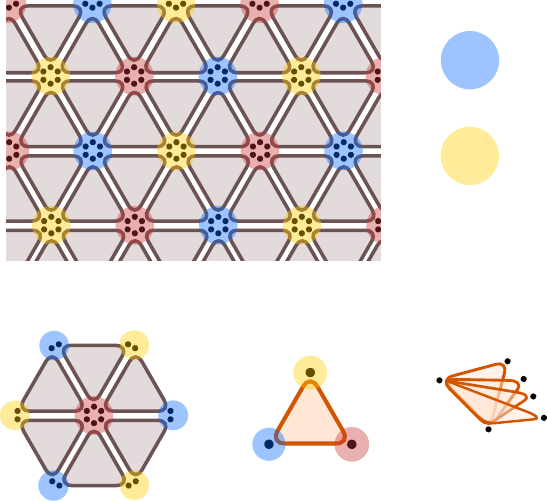}
                \put(93,79){\color{blue}{$A$}} \put(93,61){\color{orange}{$B$}}
                \put(38,15){$=$}
                \put(42,4){\color{blue}{$A$}} \put(55,28){\color{orange}{$B$}} \put(68,4){\color{red}{$C_i$}}
                \put(98,7){$\vdots$}
                \put(85,8){\scalebox{0.8}{\color{blue}{$A$}}} \put(74,21){\scalebox{0.8}{\color{orange}{$B$}}}
                \put(95,26){\scalebox{0.6}{\color{red}{$C_1$}}} \put(98,22){\scalebox{0.6}{\color{red}{$C_2$}}} \put(101,18){\scalebox{0.6}{\color{red}{$C_3$}}} \put(102,14){\scalebox{0.6}{\color{red}{$C_4$}}}
            \end{overpic}
        \end{center}
        \caption{Folding a triangular lattice into a fan.}
        \label{fig:triangular-fan}
    \end{subfigure}
    \hspace*{0.3cm}
    \begin{subfigure}{.45\linewidth}
        \begin{center}
            \begin{overpic}[width=.9\linewidth,grid=false]{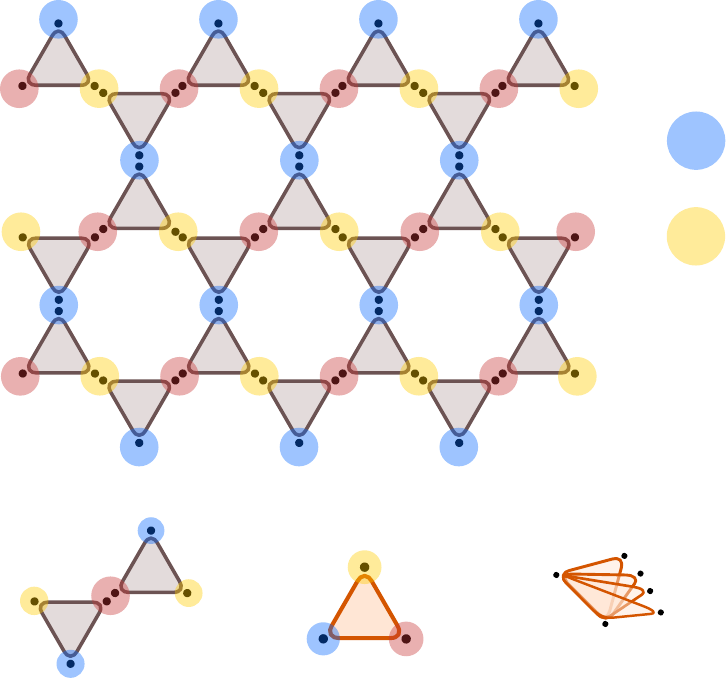}
                \put(103,72){\color{blue}{$A$}} \put(103,59){\color{orange}{$B$}}
                \put(32,10){$=$}
                \put(38,2){\scalebox{0.8}{\color{blue}{$A$}}} \put(48,19){\scalebox{0.8}{\color{orange}{$B$}}} \put(60,2){\scalebox{0.8}{\color{red}{$C_i$}}}
                \put(90,0){$\vdots$}
                \put(79,3){\scalebox{0.8}{\color{blue}{$A$}}} \put(70,12){\scalebox{0.8}{\color{orange}{$B$}}}
                \put(86,20){\scalebox{0.6}{\color{red}{$C_1$}}} \put(90,16){\scalebox{0.6}{\color{red}{$C_2$}}} \put(93,12){\scalebox{0.6}{\color{red}{$C_3$}}} \put(94,7){\scalebox{0.6}{\color{red}{$C_4$}}}
            \end{overpic}
        \end{center}
        \caption{Folding a kagome lattice into a fan.}
        \label{fig:kagome-fan}
    \end{subfigure}
    \caption{Foldings of two lattices with 3-edges. For the kagome lattice, the fan has plaquettes which consist of \emph{two} of the original plaquettes, whereas for the triangular lattice each fan plaquette consists of \emph{six} copies of the original plaquette.}
    \label{fig:folding-fan}
\end{figure}
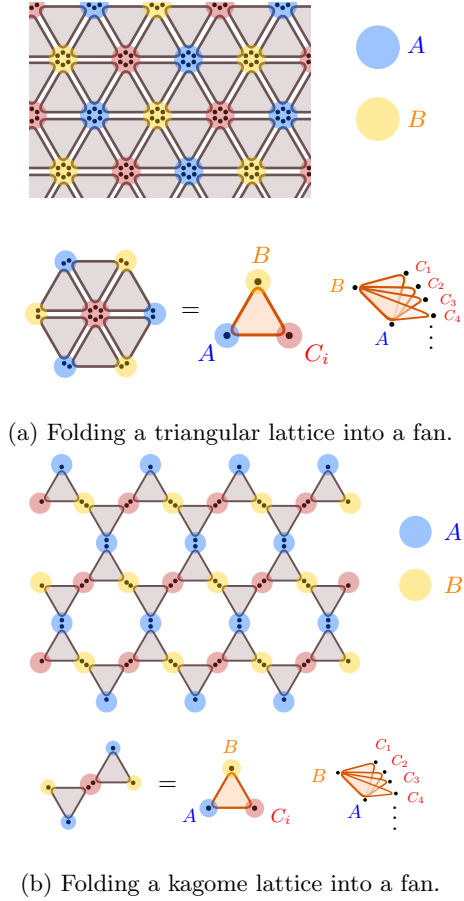

This means that once we have folded to a fan, we can apply generalized flattening bounds on individual plaquettes.
Two important examples where we can fold to a fan are the triangular lattice and the kagome lattice, see \cref{fig:folding-fan}.
We denote by $\kag(n)$ and $\trian(n)$ the kagome and triangular lattice with $n$ edges.

From these foldings and \cref{lem:folding} we see that
\begin{lem}\label{lem:folding lattice to fan}
    Suppose that $\ket{\phi}_{\kag(2n)} \geq \ket{\psi}_{\kag(2n)}$, then
    \begin{align*}
        \ket{\phi^{\ot 2}}_{\fan(n)} \geq \ket{\psi^{\ot 2}}_{\fan(n)}.
    \end{align*}
    If $\ket{\phi}_{\trian(6n)} \geq \ket{\psi}_{\trian(6n)}$, then
    \begin{align*}
        \ket{\phi^{\ot 6}}_{\fan(n)} \geq \ket{\psi^{\ot 6}}_{\fan(n)}.
    \end{align*}
    The same statements are true using degenerations or asymptotic restrictions.
\end{lem}

We can use this to prove obstructions to entanglement structure transformations, where we consider transformations from an entanglement structure where we place $\ket{\ghz_{r}(3)}$ states on each plaquette to an entanglement structure with some 3-party state $\ket{\phi}$ at each plaquette.
To make the question more concrete, let us take the kagome lattice, and study the conversion of $\ket{\ghz_{r}(3)}$ states to $\ket{\epr_D(3)}$ states.
The `naive' flattening bound, by just taking the rank at a single vertex, gives
\begin{align}\label{eq:flattening ghz epr}
    r^2 \geq D^4.
\end{align}
Asymptotic lower bounds do not perform better: if we use \cref{thm:folding asymptotic}, we see that we need
\begin{align*}
    \ket{\ghz_r(3)} \geqas \ket{\epr_D}_{\triangle},
\end{align*}
but since the best known lower bound for the matrix multiplication exponent is $\omega \geq 2$, we again get $r \geq D^2$.

\begin{thm}
    Suppose that
    \begin{align*}
        \ket{\ghz_r(3)}_{\kag(n)} \geqdeg \ket{\phi}_{\kag(n)},
    \end{align*}
    then for any generalized flattening we must have
    \begin{align*}
        r^2 \geq \frac{\rk((\ket{\phi}^{\ot 2})^{(P)})}{\CR(P)}.
    \end{align*}
    In particular, if $\ket{\phi} = \ket{\epr_D}_{\triangle}$ we must have
    \begin{align*}
        r^2 \geq 2D^4 - D^2.
    \end{align*}
    Similarly, if
    \begin{align*}
        \ket{\ghz_r(3)}_{\trian(n)} \geqdeg \ket{\phi}_{\trian(n)},
    \end{align*}
    then for any generalized flattening we must have
    \begin{align*}
        r^6 \geqdeg \frac{\rk((\ket{\phi}^{\ot 6})^{(P)})}{\CR(P)}.
    \end{align*}
    If $\ket{\phi} = \ket{\epr_D}_{\triangle}$ we must have $r^6 \geq 2D^{12} - D^6$.
\end{thm}
\begin{proof}
    Note that if we have $n$ edges on an arbitrary 3-regular hypergraph $G$
    \begin{align*}
        \BR(\ket{\ghz_r(3)}_G) \leq r^n
    \end{align*}
    and $\ket{\psi}_G \geqdeg \ket{\phi}_G$ implies $\BR(\ket{\psi}_G) \geq \BR(\ket{\phi}_G)$.
    On the other hand, by \cref{lem:multiflattening} and \cref{lem:fan flattening rank}, for any generalized flattening $P$
    \begin{align*}
        \BR(\ket{\phi}_{\fan(n)}) \geq \left(\frac{\rk((\ket{\phi}^{\ot 2})^{(P)})}{\CR(P)}\right)^n.
    \end{align*}
    The statement about a general 3-party state $\ket{\phi}$ now follows from \cref{lem:folding lattice to fan}.

    For the special case where $\ket{\phi} = \ket{\epr_D}_{\triangle}$, using Koszul flattenings one obtains \cite{landsberg2015new} (see also \cite{landsberg2017geometry})
    \begin{align*}
        \ket{\ghz_{r}(3)} \geqdeg \ket{\epr_D}_{\triangle} \Rightarrow r \geq 2D^2 - D.
    \end{align*}
    Since this bound is obtained by a generalized flattening, this implies the desired result.
\end{proof}

For instance, on the kagome lattice, if $D = 2$ this requires $r \geq 6$ while \cref{eq:flattening ghz epr} only gives $r \geq 4$, and for $D = 3$ this gives $r \geq 13$, while \cref{eq:flattening ghz epr} only requires $r \geq 9$.
On the triangular lattice, if $D = 2$ this requires $r \geq 5$ and for $D = 3$ this requires $r \geq 11$ (while again, \cref{eq:flattening ghz epr} only gives $r \geq 4$ and $r \geq 9$ respectively).

\subsection{Folding and the substitution method}\label{sec:substitution}
Another important method for proving obstructions (i.e. lower bounds in algebraic complexity) is the \emph{substitution method}.
We start by explaining the idea of the substitution method, and for convenience we again restrict to 3-tensors.
Suppose we have states $\ket{\phi_{ABC}}$ and $\ket{\psi_{ABC}}$ on three parties $A$, $B$ and $C$ and suppose there exists a restriction
\begin{align*}
    (M_A \ot M_B \ot M_C) \ket{\phi_{ABC}} = \ket{\psi_{ABC}}.
\end{align*}
In the substitution method we observe that we may `substitute' any $\ket{x} \in \H_A$ in the first factor and get a restriction
\begin{align*}
    (\bra{x} M_A \ot M_B \ot M_C) \ket{\phi_{ABC}} = (\bra{x} \ot \id_{BC}) \ket{\psi_{ABC}}.
\end{align*}
Note that if we think of the tensor as a map from $\H_A$ to $\H_B \ot \H_C$ this corresponds to substituting some value in the map.
This is typically used for the case where $\ket{\phi_{ABC}}$ is a $\ghz$ state and we want to compute the rank $\R(\ket{\psi})$.

As a concrete example, let us consider the $W$ state on three parties
\begin{align*}
    \ket{W(3)} = \ket{100} + \ket{010} + \ket{001}
\end{align*}
and bound its rank.
It is clear that there exists a restriction $\ket{\ghz_3(3)} \geq \ket{W(3)}$, since $\ket{W(3)}$ is defined as a sum of three product tensors.
So, $\R(\ket{W(3)}) \leq 3$.
Can we do better?
The substitution method shows us that this is not the case.
Let $\ket{x} = \ket{0} + \overline{x}\ket{1}$ be nonzero, with $x \in \CC$, then applying the substitution we find that
\begin{align*}
    \ket{W^{(x)}(3)} = \ket{01} + \ket{10} + x \ket{00}.
\end{align*}
Note that $\R(\ket{W^{(x)}(3)}) = \rk(\ket{W^{(x)}(3)}) = 2$ for all choices of $\ket{x}$.
On the other hand, suppose that we have a restriction $\ket{\ghz_2(3)} \geq \ket{W(3)}$, then we have
\begin{align*}
    \ket{W(3)} = \sum_{i=0,1} \ket{a_i}\ket{b_i}\ket{c_i}.
\end{align*}
At least one of the $\ket{a_i}$ must \emph{not} be proportional to $\ket{0}$, so we can choose $\ket{x}$ such that $\braket{x | a_i} = 0$. But this would imply that $\ket{W^{(x)}(3)}$ is a product state, which is a contradiction.
Therefore, there does not exist a restriction $\ket{\ghz_2(3)} \geq \ket{W(3)}$ and $\R(W(3)) = 3$.

There does exist a degeneration $\ket{\ghz_2(3)} \geqdeg \ket{W(3)}$, given by
\begin{align*}
    (\ket{0} + \eps \ket{1})^{\ot 3} - \ket{0}^{\ot 3} = \eps \ket{W(3)} + \bigO(\eps^2)
\end{align*}
so $\BR(\ket{W(3)}) = 2$ and we see that the substitution method is able to distinguish degenerations from restrictions (which is not the case for flattening ranks).

Here we will apply a more complicated version of the substitution method to prove that on the kagome lattice or a triangular lattice there is no restriction from the entanglement structure using $\ket{\ghz_2(3)}$ states to the entanglement structure using $\ket{W(3)}$ states.
This is not a direct consequence of the substitution method above, as it could be the case that the transformation is not possible on a single plaquette, but is possible on the lattice.
Note that we have $\ket{\ghz_2(3)} \geqdeg \ket{W(3)}$ and hence $\ket{\ghz_2(3)} \geqas \ket{W(3)}$.

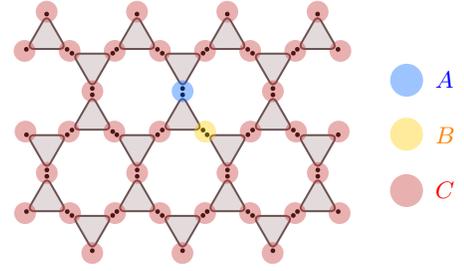
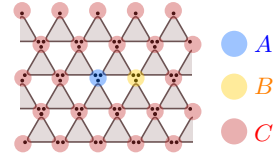
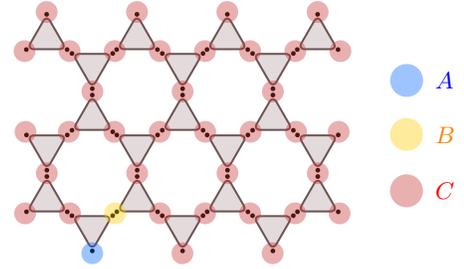
\begin{figure}
    \centering
    \begin{subfigure}{.3\linewidth}
        \begin{center}
            \begin{overpic}[width=.95\linewidth,grid=false]{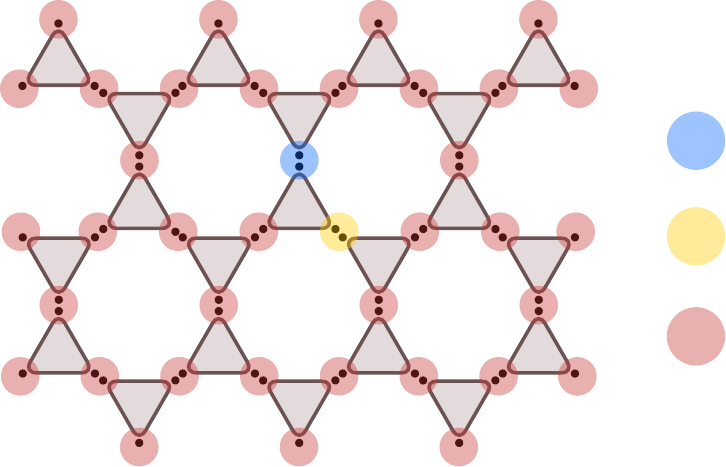}
                \put(103,43){\color{blue}{$A$}} \put(103,29.5){\color{orange}{$B$}} \put(103,16){\color{red}{$C$}}
            \end{overpic}
        \end{center}
        \caption{Folding a kagome lattice to a single plaquette.}
        \label{fig:kagome-sub}
    \end{subfigure}
    \hspace*{0.2cm}
    \begin{subfigure}{.3\linewidth}
        \begin{center}
            \begin{overpic}[width=.7\linewidth,grid=false]{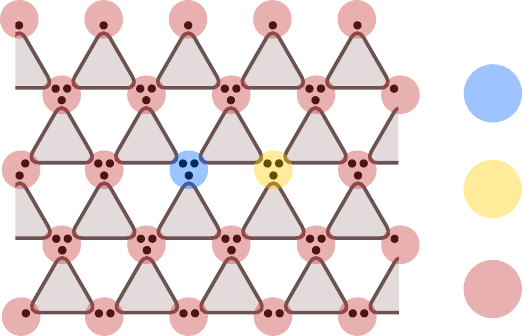}
                \put(103,43){\color{blue}{$A$}} \put(103,25){\color{orange}{$B$}} \put(103,5){\color{red}{$C$}}
            \end{overpic}
        \end{center}
        \caption{Folding a triangular lattice to a single plaquette.}
        \label{fig:triangular-sub}
    \end{subfigure}
    \hspace*{0.2cm}
    \begin{subfigure}{.3\linewidth}
        \begin{center}
            \begin{overpic}[width=.95\linewidth,grid=false]{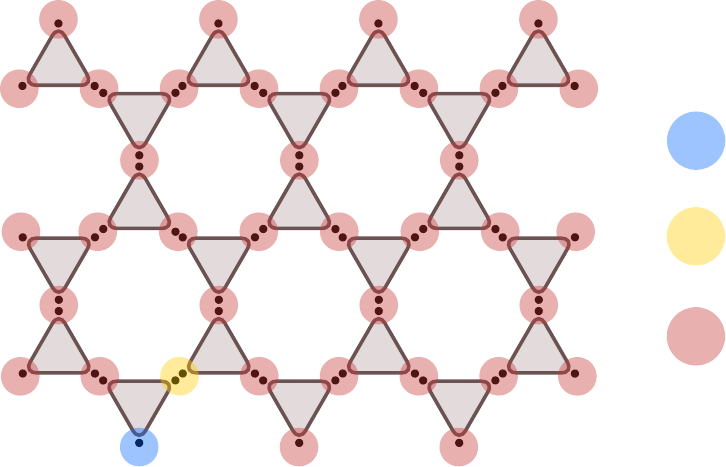}
                \put(103,43){\color{blue}{$A$}} \put(103,29.5){\color{orange}{$B$}} \put(103,16){\color{red}{$C$}}
            \end{overpic}
        \end{center}
        \caption{Folding a kagome lattice with boundary (so here the plaquette neighboring $A$ and $B$ is at the boundary of the lattice).}
        \label{fig:kagome-sub-boundary}
    \end{subfigure}
    \caption{Foldings of two lattices with 3-edges. For the kagome lattice, the fan has plaquettes which consist of \emph{two} of the original plaquettes, whereas for the triangular lattice each fan plaquette consists of \emph{six} copies of the original plaquette.}
    \label{fig:folding-subsitution}
\end{figure}

We will use a folding, where we choose one specific plaquette (because of translation invariance of the lattice it does not matter which one).
Let us denote the three vertices adjacent to this plaquette $a$, $b$ and $c$.
In the folding we choose we map $a$ to $A$, $b$ to $B$ and all other vertices to $C$. We may apply a similar folding to a (half-filled) triangular lattice, see \cref{fig:folding-subsitution}. For the remainder of this section we let $G$ either be this lattice or the kagome lattice.

For the structure with $\ket{\ghz_2(3)}$ this gives a state equivalent to the three parties sharing a $\ket{\ghz_2(3)}$ state, and level-2 $\epr$ pairs between $A$ and $C$, and between $B$ and $C$, a state we denote by $\ket{\phi_{ABC}}$ the state $\ket{\ghz_2(3)} \ot \ket{\EPR{2}}_{\wedge}$.
Similarly, for the entanglement structure using the $\ket{W(3)}$ state we get a state equivalent to the three parties sharing a $W$ state, and level-2 $\epr$ pairs between $A$ and $C$, and between $B$ and $C$, which we denote by $\ket{\psi_{ABC}} = \ket{W(3)} \ot \ket{\EPR{2}}_{\wedge}$.
We label the parties as follows:

\begin{center}
    \vspace*{0.3cm}
    \begin{overpic}[height=1.2cm,grid=false]{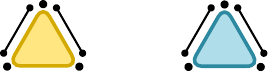}
        \put(-5,-5){\scalebox{0.8}{$A_1$}} \put(-8,5){\scalebox{0.8}{$A_2$}} \put(30,-5){\scalebox{0.8}{$B_1$}} \put(32,5){\scalebox{0.8}{$B_2$}}
        \put(4,27){\scalebox{0.8}{$C_2$}} \put(12.5,28){\scalebox{0.8}{$C_1$}} \put(22,27){\scalebox{0.8}{$C_3$}}
        \put(63,-5){\scalebox{0.8}{$A_1$}} \put(60,5){\scalebox{0.8}{$A_2$}} \put(98,-5){\scalebox{0.8}{$B_1$}} \put(100,5){\scalebox{0.8}{$B_2$}}
        \put(72,27){\scalebox{0.8}{$C_2$}} \put(80.5,28){\scalebox{0.8}{$C_1$}} \put(90,27){\scalebox{0.8}{$C_3$}}
        \put(14,8){$2$} \put(80,7){$W$}
        \put(10,-10){$\ket{\phi}$} \put(80,-10){$\ket{\psi}$}
    \end{overpic}
    \vspace*{0.5cm}
\end{center}

\begin{thm}\label{thm:w restriction}
    There is no restriction from $\ket{\phi_{ABC}}$ to $\ket{\psi_{ABC}}$.
\end{thm}

By the folding argument \cref{thm:w restriction} has as a direct consequence that the entanglement structure with the $W$ state cannot be obtained from the entanglement structure with $\ghz_2(3)$ states on the triangular or kagome lattice $G$.
This is a sharp result, since we know that there does exist a degeneration, and there does exist a restriction when one takes a direct sum of a number of copies linear in the lattice size.

\begin{cor}\label{cor:w kagome}
    If $G$ is the triangle or kagome lattice
    \begin{align*}
        \ket{\ghz_2(3)}_{G} \ngeq \ket{W(3)}_{G}.
    \end{align*}
\end{cor}

\begin{proof}[Proof of \cref{thm:w restriction}]
    We assume that there exists a restriction
    \begin{align}\label{eq:false restriction W}
        (M_A \ot M_B \ot M_C)\ket{\phi_{ABC}} = \ket{\psi_{ABC}}
    \end{align}
    and derive a contradiction.
    In this case, the tensors are all concise (their reduced density matrices have full rank) and since the systems on both sides are of equal dimension, the maps $M_A$, $M_B$ and $M_C$ must be invertible.
    We will use a version of the substitution method to show that this is not possible.

    For $\ket{x} \in (\CC^2)^{\ot 2}$, which is the Hilbert space of the $A = A_1 A_2$ system for both $\ket{\phi_{ABC}}$ and $\ket{\psi_{ABC}}$, let
    \begin{align*}
        \ket{\phi^{(x)}} = (\bra{x} \ot \id_{BC}) \ket{\phi_{ABC}}
    \end{align*}
    \begin{center}
        \begin{overpic}[height=1.2cm,grid=false]{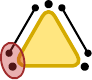}
            \put(45,30){$2$}
            \put(-20,10){\color{red}{$x$}}
            \put(20,-30){$\ket{\phi^{(x)}}$}
        \end{overpic}
        \vspace*{0.5cm}
    \end{center}
    and similarly for $\ket{\psi_{ABC}^{(x)}}$.
    These are 2-tensors on systems $B$ and $C$.
    We then define
    \begin{align*}
        X_{\phi} & = \{\ket{x} \text{ such that } \rk(\ket{\phi^{(x)}}) \leq 2 \}  \\
        X_{\psi} & = \{\ket{x} \text{ such that } \rk(\ket{\psi^{(x)}}) \leq 2 \}.
    \end{align*}
    Since the maps $M_A$, $M_B$ and $M_C$ are invertible, it is easy to see that
    \begin{align*}
        \ket{x} \in X_{\psi} \Leftrightarrow \ket{y} = M_A^{\dagger}\ket{x} \in X_{\phi}.
    \end{align*}
    Indeed,
    \begin{align*}
        \ket{\psi^{(x)}} & = (\bra{x} \ot \id_{BC})(M_A \ot M_B \ot M_C)\ket{\phi} \\
                         & = (M_B \ot M_C) \ket{\phi^{(y)}}
    \end{align*}
    so $\rk(\ket{\psi^{(x)}}) = \rk(\ket{\phi^{(y)}})$.
    So, we see that (if a restriction exists) the sets $X_{\phi}$ and $X_{\psi}$ must be related by a linear transformation.
    Now, we note that for any $\ket{x}$, $\ket{\phi^{(x)}}$ and $ \ket{\psi^{(x)}}$ still share an $\epr$ pair between $B$ and $C$.
    Denote by $\ket{\tilde \phi}$ and $\ket{\tilde \psi}$ the states where we have left out the $\epr$ pair between $B$ and $C$, so

    \begin{center}
        \vspace{0.1cm}
        \begin{overpic}[height=1.2cm,grid=false]{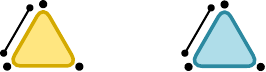}
            \put(-5,-5){\scalebox{0.8}{$A_1$}} \put(-8,5){\scalebox{0.8}{$A_2$}} \put(30,-5){\scalebox{0.8}{$B_1$}}
            \put(4,27){\scalebox{0.8}{$C_2$}} \put(12.5,28){\scalebox{0.8}{$C_1$}}
            \put(63,-5){\scalebox{0.8}{$A_1$}} \put(60,5){\scalebox{0.8}{$A_2$}} \put(98,-5){\scalebox{0.8}{$B_1$}}
            \put(72,27){\scalebox{0.8}{$C_2$}} \put(80.5,28){\scalebox{0.8}{$C_1$}}
            \put(14,8){$2$} \put(80,7){$W$}
            \put(10,-10){$\ket{\tilde\phi}$} \put(80,-10){$\ket{\tilde\psi}$}
        \end{overpic}
        \vspace*{0.5cm}
    \end{center}
    Now, for $x \in X_{\phi}$ we let
    \begin{align*}
        \ket{\tilde \phi^{(x)}} = (\bra{x} \ot \id_{BC})\ket{\tilde \phi}
    \end{align*}
    which is just $\ket{\phi^{(x)}}$ without the $\epr$ pair between $B$ and $C$, and therefore has rank $\rk(\ket{\phi^{(x)}}) = 2 \rk(\ket{\tilde \phi^{(x)}}) \leq 2$ so $\rk(\ket{\tilde \phi^{(x)}}) \leq 1$.
    The state $\ket{x}$ has either rank 1 or 2 as a state on $A_1 A_2$, so we can write a Schmidt decomposition
    \begin{align*}
        \ket{x} = \sum_{i = 1}^r \ket{f_i}_{A_1}\ket{e_i}_{A_2}
    \end{align*}
    with $r = 1$ or $r = 2$.
    Then,
    \begin{align*}
        \ket{\tilde \phi^{(x)}} = \sum_{i=1}^r \left(\braket{f_i | 0} \ket{0}_{B_1}\ket{0}_{C_1} + \braket{f_i | 1} \ket{1}_{B_1}\ket{1}_{C_1}\right) \ket{\overline{e_i}}_{C_2}
    \end{align*}
    which has rank at least $r$.
    So, $\rk(\ket{\tilde \phi^{(x)}}) \leq 1$ implies that $\ket{x}$ as a state on $A_1 A_2$ must have rank 1.
    So, write $\ket{x} = \ket{f}_{A_1}\ket{e}_{A_2}$ with $\ket{f} = f_0\ket{0} + f_1 \ket{1}$.
    Then
    \begin{align*}
        \ket{\tilde \phi^{(x)}} = \left(\bar{f_0} \ket{0}_{B_1}\ket{0}_{C_1} + \bar{f_1}\ket{1}_{B_1}\ket{1}_{C_1}\right)\ket{\overline{e}}_{C_2}
    \end{align*}
    This has rank 1 if and only $f_0 = 0$ or $f_1 = 0$, so if $\ket{x}$ can be written as
    \begin{align*}
        \ket{x} = \ket{0}\ket{e} \text{ or } \ket{x} = \ket{1}\ket{e}
    \end{align*}
    for $\ket{e} \in \CC^2$.
    In other words, $X_{\phi}$ is a union of two complex planes.
    For $\ket{\psi}$ the same reasoning holds, but with the difference that $\ket{\tilde \psi^{(x)}}$ is given by
    \begin{align*}
        \left(\bar{f_0} (\ket{0}_{B_1}\ket{1}_{C_1} + \ket{1}_{B_1}\ket{0}_{C_1}) + \bar{f_1}\ket{0}_{B_1}\ket{0}_{C_1}\right)\ket{\overline{e}}_{C_2}
    \end{align*}
    which has rank 1 if and only if $f_0 = 0$, and $X_{\psi}$ consists of all vectors
    \begin{align*}
        \ket{x} = \ket{1}\ket{e}
    \end{align*}
    for $\ket{e} \in \CC^2$.
    As $X_{\phi}$ is a union of two planes, and $X_{\psi}$ is a single plane, they can not be related by an invertible linear map $M_A$, and we conclude that no restriction as in \eqref{eq:false restriction W} exists.
\end{proof}

In this case, the proof was relatively straightforward due to the fact that the restriction had to be invertible (since the tensors were concise on Hilbert spaces of the same dimension).
However, the proof is instructive since the same strategy can be used in a more general setting, where the map $M_A$ need not be invertible.

We use this more general approach to show that there is no restriction from the entanglement structure using EPR-pairs (so placing $\ket{\EPR{2}}_{\triangle}$, the $2 \times 2$ matrix multiplication tensor on each plaquette) to the entanglement structure using the $\lambda$-tensor
\begin{align*}
    \ket{\lambda} = \sum\limits_{i,j,k = 0}^2 \epsilon_{ijk} \ket{ijk} + \ket{222}
\end{align*}
where $\epsilon_{ijk}$ is the antisymmetric tensor.
It is already known \cite{christandl2020tensor} that on the level of a single plaquette there is no restriction
\begin{align}\label{eq:no restriction lambda}
    \ket{\epr_2}_{\triangle} \ngeq \ket{\lambda}
\end{align}
while there does exist a degeneration
\begin{align*}
    \ket{\epr_2}_{\triangle} \geqdeg \ket{\lambda}.
\end{align*}
Hence, there are also asymptotic restrictions $\ket{\epr_2}_{\triangle} \geqas \ket{\lambda}$ and a lattice degeneration $\ket{\epr_2}_{G} \geqdeg \ket{\lambda}_{G}$.
Since we need to be able to separate a degeneration from a restriction, the substitution method is again a natural choice in this case.

We let $G$ be the kagome lattice (or the half-filled triangular lattice), but without periodic boundary conditions.
We perform the folding at the boundary of the lattice; see \cref{fig:kagome-sub-boundary}

In this case, for $\ket{\lambda}_{G}$ we get a state equivalent to the three parties sharing one copy of $\ket{\lambda}$ and a level-3 EPR pair between $B$ and $C$.
We denote this state by $\ket{\lambda} \ot \ket{\epr_{3}}_{BC}$.
For the state $\ket{\epr_2}_{G}$ we get a state where there are level-2 $\epr$ pairs between $A$ and $B$ and between $A$ and $C$ and a level-8 $\epr$ pair between $B$ and $C$.
One can think of this as $\ket{\epr_2}_{\triangle} \ot \ket{\epr_{4}}_{BC}$
So, we would like to show
\begin{center}
    \vspace*{0.2cm}
    \begin{overpic}[height=1.2cm,grid=false]{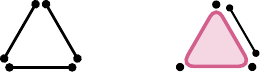}
        \put(47,12){$\ngeq$}
        \put(-5,-5){\scalebox{0.8}{$A$}} \put(30,-5){\scalebox{0.8}{$B$}}
        \put(14,29){\scalebox{0.8}{$C$}}
        \put(63,-5){\scalebox{0.8}{$A$}} \put(98,-5){\scalebox{0.8}{$B$}}
        \put(82,29){\scalebox{0.8}{$C$}}
        \put(81,8){$\lambda$}
        \put(0,14){\scalebox{0.8}{$2$}} \put(27,14){\scalebox{0.8}{$8$}} \put(14,-6){\scalebox{0.8}{$2$}}
        \put(97,14){\scalebox{0.8}{$3$}}
    \end{overpic}
    \vspace*{0.2cm}
\end{center}

\begin{restatable}{thm}{lambdarestriction}\label{thm:lambda restriction}
    There does not exist a restriction from $\ket{\epr_2}_{\triangle} \ot \ket{\epr_{4}}_{BC}$ to $\ket{\lambda} \ot \ket{\epr_{3}}_{BC}$.
\end{restatable}

The proof (based on the substitution method and generalizing the ideas in \cref{thm:w restriction}) can be found in \cref{sec:proof lambda bound}.
The folding argument implies that \cref{thm:lambda restriction} has as a direct consequence that the entanglement structure with the $\lambda$-tensor can not be represented with bond dimension 2 on any kagome lattice $G$ with a boundary, solving a main open problem from \cite{christandl2020tensor}.

\begin{cor}\label{cor:lambda kagome}
    \begin{align*}
        \ket{\epr_2}_{G} \ngeq \ket{\lambda}_{G}
    \end{align*}
\end{cor}

We conjecture that this should also not be possible with periodic boundary conditions, and explain how our methods could be used for this in \cref{sec:proof lambda bound}.

\subsection{Asymptotic restrictions and the asymptotic spectrum}
We already saw that at least for foldable hypergraphs (and permutation-invariant states), the non-existence of asymptotic restrictions is an obstruction to the existence of entanglement structure conversions.
We will now study in more depth the \emph{asymptotic conversion} of entanglement structures $\ket{\phi}_G \geqas \ket{\psi}_G$ as in \cref{sec:asymptotic tensor network conversions}.
There is a well-developed theory for asymptotic conversion of tensors.
It turns out that whether there exists an asymptotic conversion between two tensors can be decided by evaluating a compact set of appropriate functionals on the tensors, known as the \emph{asymptotic spectrum of tensors}.
Suppose we consider $k$-tensors.
Then we denote by $\mathcal{X}(k)$ the set of all functions on non-normalized quantum states (i.e. tensors)
\begin{align*}
    f:\lbrace k\text{-party quantum states}\rbrace\rightarrow \mathbb{R}_{\geq 0}
\end{align*}
which are such that they are
\begin{enumerate}
    \item monotone under restriction, so $\ket{\phi} \geq \ket{\psi}$ implies $f(\ket{\phi}) \geq f(\ket{\psi})$
    \item multiplicative under tensor products, so $f(\ket{\phi} \ot \ket{\psi}) = f(\ket{\phi})f(\ket{\psi})$
    \item additive under direct sums, so $f(\ket{\phi} \op \ket{\psi}) = f(\ket{\phi}) + f(\ket{\psi})$
    \item normalized by $f(\ket{\ghz_{r}(k)}) = r$.
\end{enumerate}
This collection of functionals $\mathcal{X}(k)$ can be given the structure of a compact topological space and is known as the asymptotic spectrum of tensors.
A function $f \in \mathcal{X}(k)$ is a \emph{point} in the asymptotic spectrum.
A cornerstone result of the algebraic theory of tensors is the following theorem by Strassen
\cite{strassen1986asymptotic, strassen1988asymptotic}.

\begin{thm}\label{thm:as spectrum}
    Let $\ket{\phi}$ and $\ket{\psi}$ be $k$-tensors.
    We have an asymptotic restriction $\ket{\phi} \geqas \ket{\psi}$ if and only if $f(\ket{\phi}) \geq f(\ket{\psi})$ for all $f \in \mathcal{X}(k)$.
\end{thm}

This gives a complete characterization of asymptotic restriction. However, we do not have complete knowledge of the asymptotic spectrum and only know a few points (note that if one could evaluate all $f \in \mathcal{X}(3)$ this would allow one directly to compute the matrix multiplication exponent $\omega$).
We will relate the asymptotic spectrum to asymptotic tensor network conversions.

Given $f \in \mathcal X(n)$ we can restrict to functionals which only act on a subset of the $n$ parties.
So, we assume we have $n$ parties $X = \{x_1,\dots,x_n\}$.
Let $k < n$ and let $Y$ be a subset of $X$ of size $\abs{Y} = k$.
Given a state $\ket{\phi_Y}$ on the $k$ parties $Y$ on a Hilbert space $\H_Y = \bigotimes_{x \in Y} \H_x$, define
\begin{align*}
    \ket{\phi_Y}_X = \ket{\phi_Y} \ot \bigotimes_{x \in X \setminus Y} \ket{0}.
\end{align*}
For any $f \in \mathcal X(n)$ we define a nonnegative function $f_Y$ on $k$ parties by
\begin{align*}
    f_Y(\ket{\phi_Y}) = f(\ket{\phi_Y}_X).
\end{align*}

\begin{lem}\label{lem:restriction of as spec}
    The functions $f_Y$ are monotone under restriction and multiplicative under tensor products.
    Moreover,
    \begin{align*}
        f_Y(\ket{\phi} \op \ket{\phi}) = \alpha f_Y(\ket{\phi})
    \end{align*}
    for
    \begin{align*}
        \alpha = f_Y(\ket{\ghz_2(k)_Y}).
    \end{align*}
\end{lem}

\begin{proof}
    The functions $f_Y$ are monotone under restriction and multiplicative under tensor products.
    Indeed, a restriction $\ket{\phi_Y} \geq \ket{\psi_Y}$ implies $\ket{\phi_Y}_X \geq \ket{\psi_Y}_X$, so using the monotonicity of $f$ we have $f_Y(\ket{\phi_Y}) \geq f_Y(\ket{\psi_Y})$.
    Similarly, $(\ket{\phi_Y} \ot \ket{\psi_Y})_X = \ket{\phi_Y}_X \ot \ket{\psi_Y}_X$, so
    \begin{align*}
        f_Y(\ket{\phi_Y} \ot \ket{\psi_Y}) & = f(\ket{\phi_Y}_X \ot \ket{\psi_Y}_X) \\
                                           & = f(\ket{\phi_Y}_X)f(\ket{\psi_Y}_X)   \\
                                           & = f_Y(\ket{\phi_Y})f_Y(\ket{\psi_Y}).
    \end{align*}
    In general, $f_Y$ need not be additive or normalized.
    For any $\ket{\phi_Y}$ we have
    \begin{align*}
        (\ket{\phi_Y} \op \ket{\phi_Y})_X \cong (\ket{\phi_Y} \ot \ket{\ghz_2(k)_{Y}})_X
    \end{align*}
    and hence, by multiplicativity
    \begin{align*}
        f_Y(\ket{\phi_Y} \op \ket{\phi_Y}) = f_Y(\ket{\phi_Y}) f_Y(\ket{\ghz_2(k)_Y}).
    \end{align*}
\end{proof}

While we do not necessarily get additivity, for the $f_Y$, multiplicativity and monotonicity are in some sense the most important properties of the points in the asymptotic spectrum, since these properties directly relate to the notion of asymptotic restriction.
We will apply this to the case where we have a graph $G$, so we may consider the asymptotic spectrum $\mathcal X(n)$ where $n = \abs{V}$.
Then as in \cref{lem:restriction of as spec} we get functions $f_e$ for each edge $e$.

\begin{thm}\label{thm:as spectrum network}
    Suppose $G$ is a hypergraph with $\abs{V} = n$ and $\ket{\phi}_G$ and $\ket{\psi}_G$ are entanglement structures.
    Then $\ket{\phi}_G \geqas \ket{\psi}_G$ if and only if for all $f \in \mathcal{X}(n)$
    \begin{align*}
        \prod_{e \in E} f_e(\ket{\phi_e}) \geq \prod_{e \in E} f_e(\ket{\psi_e})
    \end{align*}
\end{thm}

\begin{proof}
    By \cref{thm:as spectrum} we have $\ket{\phi}_G \geqas \ket{\psi}_G$ if and only if
    \begin{align*}
        f\left(\bigotimes_{e \in E} \ket{\phi_e}\right) \geq f\left(\bigotimes_{e \in E} \ket{\phi_e}\right)
    \end{align*}
    for all $f \in \mathcal{X}(n)$.
    By definition, $f \in \mathcal X(n)$ is multiplicative under tensor products.
    This gives
    \begin{align*}
        f\left(\bigotimes_{e \in E} \ket{\phi_e}\right) = \prod_{e \in E} f_e(\ket{\phi_e})
    \end{align*}
    and similarly for $\ket{\psi}_G$.
\end{proof}

Note that in \cref{thm:as spectrum network} the functions $f_e$ are not independent, as they have to derive from a global $f \in \mathcal X(n)$. So, the condition in \cref{thm:as spectrum network} is \emph{not} equivalent to $f_e(\ket{\phi_e}) \geq f_e(\ket{\psi_e})$ for all edges.
This makes sense, as in general $\ket{\phi}_G \geqas \ket{\psi}_G$ does not imply conversion on the level of individual plaquettes, as in the basic example
\begin{center}
    \begin{overpic}[height=2.5cm,grid=false]{epr-lattice}
        \put(2,14){$\ket{\phi}$} \put(31,14){$\ket{\psi}$}
        \put(47,20){$\Rightarrow$}
    \end{overpic}
\end{center}
In this case $\ket{\phi}_G$ and $\ket{\psi}_G$ are actually equivalent (and therefore asymptotically equivalent), so we have $f(\ket{\phi}_G) = f(\ket{\psi}_G)$ for all $f \in \mathcal X(n)$.
On the other hand, an appropriate rank functional clearly distinguishes single plaquettes $\ket{\phi_e}$ and $\ket{\psi_e}$.

To be able to make use of \cref{thm:as spectrum network}, we need to know concrete points in the asymptotic spectrum.
Easy examples are given by ranks across bipartitions of the parties.
This of course is equivalent to previous obstructions obtained by folding.
However, there exists a broad class of examples (in fact encompassing all known examples), which are the so-called \emph{quantum functionals} as introduced in \cite{christandl2023universal}.
These are defined as follows.
For a $k$-particle quantum state $\ket{\phi}$ we denote by $\rho^{(\phi)}_{i}$ the (mixed) quantum state resulting from tracing out all but the $i$'th system of the normalized state
\begin{align*}
    \rho^{(\phi)} = \frac{1}{\braket{\phi | \phi}}\proj{\phi}.
\end{align*}
The von Neumann entropy of the reduced state $\rho^{(\phi)}_{i}$ is given by
\begin{align*}
    H(\rho^{(\phi)}_{i}) = - \tr\mleft[ \rho^{(\phi)}_{i} \log \rho^{(\phi)}_{i} \mright]
\end{align*}
For a probability distribution $\theta = \{\theta_1,\dots, \theta_n\}$ the corresponding quantum functional is defined as $F_\theta = 2^{E_\theta}$ with
\begin{align*}
    E_\theta(\ket{\phi})=\sup_{\ket{\phi} \geqdeg \ket{\psi}} \; \sum_{i = 1}^n \theta_i H(\rho^{(\psi)}_{i}).
\end{align*}
The set that one optimizes over is closely related to the \emph{entanglement polytope} of $\ket{\phi}$, which is the set of all spectra of $\rho^{(\psi)}_{i}$ that can be obtained from a degeneration $\ket{\phi} \geqdeg \ket{\psi}$ \cite{walter2013entanglement}.
A dual characterization of the entanglement polytope is crucial in showing that the quantum functionals are multiplicative.

As an example application, consider again the $W$ state on three parties.
The $W$ state does not asymptotically restrict to the level-2 $\ghz$ state \cite{christandl2023universal} which can be seen by computing the quantum functional using the uniform distribution $\theta = \{\frac{1}{3},\frac{1}{3},\frac{1}{3}\}$, in which case
\begin{align*}
    F_{\theta}(\ket{\ghz_{2}(3)}) = 2 > F_{\theta}(\ket{W(3)}) = 2^{H(\{\frac{1}{3},\frac{2}{3}\})} \approx 1.89.
\end{align*}

The quantum functionals behave especially nice when applied as in \cref{thm:as spectrum network}.
In this case, choose a distribution $\theta = \{\theta_v : v \in V\}$.
We now see that (by multiplicativity)
\begin{align*}
    E_{\theta}(\ket{\phi}) & = \sum_{e \in E} \sum_{v \in V} E_{\theta_e}(\ket{\phi_e}) \\
                           & = \sum_{e \in E} \Theta_e E_{\theta^{(e)}}(\ket{\phi_e})
\end{align*}
where
\begin{align*}
    \Theta_e = \sum_{v \in e} \theta_v \qquad \text{and} \qquad \theta^{(e)} = \{\theta_v^{(e)} = \frac{\theta_v}{\Theta_e} : v \in e\}.
\end{align*}
So, $E_{\theta}(\ket{\phi})$ is a convex combination of quantum functionals $E_{\theta^{(e)}}(\ket{\phi_e})$.
In particular, if in \cref{thm:as spectrum network} we take $f = F_{\theta}$, then the normalized restriction to $e$ given by $f_e^{\Theta_e^{-1}}$ is again a quantum functional.
This is stronger than \cref{lem:restriction of as spec}, as it implies that (after normalization) we also have additivity of the restricted functional.

If we have a graph (so a 2-uniform hypergraph) with $n$ vertices and we place maximally entangled states of bond dimension $D_e$ on each edge giving an entanglement structure $\ket{\phi}_G$, we find that taking the uniform distribution $\theta$ with $\theta_v = \frac{1}{n}$ for $v \in V$, the quantum functional gives
\begin{align*}
    F_{\theta}(\ket{\phi}_G) = \prod_{e \in E} D_e^{\frac{2}{n}}
\end{align*}
and more generally for an arbitrary distribution $\theta$ we find
\begin{align*}
    F_{\theta}(\ket{\phi}_G) = \prod_{e = (vw) \in E} D_e^{\theta_v + \theta_w}.
\end{align*}
This can be used to lower bound the asymptotic bond dimension required for some entanglement structure of interest.

As an example application, the quantum functionals allow one to conclude that on \emph{any} hypergraph (so it does not have to be foldable), we can not have an (asymptotic) conversion of an entanglement structure using $W$ states to one using $\ghz_2(3)$ states.
A small example of such a graph which can not be folded is
\begin{center}
    \includegraphics[height=1.4cm]{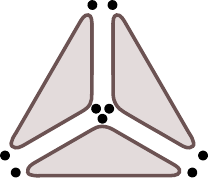}
\end{center}
as can be easily seen.

\begin{thm}
    Let $G$ be any 3-uniform hypergraph, and let $\ket{W}_G$ and $\ket{\ghz_{2}(3)}_G$ be the entanglement structures where each edge is assigned a $W$ state or $\ghz_2(3)$ state respectively.
    Then there is \emph{no} asymptotic restriction $\ket{W}_G \geqas \ket{\ghz_{2}(3)}_G$.
\end{thm}

In particular, there is also no restriction, so $\ket{W}_G \ngeq \ket{\ghz_{2}(3)}_G$.

\begin{proof}
    For the uniform distribution over the vertices, we find that each $\theta^{(e)}$ is a uniform distribution, and therefore
    \begin{align*}
        F_{\theta}(\ket{\ghz_{2}(3)}_G) > F_{\theta}(\ket{W}_G)
    \end{align*}
    which implies there is no asymptotic restriction.
\end{proof}

\section{Symmetries and ranks of entanglement structures}\label{sec:sym and rank}
We end our investigation of the resource theory of tensor networks by studying some structural properties of entanglement structures.
An important feature of tensor networks is the so-called \emph{gauge symmetry} which implies that the same quantum state can have equal bond dimension representations with different tensors applied.
This is a crucial ingredient in relating tensor network states to phases of matter.
Here, we will introduce this gauge symmetry for arbitrary entanglement structures.
This gauge symmetry has rather different properties than in the standard tensor network formalism.
For one, the gauge group is potentially a finite group. Secondly, we show that there are entanglement structures which have gauge symmetries which act on multiple edges.
We show that this does not happen for acyclic hypergraphs.
After this we turn to the \emph{tensor rank} of entanglement structures.
We work out a nontrivial example, solving an open question in \cite{chen2018tensor}.

\subsection{Stabilizers of entanglement structures}\label{subsec:stabilizers of entanglement structures}

We will now turn our attention to the stabilizer group, or gauge group, of entanglement structures.
Given an entanglement structure with Hilbert spaces $\H_v$ of dimension $d_v$ at vertex $v \in V$, we have the group $\prod_{v \in V} \GL(d_v)$ which acts on $\ket{\phi}_G$ by
\begin{align*}
    (g_v)_{v \in V} \cdot \ket{\phi}_G = \left(\bigotimes_{v \in V} g_v \right) \ket{\phi}_G.
\end{align*}

Then we define the \emph{stabilizer} of the entanglement structure as
\begin{align*}
    \stab(\ket{\phi}_G) = & \{(g_v \in \GL(d_v))_{v \in V}                                                 \\
                          & \qquad \text{ such that } (g_v)_{v \in V} \cdot \ket{\phi}_G = \ket{\phi}_G\}.
\end{align*}
A well-known example is the situation where we have maximally entangled states along edges.
If we have a maximally entangled state of level $D$ at some edge, we may act with $g \in \GL(D)$ on one side of the edge and with $g^{-\tsp} = (g^{-1})^{\tsp}$ on the other side.
\begin{center}
    \begin{overpic}[width=2cm,grid=false]{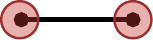}
        \put(8,-13){\color{red}{$g$}} \put(83,-13){\color{red}{$g^{-T}$}}
    \end{overpic}
    \vspace*{0.2cm}
\end{center}
This gives an element in the stabilizer since
\begin{align*}
    (g \ot g^{-T}) \sum_{i=1}^D \ket{ii} = \sum_{i=1}^D \ket{ii}.
\end{align*}
This \emph{gauge symmetry} is widely studied in the context of tensor networks and is an essential ingredient in the classification of phases and symmetries using the tensor network formalism \cite{cirac2021matrix}.
Note that it implies that for a tensor network state with tensors $T_v$, there exist transformations on the tensors $T_v$ which change the local tensors but do \emph{not} change the resulting tensor network state.
More generally, given an arbitrary entanglement structure we get elements in the stabilizer by looking at the stabilizers for single edges, that is, if $e = \{v_1,\dots,v_k\}$ we may look at $g_{e,v_i} \in \GL(\H_{e,v})$ such that
\begin{align*}
    (g_{e,v_1} \ot \dots \ot g_{e,v_k}) \ket{\phi_e} = \ket{\phi_e}.
\end{align*}
Such $g_{e,v_i}$ is called an \emph{edge stabilizer}.
Then it is clear that
\begin{align*}
    (g_v)_{v \in V}, \qquad g_v = \bigotimes_{e : v \in e} g_{e,v}
\end{align*}
is an element of $\stab(\ket{\phi}_G)$.
In the special case of an $\epr$ pair, this group was a continuous group, but in general this group can be finite \cite{gour2011necessary}.

An interesting question is whether \emph{all} symmetries of the entanglement structure arise as a symmetry of the edges states.
In other words, the question is whether every element of $\stab(\ket{\phi}_G)$ arises as a product of edge stabilizers.
The answer is \emph{no} as the following example shows.
Consider the hypergraph $G$ consisting of two 3-edges with two common vertices.
As entanglement structure we place a $W$ state on both edges.
For $q \in \CC$, let $g_q$ be defined by
\begin{align}\label{eq:symmetry W 4}
    \begin{split}
        g(q) &: (\CC^2)^{\ot 2} \to (\CC^2)^{\ot 2} \\
        g(q) &= \id + q\ket{00}\bra{11}.
    \end{split}
\end{align}
This is clearly not a tensor product operator for $q \neq 0$.
We now apply $g(q)$ and $g(-q)$ to the vertices of degree 2
\begin{center}
    \vspace{0.2cm}
    \begin{overpic}[height=1.8cm,grid=false]{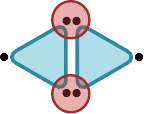}
        \put(51,85){\color{red}{$g(q)$}} \put(51,-10){\color{red}{$g(-q)$}}
        \put(23,34){$W$} \put(62,34){$W$}
    \end{overpic}
\end{center}
This stabilizes $\ket{W}_G$, but is not a product of edge stabilizers.

On the other hand, for \emph{acyclic hypergraphs} we will show that the stabilizer is just the product of the edge stabilizers.

\begin{lem}\label{lem:product stabilizer}
    Suppose $G$ has two edges $e_1$ and $e_2$, sharing a single vertex $v$.
    Let $\ket{\phi}_G = \ket{\phi_{e_1}}\ket{\phi_{e_2}}$ and $\ket{\psi}_G = \ket{\psi_{e_1}}\ket{\psi_{e_2}}$ be an entanglement structure with $\ket{\phi_1}$ and $\ket{\phi_2}$ concise, and suppose that $M_v$ is such that
    \begin{align*}
        (M_v \ot \id_{V \setminus v})\ket{\phi}_G = \ket{\psi}_G.
    \end{align*}
    Then there exist $M_i$ acting only on $\H_{e_i,v}$ such that $M = M_1 \ot M_2$.
\end{lem}
So, if we have
\begin{center}
    \begin{overpic}[height=1.2cm,grid=false]{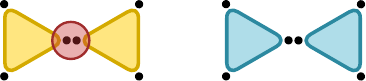}
        \put(48,10){$=$}
        \put(16,18){\color{red}{$M$}}
        \put(3,9.5){\scalebox{0.8}{$\phi_{e_1}$}} \put(29,9.5){\scalebox{0.8}{$\phi_{e_2}$}}
        \put(64,9.5){\scalebox{0.8}{$\psi_{e_1}$}} \put(88,9.5){\scalebox{0.8}{$\psi_{e_2}$}}
    \end{overpic}
\end{center}
then
\begin{center}
    \begin{overpic}[height=1.2cm]{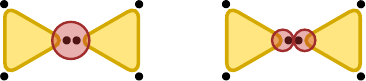}
        \put(16,18){\color{red}{$M$}}
        \put(48,10){$=$}
        \put(72,17){\scalebox{0.8}{\color{red}{$M_1$}}} \put(81.5,17){\scalebox{0.8}{\color{red}{$M_2$}}}
    \end{overpic}
\end{center}

\begin{proof}
    We group together all vertices in respectively $e_1$ and $e_2$ into parties $A$ and $B$.
    Then we consider Schmidt decompositions
    \begin{align*}
        \ket{\phi_1} = \sum_{i=1}^{r_1} s_{i} \ket{a_{i}}\ket{e_{i}} \qquad \ket{\phi_2} = \sum_{i=1}^{r_1} t_{i} \ket{b_{i}}\ket{f_{i}}
    \end{align*}
    where the $\ket{e_{i}}$ and $\ket{f_i}$ are orthonormal bases of $\H_{e_1,v}$ and $\H_{e_1,v}$ (by conciseness).
    Now, the map $M_v$ is completely defined by
    \begin{align*}
        M_v \ket{e_i}\ket{f_j} & = \Big(\id_v \ot \bra{a_i} \bra{b_j} \Big)(M_v \ot \id_{V \setminus v})\ket{\phi}_G          \\
                               & = \Big(\id_v \ot \bra{a_i} \bra{b_j}\Big) \ket{\psi}_G                                       \\
                               & = \Big((\id_v \ot \bra{a_i}) \ket{\psi_1}\Big) \Big((\id_v \ot \bra{b_j}) \ket{\psi_2}\Big).
    \end{align*}
    This is again a tensor product state, so we see that we may take
    \begin{align*}
        M_1 & = \sum_i \Big(\id_v \ot \bra{a_i}\Big) \ket{\psi_1} \\
        M_2 & = \sum_j \Big(\id_v \ot \bra{b_j}\Big) \ket{\psi_2}
    \end{align*}
\end{proof}

\begin{thm}\label{thm:stabilizer tree}
    If $G$ is an acyclic hypergraph and each $\ket{\phi_e}$ is concise, all elements of $\stab(\ket{\phi}_G)$ are products of edge stabilizers.
\end{thm}

\begin{proof}
    Take $(g_v)_{v \in V} \in \stab(\ket{\phi}_G)$.
    Pick an arbitrary vertex $v \in V$, and divide $V \setminus \{v\}$ into two groups $V_1$ and $V_2$ such that there are no edges between $V_1$ and $V_2$ (this is possible since the hypergraph is acyclic). Let $E_i$ be the set of edges $e$ for which $e \subset E_i \cup \{v\}$. The disjoint union of $E_1$ and $E_2$ is the full edge set $E$.
    Now, define a hypergraph $H$ with vertex set $V$ and edge set $F = \{f_1, f_2\}$, where $f_i = V_i \cup \{v\}$.
    Define $\ket{\phi}_H$ by
    \begin{align*}
        \ket{\phi_{f_i}} = \bigotimes_{e \in E_i} \ket{\phi_e}.
    \end{align*}
    That is, we coarse-grain the hypergraph as
    \begin{center}
        \begin{overpic}[height=2.3cm,grid=false]{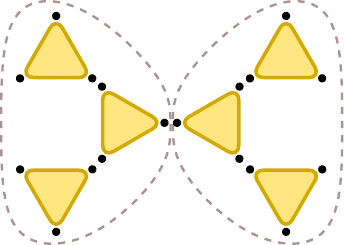}
            \put(51,40){$v$}
            \put(10,30){$f_1$}
            \put(80,30){$f_2$}
        \end{overpic}
    \end{center}
    Let
    \begin{align*}
        \ket{\psi}_H = \left(\id_v \ot \left(\bigotimes_{v \in V \setminus \{v\}} g_v\right)\right)\ket{\phi}_H.
    \end{align*}
    Note that $\ket{\psi}_H$ is again an entanglement structure on $H$.
    Then, $(g_v \ot \id_{V \setminus \{v\}})\ket{\phi}_H = \ket{\psi}_H$, and by \cref{lem:product stabilizer} there exist $g_{v,1}, g_{v,2}$ such that $g_v = g_{v,1} \ot g_{v,2}$.
    Since we showed this for an arbitrary vertex and arbitrary assignment of edges, we conclude that each $g_v$ can be written as a tensor product over the adjacent edges
    \begin{align*}
        g_v = \bigotimes_{e : v \in e} g_{e,v}.
    \end{align*}
    This decomposition is unique up to a choice of phases. After an appropriate choice of phases, in order for $(g_v)_{v \in V}$ to be a stabilizer of $\ket{\phi}_G$ we have
    \begin{align*}
        \left(\bigotimes_{v \in e} g_{e,v}\right) \ket{\phi_e},
    \end{align*}
    and we conclude that $(g_v)_{v \in V} \in \stab(\ket{\phi}_G)$ is a product of edge stabilizers.
\end{proof}

This result supplements \cref{cor:restriction tree}, which shows that on an acyclic graph $\ket{\phi}_G \geq \ket{\psi}_G$ implies $\ket{\phi_e} \geq \ket{\psi_e}$ for each edge $e \in E$.

\subsection{Ranks of entanglement structures}\label{subsec:ranks of ent struc}
We will end by making some observations regarding the tensor rank of entanglement structures. Recall that for a quantum state $\ket{\phi}$ on $k$ parties we call
\begin{equation*}
    \R(\ket{\phi}) = \min \left\{ r: \ket{\phi} = \sum_{i = 1}^{r} \ket{u_{i1}}\dots \ket{u_{ik}}, u_{ij}\in U_j \right\}
\end{equation*}
the rank of $\ket{\phi}$. In other words, it is the minimal number of product states whose linear span contains $\ket{\phi}$, or it is the minimal $r$ such that there exists a restriction $\ghz_{r}(k) \geq \ket{\phi}$.
In the case of 3-tensors it is closely related to the complexity of computing bilinear operations (e.g. the complexity of matrix multiplication is related to the tensor rank of $\ket{\epr_n}_{\triangle}$ as discussed in \cref{subsec:tensors}).
The rank of a quantum state is a natural measure of multipartite entanglement (as it is the entanglement cost from $\ghz$ states under SLOCC) \cite{chen2010tensor}, and it is interesting to study this measure for entanglement structures.

Given a hypergraph $G$ and entanglement structure $\ket{\phi}_G$ with state of $\ket{\phi_e}$ on the plaquettes of $G$ it is interesting to compare the ranks of the individual $\ket{\phi_e}$ with the rank of the entanglement structure $\ket{\phi}_G$.
It is immediate that
\begin{align}\label{eq:rankinequpsihpsi}
    \R(\ket{\phi}_G) \leq \prod_{e \in E} \R(\ket{\phi_e}).
\end{align}
This inequality can be an equality but is not so in general \cite{christandl2018tensor}.
The same is true for the border rank \cite{christandl2019border}.
The tensor rank $\R(\ket{\phi}_G)$ is well-studied in the special case where $G$ is a graph and the $\ket{\phi_e}$ are two-party states (and then we may assume without loss of generality they are $\epr$ pairs).
For example, when $G$ has no cycles, one has equality in \cref{eq:rankinequpsihpsi}.
More generally, it is easy to see that we have equality if the graph is bipartite \cite{christandl2019tensor}, since we get a sharp flattening rank lower bound from the bipartitioning of the graph. This for instance allows one to compute the rank of an entanglement structure with $\epr$ pairs on a square lattice.

On the other hand, for example for any cycle of odd length one has strict inequality in \cref{eq:rankinequpsihpsi} (with the cycle of length 3 corresponding to matrix multiplication) \cite{christandl2019tensor}.
The asymptotic version (computing $\asR(\ket{\phi}_G)$) of this question has been studied in \cite{christandl2019asymptotic} for complete graphs.

For hypergraphs with edges with $\abs{e} > 2$, in general we can even have strict inequality in \cref{eq:rankinequpsihpsi} if the hypergraph is completely disconnected \cite{christandl2018tensor}.
This occurs for instance when one takes two copies of the $W$ state, as already discussed in \cref{subsec:prior work}.
All together this leads to the following question:

\begin{center}
    \emph{Given $n$ copies of a $k$-party state $\ket{\phi}$, and a $k$-regular hypergraph $G$ with $n$ edges, how does $\R(\ket{\phi}_G)$ depend on the structure of $G$?}
\end{center}

In this work we only briefly touch on this question. As an example, we completely work out the (nontrivial) answer in the case of two $W$ states.
We have the following four possible graphs
\begin{center}
    \begin{overpic}[height=3cm,grid=false]{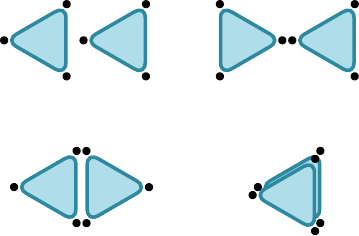}
        \put(20,33){$G_6$} \put(80,33){$G_5$}
        \put(20,-10){$G_4$} \put(80,-10){$G_3$}
    \end{overpic}
    \vspace*{0.2cm}
\end{center}
By folding we see that
\begin{align}\label{eq:folding w}
    \R(\ket{W}_{G_6}) \geq \R(\ket{W}_{G_5}) \geq \R(\ket{W}_{G_4}) \geq \R(\ket{W}_{G_3}).
\end{align}
In \cite{yu2010tensor} it was shown that
\begin{align*}
    \R(\ket{W}_{G_3}) = 7
\end{align*}
(i.e. the Kronecker product of two copies of $\ket{W}$ has rank 7).
On the other hand, in \cite{christandl2018tensor} it was shown that $\R(\ket{W}_{G_6}) \leq 8$ and in~\cite{chen2018tensor} that $\R(\ket{W}_{G_6}) \geq 8$ yielding that the tensor product of two copies of $\ket{W}$ has
\begin{align*}
    \R(\ket{W}_{G_6}) = 8.
\end{align*}
Thus, with \cref{eq:folding w} we know
\begin{align*}
    8 = \R(\ket{W}_{G_6}) & \geq \R(\ket{W}_{G_5})      \\
                          & \geq \R(\ket{W}_{G_4})      \\
                          & \geq \R(\ket{W}_{G_3}) = 7.
\end{align*}

In \cref{sec:rank w} we prove

\begin{restatable}{thm}{rankw}\label{thm:rank 2 w}
    \begin{align*}
        \R(\ket{W}_{G_4}) = 8.
    \end{align*}
\end{restatable}

This solves an open problem in \cite{chen2018tensor}, and completes the characterization of the tensor rank of two copies of the $W$ state on different hypergraphs
\begin{align*}
    \R(\ket{W}_{G_6}) & = \R(\ket{W}_{G_5}) = \R(\ket{W}_{G_4}) = 8 \\
    \R(\ket{W}_{G_3}) & = 7.
\end{align*}

In general, for acyclic hypergraphs one might expect that since their stabilizer group equals that of the separate edge, as we have shown in \cref{thm:stabilizer tree}, the rank of an entanglement structure on the acyclic hypergraph equals that of the rank one has when one just takes the tensor product. That is, given an entanglement structure $\ket{\phi}_G$ on an acyclic hypergraph $G$, one may consider the completely disconnected graph $H$ on the same number of edges (and we define $\ket{\phi}_H$ by just placing the same edge states $\ket{\phi_e}$ on these disconnected edges).
We then have the open question whether for any acyclic hypergraph
\begin{align*}
    \R(\ket{\phi}_G) \stackrel{?}{=} \R(\ket{\phi}_H).
\end{align*}

\section{Outlook and discussion}\label{sec:conclusion}
We have demonstrated that there is a rich resource theory of different ways of distributing \emph{local} multiparty entanglement over hypergraphs, such as lattices, under applying local operations. This is a resource theory of tensor networks, in the sense that it allows us to compare different entanglement structures as a resource for creating many-body tensor network states.
This resource theory establishes a framework which accommodates the previously proposed special instances of tensor networks using multiparty entangled states, extending beyond the traditional use of maximally entangled pair states.
The resource theory effectively generalizes the notion of bond dimension, allowing for a systematic comparison of different entanglement structures.

Our first main result is that this resource theory goes beyond transformations which only act on individual plaquette states and that certain transformations only become possible on a lattice.
By enabling transformations between different entanglement structures, our approach allows for the conversion between different tensor network representations of the same quantum many-body state. This may for example be used to relate representations which are natural from the physics of a many-body Hamiltonian, to representations which are practical and efficient with respect to computational implementations.
We have showcased a number of principles which can be used to construct such transformations, laying the groundwork for systematic development of transformations between entanglement structures.
In the converse direction, we provide techniques to prove obstructions to the existence of transformations of entanglement structures on a lattice.
For these methods we draw on powerful techniques developed in order to understand the computational complexity of matrix multiplication.
We have adapted generalized flattening bounds, the substitution method and asymptotic spectral functionals to lattice problems and use these to prove no-go theorems for transformations between entanglement structures.
These results illustrate that the profound mathematics developed in algebraic complexity theory has interesting applications in the theory of tensor networks.
The resource theory of tensors has been developed in different scientific communities, on the one hand to study the complexity of matrix multiplication and on the other hand to classify multipartite entanglement with respect to SLOCC transformations.
Interestingly, the resource theory of tensor networks operates at the intersection of these two viewpoints.
Indeed, our fundamental object of interest are local entanglement structures on a lattice; at the same time, tensor networks are a computational tool and we show that the resource theory of tensor networks is directly related to the complexity of tensor network contractions.

The resource theory of tensor networks opens up numerous promising avenues for future research.
In the algorithmic direction one could further advance the usage of entanglement structures in variational methods, for instance following up on the proposal \cite{christandl2021optimization}. It is also clear that the resource theory of tensor networks can be applied to infinite Projected Entangled Pair States (iPEPS) \cite{jordan2008classical} where finding good resource states could lead to more efficient algorithms.
Finally, the connection to algebraic complexity theory we have emphasized may lead to the development of improved contraction algorithms; or in the converse direction could be used to import tensor network algorithms to different problems in computer science.
Besides this it would also be interesting to explore the relevance of our resource theory to the theory of condensed matter systems, for instance relating to (symmetry protected) topological phases and using symmetries of entanglement structures to define new canonical forms of tensor networks, for instance using the framework of \cite{acuaviva2023minimal}. Finally, we believe that our viewpoint on tensor networks may also be of benefit outside the realm of quantum many-body states and quantum computation, in particular in more data-driven subjects such as machine learning.

\section*{Acknowledgement}
We acknowledge financial support from the European Research Council (ERC Grant Agreement No. 818761), VILLUM FONDEN via the QMATH Centre of Excellence (Grant No.10059) and the Villum Young Investigator program (Grant No. 25452)  and the Novo Nordisk Foundation (grant NNF20OC0059939 ‘Quantum for Life’).
MC thanks the National Center for Competence in Research SwissMAP of the Swiss National Science Foundation and the Section of Mathematics at the University of Geneva for their hospitality.
VL additionally acknowledges financial support from the European Union (ERC Grant No. 101040907). Views and opinions expressed are however those of the author(s) only and do not necessarily reflect those of the European Union or the European Research Council Executive Agency. Neither the European Union nor the granting authority can be held responsible for them.

\bibliographystyle{plainnat}
\bibliography{transformingentanglementstructure}
\newpage
\begin{appendix}

    \section{Exact tensor network representations from degenerations}\label{sec:degenerations appendix}

    In the resource theory of tensors it is often useful to consider degenerations rather than restrictions.
    The same is true for the resource theory of tensor networks.
    This notion is not standard in the tensor network literature.
    However, as already alluded to in \cref{subsec:prior work} in important cases one can convert degenerations into restrictions at small overhead.
    Here we briefly review this result, which is Theorem 14 in \cite{christandl2020tensor}, see also \cite{bini1980approximate,christandl2018tensor}.

    \begin{thm}
        Let $\ket{\phi}_G$ and $\ket{\psi}_G$ be entanglement structures and let
        \begin{align*}
            \ket{\Psi} = \left(\bigotimes_{v \in V} M_v \right)\ket{\psi}_G
        \end{align*}
        be a tensor network state using entanglement structure $\ket{\psi}_G$ (i.e. $\ket{\Psi} \leq \ket{\psi}_G$).
        \begin{enumerate}
            \item If $\ket{\phi}_G \geqdeg^d \ket{\psi}_G$ then
                  \begin{align*}
                      \ket{\Psi} = \sum_{i = 0}^{d} \ket{\Psi_i}
                  \end{align*}
                  where each of the $\ket{\Psi_i}$ has a tensor network representation using $\ket{\phi}_G$, i.e. $\ket{\Psi_i} \leq \ket{\phi}_G$.
                  Moreover, there exist $2d + 1$ tensor network states $\ket{\Phi_i} \leq \ket{\phi}_G$ and constants $\gamma_i$ such that for any observable $O$ we have
                  \begin{align*}
                      \bra{\Psi} O \ket{\Psi} = \sum_{i = 0}^{2d} \gamma_i \bra{\Phi_i} O \ket{\Phi_i}.
                  \end{align*}
                  \item\label{it:single plaquette degeneration} In the special case where each plaquette satisfies $\ket{\phi_e} \geqdeg^d \ket{\psi_e}$, we have
                  \begin{align*}
                      \ket{\phi}_G \geqdeg^{d\abs{E}} \ket{\psi}_G.
                  \end{align*}
        \end{enumerate}
    \end{thm}

    This means that if we have a single-plaquette degeneration as in \ref{it:single plaquette degeneration}, and $\ket{\Psi} \leq \ket{\psi}_G$ is a tensor network state using entanglement structure $\ket{\psi}_G$, then we can reduce to \emph{exact} tensor network computations using the entanglement structure $\ket{\phi}_G$ with an overhead which is only \emph{linear} in the system size (which is modest compared to potential gains from getting more efficient representations).

    \section{\texorpdfstring{Proof of \cref{thm:lambda restriction}}{Proof of Theorem 17}}\label{sec:proof lambda bound}

    \cref{thm:lambda restriction} states that there is no restriction
    \begin{center}
        \vspace*{0.2cm}
        \begin{overpic}[height=1.2cm,grid=false]{folded-lambda-boundary}
            \put(47,12){$\ngeq$}
            \put(-5,-5){\scalebox{0.8}{$A$}} \put(30,-5){\scalebox{0.8}{$B$}}
            \put(14,29){\scalebox{0.8}{$C$}}
            \put(61,-5){\scalebox{0.8}{$A'$}} \put(98,-5){\scalebox{0.8}{$B'$}}
            \put(82,29){\scalebox{0.8}{$C'$}}
            \put(81,8){$\lambda$}
            \put(0,14){\scalebox{0.8}{$2$}} \put(27,14){\scalebox{0.8}{$8$}} \put(14,-6){\scalebox{0.8}{$2$}}
            \put(97,14){\scalebox{0.8}{$3$}}
        \end{overpic}
        \vspace*{0.2cm}
    \end{center}
    where we have labelled the parties on the right-hand side with a prime to indicate that they have different Hilbert spaces.
    We will follow a strategy which is a generalization of the substitution method we used for \cref{thm:w restriction}.

    Consider the general situation where we have 3-tensors $\ket{\phi_{ABC}}$ and $\ket{\psi_{A'B'C'}}$ on three parties.
    The Hilbert spaces are potentially different; we will assume that $\dim(\H_A) \geq \dim(\H_{A'})$.

    For $\ket{x} \in \H_A$, similar to before we let
    \begin{align*}
        \ket{\phi^{(x)}} = (\bra{x} \ot \id_{BC}) \ket{\phi_{ABC}}
    \end{align*}
    and similarly for $\ket{\psi_{ABC}^{(x)}}$ for $\ket{x} \in \H_{A'}$.
    These are 2-tensors on systems $B$ and $C$ (or $B'$ and $C'$).
    As before we define
    \begin{align*}
        X_{\phi}^{(k)} & = \{\ket{x} \text{ such that } \rk(\ket{\phi^{(x)}}) \leq k \}  \\
        X_{\psi}^{(k)} & = \{\ket{x} \text{ such that } \rk(\ket{\psi^{(x)}}) \leq k \}.
    \end{align*}

    We now assume that $\ket{\phi} \geq \ket{\psi}$, so there exists a transformation $(M_A \ot M_B \ot M_C)\ket{\phi} = \ket{\psi}$. If this is the case, we can relate $X_{\phi}^{(k)}$ and $X_{\psi}^{(k)}$ by the following two lemmas.

    \begin{lem}\label{lem:M dagger}
        Suppose that $\ket{\phi} \geq \ket{\psi}$.
        If for $\ket{y} \in \H_{A'}$ we have $M_A^{\dagger}\ket{y} \in X_{\phi}^{(k)}$, then $\ket{y} \in X_{\psi}^{(k)}$.
    \end{lem}

    \begin{proof}
        If $M_A^{\dagger}\ket{y} \in X_{\phi}^{(k)}$, and we let $\ket{x} = M_A^{\dagger}\ket{y}$ then
        \begin{align*}
            \rk(\ket{\psi}^{(y)}) & = \rk((\bra{y}M_A \ot M_B \ot M_C)\ket{\phi}) \\
                                  & = \rk((M_B \ot M_C) \ket{\phi^{(x)}})         \\
                                  & \leq \rk(\ket{\phi^{(x)}}) \leq k
        \end{align*}
        so $\ket{y} \in X_{\psi}^{(k)}$.
    \end{proof}

    \begin{lem}\label{lem:subspace}
        Suppose that $\ket{\phi} \geq \ket{\psi}$ and suppose that the reduced density matrix of $\ket{\psi}$ has full rank on $A'$. Let $U \subseteq \H_{A}$ be the subspace given by $M_A^{\dagger} \H_{A'}$. Then $\dim(U) = \dim(\H_{A'})$ and there exists an invertible map $N_A : U \to \H_{A'}$ such that if $\ket{x} \in X_{\phi}^{(k)} \cap U$, then we have $N_A \ket{x} \in X_{\psi}^{(k)}$.
    \end{lem}

    \begin{proof}
        The fact that the reduced density matrix of $\ket{\psi}$ has full rank on $A'$ means that $M_A$ must be surjective, and hence $M_A^{\dagger}$ is injective.
        Therefore, restricting it to its image $U$, it is invertible and we let $N_A$ be its inverse.
        If $\ket{x} \in X_{\phi}^{(k)} \cap U$, then by \cref{lem:M dagger} we find $N_A \ket{x} \in X_{\psi}^{(k)}$.
    \end{proof}

    Now we specialize to the situation in \cref{eq:no restriction lambda} we are interested in. We assign subsystems as follows:
    \begin{center}
        \vspace*{0.2cm}
        \begin{overpic}[height=1.2cm,grid=false]{folded-lambda-boundary}
            \put(47,12){$\ngeq$}
            \put(-10,5){\scalebox{0.8}{$A_2$}} \put(-5,-5){\scalebox{0.8}{$A_1$}}  \put(32,5){\scalebox{0.8}{$B_2$}} \put(30,-5){\scalebox{0.8}{$B_1$}}
            \put(5,29){\scalebox{0.8}{$C_1$}} \put(18,29){\scalebox{0.8}{$C_2$}}
            \put(60,-2){\scalebox{0.8}{$A'$}} \put(98,-5){\scalebox{0.8}{$B_1'$}} \put(102,5){\scalebox{0.8}{$B_2'$}}
            \put(80,29){\scalebox{0.8}{$C_1'$}} \put(89,27){\scalebox{0.8}{$C_2'$}}
            \put(82,8){$\lambda$}
            \put(1,14){\scalebox{0.8}{$2$}} \put(27,14){\scalebox{0.8}{$8$}} \put(14,4){\scalebox{0.8}{$2$}}
            \put(97,14){\scalebox{0.8}{$3$}}
            \put(10,-10){$\ket{\phi}$} \put(79,-10){$\ket{\psi}$}
        \end{overpic}
        \vspace*{0.5cm}
    \end{center}

    We are now ready to prove \cref{thm:lambda restriction} from the main text:

    \lambdarestriction*

    \begin{proof}
        For our purposes we take $k = 8$ and write $X_{\phi}^{(8)} = X_{\phi}$ and $X_{\psi}^{(8)} = X_{\psi}$.
        We will identify the sets $X_{\phi}$ and $X_{\psi}$ and show that no subspace as in \cref{lem:subspace} can exist, from which we conclude that there can be no restriction.

        First of all, similar to the way we reasoned in the proof of \cref{thm:w restriction}
        \begin{align*}
            \rk(\ket{\phi^{(x)}}) = 8\rk(\ket{x})
        \end{align*}
        for any $\ket{x} \in \H_{A_1} \ot \H_{A_2} = \CC^2 \ot \CC^2$.
        Thus, if $\ket{x} \in X_{\phi}$ it must have $\rk(\ket{x}) \leq 1$ and
        \begin{align*}
            X_{\phi} = \{\ket{x_1} \ket{x_2}, \; \ket{x_i} \in \CC^2 \}.
        \end{align*}
        Next, for $X_{\psi}$, note that
        \begin{align*}
            \rk\mleft(\ket{\psi^{(x)}} \mright) & = \rk\mleft( (\bra{x} \ot \id_{B'C'})\ket{\psi_{A'B'C'}} \mright)                       \\
                                                & = \rk\mleft( (\bra{x} \ot \id_{B_1'C_1'})\ket{\lambda_{A'B_1'C_1'}} \mright)\rk(\epr_3) \\
                                                & = 3 \rk\mleft( (\bra{x} \ot \id_{B_1'C_1'})\ket{\lambda_{A'B_1'C_1'}} \mright).
        \end{align*}
        Therefore, $\ket{x} \in X_{\psi}$ if and only if
        \begin{align*}
            \rk\mleft( (\bra{x} \ot \id_{B_1'C_1'})\ket{\lambda_{A'B_1'C_1'}} \mright) \leq 2.
        \end{align*}
        Suppose
        \begin{align*}
            \ket{x} = x_0 \ket{0} + x_1\ket{1} + x_2\ket{2}
        \end{align*}
        then writing $\lambda^{(x)} = (\bra{x} \ot \id_{B'C_1'})\ket{\lambda_{A'B'C_1'}}$ as a $3 \times 3$ matrix
        \begin{align*}
            \Lambda^{(x)} = \begin{pmatrix} 0 & \overline{x_2} &  -\overline{x_1} \\  -\overline{x_2} & 0 & \overline{x_0} \\  \overline{x_1} &  -\overline{x_0} & \overline{x_2}\end{pmatrix}.
        \end{align*}
        We have $\rk(\lambda^{(x)}) \leq 2$ if and only if this matrix has determinant zero.
        This is the case if and only if $x_2 = 0$ since $\det(\Lambda^{(x)}) = (\overline{x_2})^{3}$.
        We conclude that
        \begin{align*}
            X_{\psi} = \Span\{\ket{0}, \ket{1}\}.
        \end{align*}
        Now we assume that a restriction exists, and therefore we find a subspace $U \subset \H_A = \CC^2 \ot \CC^2$ of dimension $\dim(\H_{A'}) = 3$ from \cref{lem:subspace} (note that the tensors are concise, so the full rank condition is satisfied).
        Since $\dim(\H_A) - \dim(U) = 1$, the subspace $U$ must be described a by a single linear equation.
        We may assume without loss of generality (after a choice of basis on $\H_{A_2}$) that $U$ is given by
        \begin{align*}
            U = \{\ket{x} = & x_{00}\ket{00} + x_{01}\ket{01} + x_{10}\ket{10}              \\
                            & + (ax_{00} + bx_{01} + cx_{10})\ket{11}, \; x_{ij} \in \CC\}.
        \end{align*}
        We claim that $\Span(U \cap X_{\phi}) = U$. Recall that $X_{\phi}$ is the set of product states.
        The following vectors are elements of $U \cap X_{\phi}$:
        \begin{align*}
            \ket{x_1} & = \ket{1}(\ket{0} + c\ket{1})   \\
            \ket{x_2} & = (\ket{0} + b\ket{1})\ket{1}   \\
            \ket{x_3} & = \ket{0}(b\ket{0} - a\ket{1})  \\
            \ket{x_4} & = (c\ket{0} - a\ket{1})\ket{0}.
        \end{align*}
        E.g., for $\ket{x_3}$, we have $x_00 = b$, $x_{01} = -a$, $x_{10} = 0$, and the coefficient of $\ket{11}$ is given by $ax_{00} + bx_{01} + cx_{10} = 0$ so $\ket{x_3} \in U$.
        For any choice of $a,b,c$  for which we do not have $b = c = 0$, these vectors span a three-dimensional space, so $\Span(U \cap X_{\phi}) = U$.
        In the case where $b = c = 0$, the following three vectors in $U \cap X_{\phi}$
        \begin{align*}
            \ket{x_1} & = \ket{1}\ket{0}                          \\
            \ket{x_2} & = \ket{0}\ket{1}                          \\
            \ket{x_3} & = (\ket{0} + \ket{1})(\ket{0} + a\ket{1})
        \end{align*}
        are linearly independent, so $\Span(U \cap X_{\phi}) = U$ in this case as well.
        However, this leads to a contradiction: we saw that $\dim(X_{\psi}) = 2$, but by \cref{lem:subspace} we have a linear invertible map $N_A$ mapping $U \cap X_{\phi}$ (and hence this maps $\Span(U \cap X_{\phi})$) into $X_{\psi}$ but this is not possible if $\dim(\Span(U \cap X_{\phi})) = 3$.
    \end{proof}

    As a consequence of \cref{thm:lambda restriction}, there is no bond dimension $D = 2$ representation for the $\lambda$ tensor on a kagome lattice (\cref{cor:lambda kagome}). However, we proved this with a folding using open boundary conditions.
    We conjecture that the same is true with periodic boundary conditions.
    One way to prove this is to use the same folding as in \cref{cor:w kagome}.
    In this case, it would suffice to show
    \begin{center}
        \begin{overpic}[height=1.2cm,grid=false]{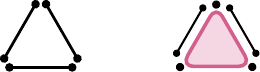}
            \put(47,12){$\ngeq$}
            \put(81,8){$\lambda$}
            \put(1,14){\scalebox{0.8}{$8$}} \put(27,14){\scalebox{0.8}{$8$}} \put(14,-6){\scalebox{0.8}{$2$}}
            \put(97,14){\scalebox{0.8}{$3$}} \put(66,14){\scalebox{0.8}{$3$}}
        \end{overpic}
        \vspace*{0.1cm}
    \end{center}
    We have obtained numerical evidence that there is indeed no such restriction, but were unable to give a complete proof, so we leave this as a conjecture.

    \section{Tensor rank of two copies of the \texorpdfstring{$W$}{W} state}\label{sec:rank w}
    The goal of this section is to prove \cref{thm:rank 2 w}, which determines the tensor rank of two copies of the $W$ state
    \begin{equation*}
        \ket{W} = \ket{001} + \ket{010} + \ket{100} \in \left( \mathbb{C}^{2} \right)^{\otimes 3}
    \end{equation*}
    on the hypergraph $G_4$.

    \rankw*

    \begin{proof}
        Our proof strategy is inspired by \cite{chen2018tensor}. We consider $\ket{W}_{G_4}$ with subsystems $A$, $B$, $C$, $D$ as follows:
        \begin{center}
            \vspace*{0.2cm}
            \begin{overpic}[height=1.5cm,grid=false]{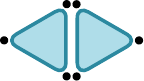}
                \put(-15,25){$A$} \put(105,25){$D$} \put(45,60){$B$} \put(45,-15){$C$}
            \end{overpic}
            \vspace*{0.2cm}
        \end{center}

        Suppose that $\ket{W}_{G_4}$ has rank 7 or less, then there must exist a decomposition
        \begin{align*}
            \ket{W}_{G_4} = \sum_{i=1}^7 \ket{a_i} \ket{b_i} \ket{c_i} \ket{d_i}
        \end{align*}
        with $\ket{a_i},\ket{d_i} \in \CC^2$ and $\ket{b_i},\ket{c_i} \in \CC^2 \ot \CC^2$.
        Since $(\id_{ACD} \ot \bra{11}_B)\ket{W}_{G_4} \neq 0$ we may assume without loss of generality that $\braket{11 | b_{7}} = 1$.
        So, we expand
        \begin{align*}
            \ket{b_7} = x_{00}\ket{00} + x_{01} \ket{01} + x_{10}\ket{10} + \ket{11}
        \end{align*}
        for some $x_{ij} \in \CC$.
        We will now use a symmetry of a single copy of the $W$ state.
        If we define
        \begin{align*}
            h(p) : \CC^2 & \to \CC^2                                           \\
            \ket{0}      & \mapsto \ket{0}, \ket{1} \mapsto \ket{1} + p\ket{0}
        \end{align*}
        then
        \begin{align*}
            (h(p) \ot h(q) \ot h(r))\ket{W} = \ket{W}
        \end{align*}
        if $p + q + r = 0$.
        We now apply the map
        \begin{align*}
            \id_A \ot (h(-x_{01}) \ot h(-x_{10}))_B \ot (h(x_{01}) \ot h(x_{10}))_C \ot \id_D
        \end{align*}
        to $\ket{W}_{G_4}$. It leaves the state invariant, but we get a new decomposition
        \begin{align*}
            \ket{W}_{G_4} = \sum_{i=1}^7 \ket{a'_i} \ket{b'_i} \ket{c'_i} \ket{d'_i}
        \end{align*}
        which now has
        \begin{align*}
            \ket{b'_7} = x_{00}'\ket{00} + \ket{11}
        \end{align*}
        for some $x_{00}' \in \CC$.
        Next, we use the symmetry given by
        \begin{align*}
            \id_A \ot g(-q)_B \ot g(q)_C \ot \id_D
        \end{align*}
        for $g(q)$ with any $q \in \CC$ as in \cref{eq:symmetry W 4}.
        This allows us, using $q = x_{00}'$, to transform to a decomposition
        \begin{align*}
            \ket{W}_{G_4} = \sum_{i=1}^7 \ket{a_i''} \ket{b_i''} \ket{c_i''} \ket{d_i''}
        \end{align*}
        with
        \begin{align*}
            \ket{b_7''} = \ket{11}.
        \end{align*}
        Next, we observe that
        \begin{align*}
            (\id_{ACD} \ot (\bra{0} \ot \id)_B) \ket{W}_{G_4} = \ket{\phi}\ket{W}
        \end{align*}
        where $\ket{\phi} = \ket{01} + \ket{10}$.

        \begin{center}
            \vspace*{0.3cm}
            \begin{overpic}[height=1.2cm,grid=false]{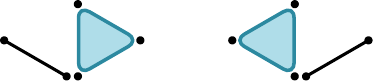}
                \put(12,-15){$\ket{\phi}\ket{W}$}
                \put(70,-15){$\ket{W}\ket{\phi}$}
                \put(-7,9){$A$} \put(42,10){$D$} \put(18,23){$B$} \put(18,-7){$C$}
                \put(53,10){$A$} \put(103,10){$D$} \put(76,23){$B$} \put(76,-7){$C$}
            \end{overpic}
            \vspace*{0.5cm}
        \end{center}
        Similarly,
        \begin{align*}
            (\id_{ACD} \ot (\id \ot \bra{0})_B) \ket{W}_{G_4} = \ket{W}\ket{\phi}.
        \end{align*}
        By applying the same map to our rank 7 decomposition, we see that these tensors have rank at most 6 (since $\ket{b_7''} = \ket{11}$):
        \begin{align*}
            \ket{\phi}\ket{W} & = \sum_{i=1}^6  \ket{a_i''} (\bra{0} \ot \id) \ket{b_i''} \ket{c_i''} \ket{d_i''}  \\
            \ket{W}\ket{\phi} & = \sum_{i=1}^6  \ket{a_i''} (\id \ot \bra{0}) \ket{b_i''} \ket{c_i''} \ket{d_i''}.
        \end{align*}
        Now, grouping $D$ together with $B$ turns $\ket{W}\ket{\phi}$ into a 3-tensor
        \begin{center}
            \vspace*{0.3cm}
            \begin{overpic}[height=1.2cm,grid=flase]{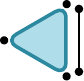}
                \put(-30,40){$A$} \put(95,105){$BD$} \put(95,-20){$C$}
            \end{overpic}
            \vspace*{0.1cm}
        \end{center}
        of rank 6 (this can be shown using the substitution method).
        In particular, the vectors $\ket{\alpha_i} = \ket{a_i''} \ot \ket{c_i''}$ for $i=1,\dots,6$ are linearly independent.
        We claim that the vectors $\ket{\beta_i} = (\bra{0} \ot \id)\ket{b_i''} \ot \ket{d_i''}$ for $i=1,\dots,6$ must span at least a dimension 3 subspace.
        To see this, from the decomposition of $\ket{\phi}\ket{W}$ and letting
        \begin{align*}
            \ket{\tilde \alpha_i} = (\bra{\phi} \ot \id)\ket{\alpha_i}
        \end{align*}
        on $C$ (since $(\bra{\phi} \ot \id)(\ket{\phi} \ket{W}) = \ket{W}$) we obtain a decomposition
        \begin{align*}
            \ket{W} = \sum_{i=1}^6 \ket{\tilde \alpha_i} \ket{\beta_i}
        \end{align*}
        where the $\ket{\beta_i}$ are product states on $BD$.
        This implies that if there were at most two linearly independent elements amongst the $\ket{\beta_i}$, $\ket{W}$ would have tensor rank at most two, but we know that $\R(\ket{W}) = 3$.
        Now we consider $\ket{\phi}\ket{W}$ as a bipartite tensor between $AC$ and $BD$. It is clear that it has rank 2:
        \begin{center}
            \vspace*{0.3cm}
            \begin{overpic}[height=1.2cm,grid=false]{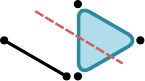}
                \put(-17,25){$A$} \put(105,25){$D$} \put(45,60){$B$} \put(45,-16){$C$}
            \end{overpic}
            \vspace*{0.2cm}
        \end{center}

        However, we can write
        \begin{align*}
            \ket{\phi} \ket{W}= \sum_{i = 1}^6 \ket{\alpha_i}_{AC} \ot \ket{\beta_i}_{BD}.
        \end{align*}
        The existence of three independent vectors amongst the $\ket{\beta_i}$, and the linear independence of the $\ket{\alpha_i}$ implies that this has rank at least 3 as a 2-tensor, which is a contraction with the fact that it has rank 2.
    \end{proof}

    \section{Hardness of tensor network contraction}\label{sec:vnp}

    We will now discuss a hardness result for the complexity of tensor network contraction in the arithmetic circuit model.
    Recall that in this model of computation, the goal is to compute some polynomial through an arithmetic circuit.
    We will first define the complexity classes $VP$ and $VNP$.
    A sequence of functions $f_n$, $n = 1,2,\dots$ is in $VP$ if $f_n$ are polynomials in a number of variables and with a degree that is polynomial in $n$, and moreover there exists a family of arithmetic circuits of polynomial size in $n$ computing $f_n$.
    A representative example is the $n \times n$ \emph{determinant} polynomial.
    Next we define the class of $VNP$, which is the analog of $NP$.
    A sequence of functions $f_n$, $n = 1,2,\dots$ is in $VNP$ if $f_n$ has a number of variables $v(n)$ and degree which are polynomial in $n$, and moreover there exists a number $m(n)$ and a sequence of polynomials $g_n$ with $v(n) + m(n)$ variables such that $g_n$ is in $VP$ and $f_n(x_1,\dots,x_{v(n)})$ can be computed as
    \begin{align*}
        \sum_{y \in \{0,1\}^{m(n)}} g_n(x_1,\dots,x_{v(n)},y_1,\dots,y_{m(n)}).
    \end{align*}
    The paradigmatic example is the $n \times n$ \emph{permanent} polynomial. It is conjectured that $VP \neq VNP$.
    There are connections to the problem of $P$ versus $NP$; for instance it is known that $VP = VNP$ over a field of characteristic zero, together with the generalized Riemann hypothesis, would imply $P/ \poly = NP/ \poly$, see Corollary 4.6 in \cite{burgisser2000completeness}.
    We now formally state our result on the hardness of tensor network contraction, showing that tensor network contraction is $VNP$-complete.

    \begin{thm}\label{thm:hardness}
        Let $f_{\ket{\phi}_G}$ be defined as in \cref{eq:tensor network coeff} for an arbitrary hypergraph $G$ and entanglement structure $\ket{\phi}_G$. We denote by \begin{align*}
            f_n = f_{\ket{\phi}_{G_n}}
        \end{align*}
        the family of polynomials where we take $G_n$ to be the $n \times n$ square lattice graph (so each edge is a 2-edge), and the entanglement structure $\ket{\phi}_{G_n}$ is given by placing level-2 $\epr$ pairs on each edge.
        \begin{enumerate}
            \item\label{it:vnp} The problem of computing tensor network contraction coefficients, given by the polynomial $f_{\ket{\phi}_G}$ on hypergraphs with $n$ edges and constant degree, and local Hilbert spaces of polynomial size in $n$ with an arbitrary entanglement structure $\ket{\phi}_G$, is in $VNP$.
            \item\label{it:vnp hard} The problem of computing tensor network contraction coefficients with bond dimension $D = 2$ on a square lattice, given by the polynomial $f_n$, is $VNP$-hard.
        \end{enumerate}
    \end{thm}

    \begin{proof}
        Note that in \cref{eq:tensor network coeff}, the argument of $f_{\ket{\phi}_{G_n}}$ consists of $(T_v)_{v \in V}$, and each $T_v$ consists of $\dim(\H_v)$ variables, so if all local Hilbert spaces have polynomial dimension and the vertices have constant degree, then $f_{\ket{\phi}_{G_n}}$ has as polynomial number of variables and polynomial degree (it is linear in each of the variables).
        Given a hypergraph $G$ with $n$ edges, and plaquette states $\ket{\phi_e}$, for edges $e \in E$ on Hilbert spaces of polynomial dimension, we can find a restriction $\ket{\psi}_G \geq \ket{\phi}_G$ with plaquette states $\ket{\psi_e}$ on edge Hilbert spaces of polynomial dimension and which consist of a collection of level-2 $\epr$ pairs.
        By the argument in \cref{thm:restriction vs complexity} this implies that it suffices to prove the result for graphs $G$ with only 2-edges, with edge states $\ket{\phi_e}$ which are level-2 $\epr$ pairs.
        For each edge $e \in E$ we may define
        \begin{align*}
            \ket{\phi_{e}(y_e)} = \left((1 - y_e)\ket{0} + y_e\ket{1}\right)^{\ot 2}
        \end{align*}
        where $y_e$ is a variable on the edge $e$.
        We let $\ket{\phi(y)}_{G_n}$ be the entanglement structure where we place these states on $G_n$, which is a state depending on the variables $y = (y_e)_{e \in E}$.
        Then we define the following polynomial:
        \begin{align*}
            g_{\ket{\phi}_G}((T_v)_{v \in V}, (y_e)_{e \in E}) = \left(\bigotimes_{v \in V} T_{v} \right) \ket{\phi(y)}_{G_n}.
        \end{align*}
        The $g_{\ket{\phi}_G}$ are in $VP$, since they can be computed as
        \begin{align*}
            \prod_{v \in V}\left( T_v \bigotimes_{e, v \in e} \left((1 - y_e)\ket{0} + y_e\ket{1}\right) \right)
        \end{align*}
        i.e. since the edge states are product states, the computation factorizes over the vertices.
        Moreover, the tensor network contraction $f_{\ket{\phi}_G}((T_v)_{v \in V})$ is computed by
        \begin{align*}
            \sum_{(y_e \in \{0,1\})_{e \in E}} g_{\ket{\phi}_G}((T_v)_{v \in V}, (y_e)_{e \in E})
        \end{align*}
        so we conclude $f_{\ket{\phi}_G}$ is in $VNP$, proving \ref{it:vnp}.

        We proceed to prove \ref{it:vnp hard}.
        We will do so by reducing the computation of a partition function of matchings on a square lattice graph to the tensor network contraction $f_n$.
        We will then use that computing the partition function of weighted (non-perfect) matchings on a square lattice graph is $VNP$-hard.
        That is, one considers the graph $G_n$. A matching is a subset of edges $M \subseteq E$ such that the edges in $M$ have no common vertices.
        We let
        \begin{align*}
            x = \{x(e)\}_{e \in E}
        \end{align*}
        be set of variables assigned to the edges $e \in E$. We then define a partition function as
        \begin{align*}
            Z_n(x) = \sum_{M} \prod_{e \in M} x(e)
        \end{align*}
        where the sum is over all matchings $M$ of $G_n$.
        By Theorem 3.7 in \cite{burgisser2000completeness}, based on \cite{jerrum1987two}, computing this polynomial is $VNP$-hard.
        We now argue that it can be computed as a tensor network contraction.
        To this end, we need to assign the edge variables to vertices.
        We will do so by assigning them to the vertex to the right or to the vertex below.
        We can then enforce the matching constraint locally at the vertices.
        Consider a vertex $v$ which has edges $e_1$, $e_2$, $e_3$, $e_4$ (starting from the left edge in clockwise order).
        We then define the tensor $T_v$ as follows:
        \begin{align*}
            T_v(x_{e_1},x_{e_2}) & = x_{e_1}\bra{1}_{e_1}\bra{0}_{e_2}\bra{0}_{e_3}\bra{0}_{e_4}        \\
                                 & \qquad + x_{e_2}\bra{0}_{e_1}\bra{1}_{e_2}\bra{0}_{e_3}\bra{0}_{e_4} \\
                                 & \qquad + \bra{0}_{e_1}\bra{0}_{e_2}\bra{1}_{e_3}\bra{0}_{e_4}        \\
                                 & \qquad + \bra{0}_{e_1}\bra{0}_{e_2}\bra{0}_{e_3}\bra{1}_{e_4}        \\
                                 & \qquad + \bra{0}_{e_1}\bra{0}_{e_2}\bra{0}_{e_3}\bra{0}_{e_4}
        \end{align*}
        or in diagrammatic notation
        \begin{center}
            \begin{overpic}[height=2.7cm,grid=false]{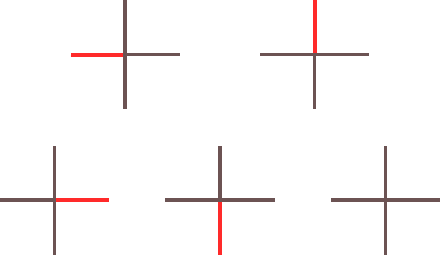}
                \put(6,44){$x_{e_1}$} \put(43,44){$+ x_{e_2}$}
                \put(-8,10.5){$+$} \put(28,10.5){$+$} \put(67,10.5){$+$}
                \put(18,47){\scalebox{0.5}{$1$}} \put(30,53){\scalebox{0.5}{$0$}} \put(37,41.5){\scalebox{0.5}{$0$}} \put(25,35){\scalebox{0.5}{$0$}}
                \put(62,47){\scalebox{0.5}{$0$}} \put(73,53){\scalebox{0.5}{$1$}} \put(80,41.5){\scalebox{0.5}{$0$}} \put(68,35){\scalebox{0.5}{$0$}}
                \put(2,14){\scalebox{0.5}{$0$}} \put(14,20){\scalebox{0.5}{$0$}} \put(22,8){\scalebox{0.5}{$1$}} \put(9,2){\scalebox{0.5}{$0$}}
                \put(40,14){\scalebox{0.5}{$0$}} \put(51.5,20){\scalebox{0.5}{$0$}} \put(59,8){\scalebox{0.5}{$0$}} \put(47,2){\scalebox{0.5}{$1$}}
                \put(78,14){\scalebox{0.5}{$0$}} \put(89,20){\scalebox{0.5}{$0$}} \put(96,8){\scalebox{0.5}{$0$}} \put(84,2){\scalebox{0.5}{$0$}}
            \end{overpic}
        \end{center}
        We find that in the resulting tensor network contraction, we sum over all assignments of $0,1$ to the edges where there is at most one $1$ neighboring each vertex, and we assign variable $x_e$ to each edge which is labelled by $1$:
        \begin{center}
            \begin{overpic}[height=2.5cm,grid=false]{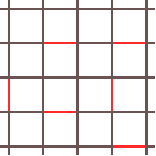}
                \put(-60,45){$\sum\limits_{\text{matchings}}$}
                \put(35,77){\scalebox{0.7}{$x_{e_1}$}} \put(79,77){\scalebox{0.7}{$x_{e_2}$}}
                \put(10,37){\scalebox{0.7}{$x_{e_3}$}} \put(76,37){\scalebox{0.7}{$x_{e_4}$}}
                \put(35,33){\scalebox{0.7}{$x_{e_5}$}}
                \put(79,10){\scalebox{0.7}{$x_{e_6}$}}
            \end{overpic}
        \end{center}
        The resulting polynomial over the variables $x_e$ is precisely $Z_n(x)$.
    \end{proof}

\end{appendix}

\end{document}